\documentclass[10pt]{amsart}

\usepackage{amsmath,amsfonts,amscd,amssymb,epsf, epsfig}
\usepackage[small]{caption}

\theoremstyle{definition}

\theoremstyle{remark} 
\numberwithin{equation}{section}

 \DeclareMathOperator{\Res}{Res}
\DeclareMathOperator{\PP}{PP}

\newcommand{\Z}{{\mathbb{Z}}}

\newcommand{\C}{\mathbb{C}}

\newcommand{\R}{\mathbb{R}}

\newcommand{\pa}{\partial}

\begin{document}

\title[Self--interacting fractional Klein--Gordon field]{Topological symmetry breaking of
self--interacting fractional Klein--Gordon field on toroidal
spacetime }

\author{S.C. Lim$^1$}\email{$^1$sclim@mmu.edu.my}\author{L.P.
Teo$^{2}$}\email{$^2$lpteo@mmu.edu.my}

\maketitle

\noindent {\scriptsize \hspace{1cm}$^1$Faculty of Engineering,
Multimedia University, Jalan Multimedia, }

\noindent {\scriptsize \hspace{1.1cm} Cyberjaya, 63100, Selangor
Darul Ehsan, Malaysia.}

\noindent {\scriptsize \hspace{1cm} $^2$Faculty of Information
Technology, Multimedia University, Jalan Multimedia,}

\noindent{\scriptsize \hspace{1.1cm} Cyberjaya, 63100, Selangor
Darul Ehsan, Malaysia.}
\begin{abstract}
Quartic self--interacting fractional Klein--Gordon scalar massive
and massless field theories on toroidal spacetime are studied. The
effective potential and topologically generated mass are determined
using zeta function regularization technique. Renormalization of
these quantities are derived. Conditions for symmetry breaking are
obtained analytically. Simulations are carried out to illustrate
regions or values of compactified dimensions where symmetry breaking
mechanisms appear.

\vspace{0.3cm}

\end{abstract}

\section{Introduction}
The concept of fractal has permeated virtually all branches of
natural sciences since its was first introduced by Mandelbrot about
three decades ago \cite{c1}. The first fractal process encountered
in physics is the Brownian motion, whose paths have been used in
Feynman path integral approach to (Euclidean) quantum mechanics
\cite{c2}. Based on the path integral method, Abbot and Wise
\cite{c3} showed that the quantum trajectories of a point-like
particle is a fractal of Hausdorff dimension two. Brownian motion
also played an important role in stochastic mechanics \cite{n1, n2},
which was an attempt to give an alternative formulation to quantum
mechanics. Early applications of fractal geometry in quantum field
theory focused mainly on the studies of quantum field models in
fractal sets and fractal spacetime, and quantum field theory of spin
systems such as Ising spin model (see \cite{c4} for a review on
fractal geometry in quantum theory). The applications were
subsequently extended to fractal Wilson loops in lattice gauge
theory \cite{c5}, and fractal geometry of random surfaces in quantum
gravity \cite{c6}. There exist models of quantum gravity such as
Quantum Einstein Gravity model which predicts that spacetime is
fractal with fractal or Hausdorff dimension two at sub-Planckian
distance \cite{c7}.

    The next important step in the applications of fractal
geometry in physics is the realization of the close connection
between fractional calculus \cite{c8, c9, c10, c11} and processes
and phenomena which exhibit fractal behavior. Such an association
allows the use of fractional differential equations to describe
fractal phenomena. Applications of fractional differential equations
in physics have spread rapidly, in particular in condensed matter
physics, where fractional differential equations are well-suited to
describe anomalous transport processes such as anomalous diffusion,
non-Debye relaxation process, etc \cite{c12,c13, c14, c15, c16}.
More recently, such applications have been extended to quantum
mechanics. Analogous to the fractional diffusion equations, various
versions of fractional Schr\"odinger equations (the
space--fractional, time--fractional and spacetime--fractional
Schr\"odinger equations) have been studied \cite{c17, c18, c19,c20,
c21, c22, c23}. Based on the fractional Euler-Lagrange equation in
the presence of Grassmann variables, Baleanu and Muslih \cite{c24}
have considered supersymmetric quantum mechanics using the path
integral method.

    It is interesting to note that the works on fractional Klein--Gordon
equation have been carried out nearly a decade before that on
fractional Schr\"odinger equations. The square--root and cubic--root
Klein--Gordon equations, Klein--Gordon equation of arbitrary
fractional order, and fractional Dirac equation have been studied by
various authors \cite{c25, c26, c27, c28, c29, c30}. Canonical
quantization of fractional Klein--Gordon field has been considered
by Amaral and Marino \cite{c31}, Barcci, Oxman and Rocca \cite{c32},
and stochastic quantization of fractional Klein--Gordon field and
fractional abelian gauge field have been studied by Lim and Muniandy
\cite{c33}. There are also works in constructive field theory
approach to fractional Klein--Gordon field, where the analytic
continuation of the Euclidean (Schwinger) $n$-point functions to the
corresponding $n$-point Wightman functions are studied \cite{a,b}.
More recently, results on finite temperature fractional
Klein--Gordon field \cite{c34}, and the Casimir effect associated
with the massive and massless fractional fields at zero and finite
temperature with fractional Neumann boundary conditions have been
obtained \cite{c35}. We would like to point out that until now all
these studies consider only free fractional fields. Therefore it
would be interesting and important to study a simple model of
interacting fractional field. This is exactly the main objective of
our paper.

    In this paper we consider for the first time the model
of   scalar massive and massless fractional Klein--Gordon fields
with quartic self--interaction. It is well known that in the
ordinary field theory, $\varphi^4$  theory is an important and
useful model because it has applications in Weinberg-Salam model of
weak interactions \cite{c36}, inflationary models of early universe
\cite{c37}, solid state physics \cite{c38} and soliton theory
\cite{c39}, etc. In addition, it is also known that a massless field
can develop a mass as a result of both self--interaction and
nontrivial spacetime topology, and such a phenomenon is known as
topological mass generation \cite{c40, c41, c42}. The main aim of
this paper is to study the possibilities of topological mass
generation and symmetry breaking mechanism for a fractional scalar
field with interaction in a toroidal spacetime. In the case of
ordinary quantum fields, topological mass generation in toroidal
spacetime has been studied by Actor \cite{c43}, Kirsten \cite{c44},
Elizalde and Kirsten \cite{EK1} by using the zeta function
regularization technique \cite{a1, E1, ET, K}. We shall show that
with some modifications, the zeta function method can also be
employed to study the  topological mass generation and symmetry
breaking mechanism in the fractional $\varphi^4$ theory.

    In section 2 we discuss the fractional scalar
Klein--Gordon massive and massless fields with quartic
self-interaction on the toroidal spacetime. The effective potential
of this fractional $\varphi^4$  model is determined up to one loop
quantum effects using the zeta function regularization method.
Section 3 contains the renormalization of the effective potential.
The derivation of the renormalized topologically generated mass and
symmetry breaking mechanism will be given in section 4. The final
section gives a summary of main results obtained, and perspective
for further work. We also include  simulations to illustrate the
dependency of symmetry breaking mechanism and renormalized
topologically generated mass on spacetime dimensions, fractional
order of the Klein--Gordon field, etc.

\section{One--loop effective potential of  fractional scalar field with $\lambda \varphi^4$ interaction}

In this section, we compute the one loop effective potential of the
real  fractional scalar Klein--Gordon field with $\lambda \varphi^4$
interaction in a $d$ -- dimensional spacetime. In this paper, the
spacetime we consider is the toroidal manifold $T^{p}\times T^{q}$,
$q:=d-p$, with compactification lengths $L_1, \ldots, L_p, L_{p+1},
\ldots, L_d$, where $L_{p+1}=\ldots=L_d=L$ and $L_i, 1\leq i\leq p$
are assumed to be much smaller than $L$. We will take the limit
$L\rightarrow \infty$, which results in the limiting toroidal
spacetime $T^p\times\R^{q}$. In this spacetime, the scalar field
$\varphi(\mathbf{x})$ can be regarded as a function of
$\mathbf{x}\in \R^{d}$ which satisfies the periodic boundary
conditions with period $L_j$, $1\leq j\leq d$, in the $x_j$
direction. The Lagrangian of the theory is
\begin{align*}
\mathcal{L}=-\frac{1}{2}\varphi(\mathbf{x})\left(-\Delta+m^2\right)^{\alpha}\varphi(\mathbf{x})
-\frac{\lambda}{4!}\varphi(\mathbf{x})^4, \hspace{1cm}\alpha>0,
\end{align*}where $\Delta$ is the Laplace operator
$\Delta=\frac{\pa^2}{\pa x_1^2}+\ldots+\frac{\pa^2}{\pa x_{d}^2}$.
 For a function
$$f(\mathbf{x})=\sum_{\mathbf{k}\in\mathbb{Z}^{d}} a_\mathbf{k}
e^{2\pi i \sum_{j=1}^{d} \frac{k_j x_j}{L_j}}$$expanded with respect
to the basis $\left\{e^{2\pi i \sum_{j=1}^{d} \frac{k_j
x_j}{L_j}}\,:\, \mathbf{k}\in \Z^d\right\}$ of functions on the
toroidal spacetime $T^p\times T^{q}$, the fractional differential
operator $(-\Delta+m^2)^{\alpha}$ acts on $f$  by the formula
\begin{align*} \left[(-\Delta+m^2)^{\alpha} f\right](x) =
\sum_{\mathbf{k}\in\mathbb{Z}^{d}} a_\mathbf{k}
\left(\sum_{j=1}^{d}\left[\frac{2\pi
k_j}{L_j}\right]^2+m^2\right)^{\alpha} e^{2\pi i \sum_{j=1}^{d}
\frac{k_j x_j}{L_j}}.
\end{align*}
The partition function of the theory is given by
\begin{align*}
\mathcal{Z}=\int\mathcal{D}\varphi(\mathbf{x})
\exp\left(-\int\limits_{T^p\times
T^{q}}\left\{\frac{1}{2}\varphi(\mathbf{x})\left(-\Delta+m^2\right)^{\alpha}\varphi(\mathbf{x})
+\frac{\lambda}{4!}\varphi(\mathbf{x})^4\right\}d^{d}\mathbf{x}\right).
\end{align*}In the toroidal spacetime, we can assume a constant classical background field
$\widehat{\varphi}$. The quantum fluctuations around this background
field is defined to be $\phi=\varphi-\widehat{\varphi}$. Then to the
one--loop order, we have
\begin{align*}
\log\mathcal{Z}=-\frac{1}{2}m^{2\alpha}\widehat{\varphi}^2-\frac{\lambda}{4!}\widehat{\varphi}^4-V_Q,
\end{align*}where $V_{Q}$ is the functional determinant (called the quantum
potential),
\begin{align*}
V_Q=\frac{1}{2 \mathcal{V}_{d} }\log\det
\left(\frac{(-\Delta+m^2)^{\alpha}+\frac{1}{2}\lambda\widehat{\varphi}^2}{\mu^2}\right).
\end{align*}
 Here $\mathcal{V}_{d}=L^{q}\prod_{i=1}^p
L_i=L^{q}\mathcal{V}_p$ is the volume of spacetime and $\mu$ is a
scaling length. The effective potential including one--loop quantum
effects is then given by
\begin{align*}
V_{\text{eff}}(\widehat{\varphi})
=\frac{1}{2}m^{2\alpha}\widehat{\varphi}^2+\frac{\lambda}{4!}\widehat{\varphi}^4+V_Q.\end{align*}
To calculate $V_Q$, we use the zeta function prescription \cite{a1,
E1, ET, K}. By taking $L\rightarrow \infty$ limit, $V_Q$ is equal to
\begin{align}\label{eq830_3}
V_Q=\frac{1}{2(2\pi)^{q}\mathcal{V}_p}\left[\zeta(0)\log\mu^2-\zeta'(0)\right],
\end{align}where the zeta function $\zeta(s)$ is defined as
\begin{align}\label{eq723_1}\zeta(s)=&
\int\limits_{\R^{q}}\sum_{\mathbf{k}\in
\mathbb{Z}^p}\left\{\left(|w|^2+\sum_{i=1}^p\left[\frac{2\pi
k_i}{L_i}\right]^2+m^2\right)^{\alpha}
+\frac{\lambda}{2}\widehat{\varphi}^2\right\}^{-s}d^{q}
w\\=&\frac{2\pi^{\frac{q}{2}}}{\Gamma\left(\frac{q}{2}\right)}\int_0^{\infty}w^{q-1}\sum_{\mathbf{k}\in
\mathbb{Z}^p}\left\{\left(w^2+\sum_{i=1}^p\left[\frac{2\pi
k_i}{L_i}\right]^2+m^2\right)^{\alpha}
+\frac{\lambda}{2}\widehat{\varphi}^2\right\}^{-s}dw\nonumber,
\end{align}when $\text{Re}\;s>d/(2\alpha)$. When $p=d$ or equivalently when $q=0$, we understand
that
$$\frac{2\pi^{\frac{q}{2}}}{\Gamma\left(\frac{q}{2}\right)}\int_0^{\infty}
w^{q-1} f(w) dw=f(0).$$ We need to find an analytic continuation of
$\zeta(s)$ to evaluate $\zeta(0)$ and $\zeta'(0)$. Let
$$a_i=\frac{2\pi}{L_i}, \;\;\;\;1\leq i\leq
p,\hspace{1cm}\text{and}\hspace{1cm}b=\sqrt{\frac{\lambda}{2}}\widehat{\varphi}.$$
Using standard techniques, we
have\begin{align}\label{eq1_8_4}\zeta(s)
=&\frac{1}{\Gamma(s)}\int_0^{\infty}t^{s-1}K(t)dt,
\end{align}where $$K(t):=\frac{2\pi^{\frac{q}{2}}}{\Gamma\left(\frac{q}{2}\right)}\int_0^{\infty}w^{q-1}dw
\sum_{\mathbf{k}\in\mathbb{Z}^p
}\exp\left(-t\left\{\left(w^2+\sum_{i=1}^p
\left[a_ik_i\right]^2+m^2\right)^{\alpha}+b^2\right\}\right)$$ is
called the \emph{global} heat kernel. Now we have  to find the
asymptotic behavior of $K(t)$ when $t\rightarrow 0$. For this
purpose, we rewrite
\begin{align}
\label{eq1_8_11}K(t)=A(t)e^{-tb^2},
\end{align}where
\begin{align*}
A(t):=&\frac{2\pi^{\frac{q}{2}}}{\Gamma\left(\frac{q}{2}\right)}\int_0^{\infty}w^{q-1}dw
\sum_{\mathbf{k}\in\mathbb{Z}^p
}\exp\left(-t\left\{\left(w^2+\sum_{i=1}^p
\left[a_ik_i\right]^2+m^2\right)^{\alpha}\right\}\right);
\end{align*} and employ the Mellin--Barnes
integral representation of exponential function (see e.g.
\cite{EKZ})
\begin{align}\label{eq1_8_5}
e^{-z}=\frac{1}{2\pi i}\int_{u-i\infty}^{u+i\infty} dv
\Gamma(v)z^{-v}, \hspace{1cm} u\in \R^+,
\end{align}to $A(t)$. However, in the massless (i.e. $m=0$) case,  the $\mathbf{k}=\mathbf{0}$ term has
to be treated differently. Therefore, we discuss the results for the
massive ($m>0$) case and the massless ($m=0$) case separately.

\subsection{The massive case $m>0$}
Using \eqref{eq1_8_5}, we have
\begin{align}\label{eq1_8_6_1}
A(t)
=&\frac{2\pi^{\frac{q}{2}}}{\Gamma\left(\frac{q}{2}\right)}\int_0^{\infty}w^{q-1}dw\frac{1}{2\pi
i}\int_{u-i\infty}^{u+i\infty} dv \Gamma(v)
t^{-v}\sum_{\mathbf{k}\in\mathbb{Z}^p }\left(w^2+\sum_{i=1}^p
\left[a_ik_i\right]^2+m^2\right)^{-\alpha v}\\
=& \frac{\pi^{\frac{q}{2}}}{2\pi i}\int_{u-i\infty}^{u+i\infty} dv
\Gamma(v) t^{-v}\frac{\Gamma\left(\alpha
v-\frac{q}{2}\right)}{\Gamma(\alpha v)}Z_{E,p}\left(\alpha
v-\frac{q}{2}; a_1, \ldots, a_p; m\right),\nonumber\end{align} with
$u>\frac{d}{2\alpha}$. Here for a positive integer $p$ and positive
real numbers, $a_1,\ldots, a_p$, $m$, $ Z_{E,p}(s; a_1,\ldots, a_p;
m)$ is the inhomogeneous Epstein zeta function defined by
\begin{align*}
Z_{E,p}(s; a_1, \ldots,
a_p;m)=\sum_{\mathbf{k}\in\mathbb{Z}^p}\left(\sum_{i=1}^p
[a_ik_i]^2+m^2\right)^{-s}
\end{align*}when $\text{Re}\, s>\frac{p}{2}$. For $p=0$, we use the convention $Z_{E,0}(s;m)=m^{-2s}$. Some facts
about the function $Z_{E,p}(s; a_1, \ldots, a_p;m)$ are summarized
in appendix A. In particular,  $\Gamma(s)Z_{E,p}(s;a_1,\ldots,
a_p;m)$ has simple poles at $s=\frac{p}{2}-j$,
$j\in\mathbb{N}\cup\{0\}$, with residues
\begin{align*}
\text{Res}_{s=\frac{p}{2}-j}\left\{\Gamma(s)Z_{E,p} (s;a_1,\ldots,
a_p;
m)\right\}=\frac{(-1)^j}{j!}\frac{\pi^{\frac{p}{2}}}{\left[\prod_{i=1}^p
a_i\right]}m^{2j}.
\end{align*} When $p=0$, this formula is
still valid, where $\prod_{i=1}^p a_i$ is understood as $1$.
Applying residue calculus to \eqref{eq1_8_6_1}, we find that when
$t\rightarrow 0$,
\begin{align}\label{eq1_8_7}
A(t)\sim \frac{\pi^{\frac{d}{2}}}{\left[\prod_{i=1}^p
a_i\right]}\sum_{j=0}^{\left[\frac{d}{2}\right]}\frac{(-1)^j}{j!}\frac{\Gamma\left(\frac{d-2j}{2\alpha}\right)}{
\alpha\Gamma\left(\frac{d}{2}-j\right)}m^{2j}t^{-\frac{d-2j}{2\alpha}}+O\left(t^{\frac{1}{2\alpha}}\right).
\end{align}Here $[x]$ denotes the largest integer not more than $x$, and we understand that
\begin{align*}
\left.\frac{\Gamma(z/\alpha)}{\Gamma(z)}\right|_{z=0}=\lim_{z\rightarrow
0}\frac{\Gamma(z/\alpha)}{\Gamma(z)}=\alpha.
\end{align*}
From \eqref{eq1_8_11} and \eqref{eq1_8_7}, we have
\begin{align}\label{eq1_8_9}
 K(t)=A(t)e^{-tb^2}\sim
 B(t)+O(t^{\frac{1}{2\alpha}})\hspace{1cm}\text{as}\;\;t\rightarrow 0,
\end{align} where\begin{align*}
B(t)=&\frac{\pi^{\frac{d}{2}}}{\left[\prod_{i=1}^p
a_i\right]}\sum_{j=0}^{\left[\frac{d}{2}\right]}\frac{(-1)^j}{j!}\frac{\Gamma\left(\frac{d-2j}{2\alpha}\right)}{
\alpha\Gamma\left(\frac{d}{2}-j\right)}m^{2j}t^{-\frac{d-2j}{2\alpha}}e^{-tb^2}.
\end{align*}Now  $\zeta(s)$ given by \eqref{eq1_8_4} can be rewritten  as
\begin{align*}
\zeta(s) =&\frac{1}{\Gamma(s)}\left(\int_0^{\infty} t^{s-1} B(t) dt
+ \int_0^{\infty}t^{s-1}\left(K(t)-B(t)\right)dt \right).
\end{align*}
Integrating the first term gives \begin{align}\label{eq1_7_1}
\zeta_1(s):=&\frac{1}{\Gamma(s)}\int_0^{\infty} t^{s-1} B(t) dt\\
=&\frac{\pi^{\frac{d}{2}}}{\left[\prod_{i=1}^p
a_i\right]}\sum_{j=0}^{\left[\frac{d}{2}\right]}\frac{(-1)^j}{j!}\frac{\Gamma\left(\frac{d-2j}{2\alpha}\right)}{
\alpha\Gamma\left(\frac{d}{2}-j\right)}m^{2j}\frac{\Gamma\left(s-\frac{d-2j}{2\alpha}\right)}{\Gamma(s)}
b^{\frac{d-2j}{\alpha}-2s}.\nonumber
\end{align}Clearly, $\zeta_1(s)$ defines a meromorphic function on $\C$. On the other hand, $K(t)-B(t)$
decays exponentially as $t\rightarrow \infty$, whereas by
\eqref{eq1_8_9}, $K(t)-B(t)=O(t^{\frac{1}{2\alpha}})$ as
$t\rightarrow 0$. Therefore, the function
\begin{align}\label{eq1_8_1} \int_0^{\infty}
t^{s-1}\left(K(t)-B(t)\right)dt
\end{align}is an analytic function
 for $\text{Re}\,s>-1/(2\alpha)$. Consequently, the function \begin{align*}\zeta_2(s)=\frac{1}{\Gamma(s)}
 \int_0^{\infty}t^{s-1}(K(t)-B(t))dt
\end{align*}is also an analytic function
 for $\text{Re}\,s>-1/(2\alpha)$. Combining with $\zeta_1(s)$,  we find that
 $\zeta_1(s)+\zeta_2(s)$ gives an analytic continuation of
$\zeta(s)$ to the domain $\text{Re}\, s>-1/(2\alpha)$. This allows
us to find $\zeta(0)$ and $\zeta'(0)$. Specifically, since
$\Gamma(z)$ has simple poles at $z=-j$, $j\in \mathbb{N}\cup\{0\}$
with residues $(-1)^j/j!$, \eqref{eq1_7_1} gives
\begin{align*}\zeta_1(0)=&\frac{\pi^{\frac{d}{2}}}{\left[\prod_{i=1}^p
a_i\right]}\sum_{j=0}^{\left[\frac{d}{2}\right]}\eta_{j,\Lambda_{\alpha,d}}\frac{(-1)^j}{j!}\frac{(-1)^{\frac{d-2j}{2\alpha}}
m^{2j}}{ \Gamma\left(\frac{d}{2}-j+1\right)}
 b^{\frac{d-2j}{\alpha}},
\end{align*}\begin{align*}
\zeta_1'(0)=&\frac{\pi^{\frac{d}{2}}}{\left[\prod_{i=1}^p
a_i\right]}\sum_{j=0}^{\left[\frac{d}{2}\right]}\eta_{j,\Lambda_{\alpha,d}}\frac{(-1)^j}{j!}\frac{(-1)^{\frac{d-2j}{2\alpha}}
m^{2j}}{ \Gamma\left(\frac{d}{2}-j+1\right)}
 b^{\frac{d-2j}{\alpha}}\left\{\psi\left(\frac{d-2j}{2\alpha}+1\right)-\psi(1)-\log b^2\right\}\\
&+\frac{\pi^{\frac{d}{2}}}{\left[\prod_{i=1}^p
a_i\right]}\sum_{j=0}^{\left[\frac{d}{2}\right]}\left(1-\eta_{j,\Lambda_{\alpha,d}}\right)\frac{(-1)^j}{j!}\frac{\Gamma\left(\frac{d-2j}{2\alpha}\right)}{
\alpha\Gamma\left(\frac{d}{2}-j\right)}m^{2j}\Gamma\left(-\frac{d-2j}{2\alpha}\right)
b^{\frac{d-2j}{\alpha}}.
\end{align*}Here $\Lambda_{\alpha,d}$ is the set $$\Lambda_{\alpha,d}=\left\{
j\in \mathbb{N}\cup\{0\}\,:\, \frac{d-2j}{2\alpha}\in
\mathbb{N}\cup\{0\}\right\};$$$\eta_{\alpha,\Lambda_{\alpha,d}}$ is
defined by
$$\eta_{j,\Lambda_{\alpha,d}}=\begin{cases} 1,\hspace{1cm} \text{if}\;\; j\in\Lambda_{\alpha,d},\\
0,\hspace{1cm}\text{otherwise},\end{cases};$$  $\psi(z)$ is the
logarithmic derivative of gamma function, i.e.
$\psi(z)=\Gamma'(z)/\Gamma(z)$.
 On the other hand, since $1/\Gamma(s)=s/\Gamma(s+1)$, $\Gamma(1)=1$ and the
 function defined by \eqref{eq1_8_1} is analytic at $s=0$, we have
 $\zeta_2(0)=0$ and
\begin{align*} \zeta_2'(0)= \int_0^{\infty}
t^{-1}\left(K(t)-B(t)\right)dt.
\end{align*}Gathering the results, we obtain
\begin{align}\label{eq1_8_2}
\zeta(0)=\zeta_1(0)=\frac{\pi^{\frac{d}{2}}}{\left[\prod_{i=1}^p
a_i\right]}\sum_{j=0}^{\left[\frac{d}{2}\right]}\eta_{j,\Lambda_{\alpha,d}}\frac{(-1)^j}{j!}\frac{(-1)^{\frac{d-2j}{2\alpha}}
m^{2j}}{ \Gamma\left(\frac{d}{2}-j+1\right)}
 b^{\frac{d-2j}{\alpha}},
\end{align}and
\begin{align}\label{eq1_8_10}
\zeta'(0)=&\zeta_1'(0)+\zeta_2'(0)\\\nonumber=&\frac{\pi^{\frac{d}{2}}}{\left[\prod_{i=1}^p
a_i\right]}\sum_{j=0}^{\left[\frac{d}{2}\right]}\eta_{j,\Lambda_{\alpha,d}}\frac{(-1)^j}{j!}\frac{(-1)^{\frac{d-2j}{2\alpha}}
m^{2j}}{ \Gamma\left(\frac{d}{2}-j+1\right)}
 b^{\frac{d-2j}{\alpha}}\left\{\psi\left(\frac{d-2j}{2\alpha}+1\right)-\psi(1)-\log b^2\right\}\\
&+\frac{\pi^{\frac{d}{2}}}{\left[\prod_{i=1}^p
a_i\right]}\sum_{j=0}^{\left[\frac{d}{2}\right]}\left(1-\eta_{j,\Lambda_{\alpha,d}}\right)\frac{(-1)^j}{j!}\frac{\Gamma\left(\frac{d-2j}{2\alpha}\right)}{
\alpha\Gamma\left(\frac{d}{2}-j\right)}m^{2j}\Gamma\left(-\frac{d-2j}{2\alpha}\right)
b^{\frac{d-2j}{\alpha}}\nonumber\\
&+ \int_0^{\infty} t^{-1}\left(K(t)-B(t)\right)dt.\nonumber
\end{align}The quantum potential $V_Q$ can then be determined by substituting $\zeta(0)$ and $\zeta'(0)$ from
\eqref{eq1_8_2} and \eqref{eq1_8_10} into \eqref{eq830_3}. Since the
quantum potential does not depend on the arbitrary normalization
constant $\mu$ if and only if $\zeta(0)=0$, we find from
\eqref{eq1_8_2} that this is the case if $d$ is odd and
$\alpha\notin \mathcal{C}_d$,
where\begin{align}\label{eq830_6}\mathcal{C}_{d}=\begin{cases}
\left\{ \frac{u}{v}\,:\, u,v\in\mathbb{N}, (u,v)=1 , u\leq
\frac{d}{2}\right\}, \hspace{1cm} &\text{if}\;\; d \;\text{is
even},\\\left\{ \frac{u}{2v}\,:\, u,v\in\mathbb{N}, (u,2v)=1 , u\leq
d\right\}, \hspace{1cm} &\text{if}\;\; d \;\text{is odd}.
\end{cases}
\end{align}

It is not easy to study the properties of the quantum potential
$V_Q$ from \eqref{eq1_8_2} and \eqref{eq1_8_10}. For most practical
purposes, it is desirable to expand $V_Q$ as a power series in
$b^2=\lambda\widehat{\varphi}^2/2$ when $b$ is small enough. For
this, we use the expansion
$$e^{-tb^2}=\sum_{j=0}^{\infty}\frac{(-1)^j}{j!} t^jb^{2j}$$in
\eqref{eq723_1}, which for $b<m^{\alpha}$, gives us
\begin{align}\label{eq829_1}
\zeta(s) =&
\frac{1}{\Gamma(s)}\sum_{j=0}^{\infty}\frac{(-1)^j}{j!}b^{2j}\int_0^{\infty}
t^{s+j-1}A(t)dt\\
=&\pi^{\frac{q}{2}}\sum_{j=0}^{\infty}
\frac{(-1)^j}{j!}b^{2j}\frac{\Gamma(s+j)\Gamma\left(\alpha(s+j)-\frac{q}{2}\right)
} {\Gamma(s)\Gamma(\alpha(s+j))
}Z_{E,p}\left(\alpha(s+j)-\frac{q}{2};a_1,\ldots, a_p;
m\right)\nonumber.
\end{align}
  The meromorphic continuation of
$\Gamma(s)Z_{E,p}(s; a_1, \ldots, a_p;m)$ gives a meromorphic
continuation of $\zeta(s)$ to $\C$ with
\begin{align}\label{eq26_27}\zeta(0)=&\pi^{\frac{q}{2}}\sum_{j=0
}^{\infty}\frac{(-1)^jb^{2j}}{\Gamma(\alpha j+1)}
\text{Res}_{s=\alpha
j-\frac{q}{2}}\Bigl\{\Gamma(s)Z_{E,p}\left(s;a_1,\ldots, a_p;
m\right)\Bigr\}, \end{align} and
\begin{align}\label{eq829_2} \zeta'(0)=&\pi^{\frac{q}{2}}\sum_{j=0}^{\infty}
\frac{(-1)^jb^{2j}}{\Gamma(\alpha j+1)} \Biggl( \alpha \PP_{s=\alpha
j-\frac{q}{2}}\Bigl\{\Gamma(s)Z_{E,p}(s;a_1,\ldots,
a_p;m)\Bigr\}\\+&\left(\psi(j+1) -\psi(1)-\alpha\psi(\alpha
j+1)\right)\Res_{s=\alpha
j-\frac{q}{2}}\Bigl\{\Gamma(s)Z_{E,p}(s;a_1,\ldots,
a_p;m)\Bigr\}\Biggr).\nonumber
\end{align}Here for a meromorphic function $h(s)$ on $\C$ with at most a simple pole
at a point $s_0\in\C$, we use the notation
\begin{align*} \Res_{s=s_0} h(s) =& \lim_{s\rightarrow s_0}
\left( (s-s_0)
h(s)\right),\\
\PP_{s=s_0} h(s)=& \lim_{s\rightarrow s_0} \left(
h(s)-\frac{\Res_{s=s_0}h(s)}{s-s_0}\right),
\end{align*}so that
\begin{align*}
h(s) =\frac{\Res_{s=s_0}h(s)}{s-s_0}+\PP_{s=s_0} h(s)+O(s-s_0)
\end{align*}as $s\rightarrow s_0$. If $h(s)$ is regular at $s=s_0$,
then $\Res_{s=s_0}h(s)=0$ and $\PP_{s=s_0} h(s)=h(s_0)$. It is easy
to check that \eqref{eq26_27} agrees with \eqref{eq1_8_2}.

\subsection{The massless case $m=0$}In this case, we write $A(t)$ as
$A_0(t)+A_1(t)$, where $A_0(t)$ corresponds to the
$\mathbf{k}=\mathbf{0}$ term, i.e.
$$A_0(t)=\frac{2\pi^{\frac{q}{2}}}{\Gamma\left(\frac{q}{2}\right)}\int_0^{\infty}
w^{q-1}e^{-tw^{2\alpha}}
dw=\frac{\pi^{\frac{q}{2}}}{\alpha}\frac{\Gamma\left(\frac{q}{2\alpha}\right)}{\Gamma\left(\frac{q}{2}\right)}
t^{-\frac{q}{2\alpha}};$$and
\begin{align*}
A_1(t):=&\frac{2\pi^{\frac{q}{2}}}{\Gamma\left(\frac{q}{2}\right)}\int_0^{\infty}w^{q-1}dw
\sum_{\mathbf{k}\in\mathbb{Z}^p\setminus\{\mathbf{0}\}
}\exp\left(-t\left\{\left(w^2+\sum_{i=1}^p
\left[a_ik_i\right]^2\right)^{\alpha}\right\}\right).
\end{align*}As in the massive case, we find that
\begin{align}\label{eq1_8_6}
A_1(t) =& \frac{\pi^{\frac{q}{2}}}{2\pi
i}\int_{u-i\infty}^{u+i\infty} dv \Gamma(v)
t^{-v}\frac{\Gamma\left(\alpha v-\frac{q}{2}\right)}{\Gamma(\alpha
v)}Z_{E,p}\left(\alpha v-\frac{q}{2}; a_1, \ldots, a_p\right),
\end{align} with $u>\frac{d}{2\alpha}$. Here for a positive integer
$p$ and positive real numbers $a_1,\ldots, a_p$, $ Z_{E,p}(s;
a_1,\ldots, a_p)$ is the homogeneous Epstein zeta function defined
by
\begin{align*}
Z_{E,p}(s; a_1, \ldots,
a_p)=\sum_{\mathbf{k}\in\mathbb{Z}^p\setminus\{\mathbf{0}\}}\left(\sum_{i=1}^p
[a_ik_i]^2\right)^{-s}
\end{align*}when $\text{Re}\, s>\frac{p}{2}$. For $p=0$, we use the convention $Z_{E,0}(s)=0$. Some facts
about the function $Z_{E,p}(s; a_1, \ldots, a_p)$ are summarized in
appendix A. In particular, for $p\geq 1$, $\Gamma(s)Z_{E,p}(s; a_1,
\ldots, a_p)$ only has simple poles at $s=0$ and $s=p/2$. As in the
massive case, \eqref{eq1_8_6} gives
\begin{align*}A_1(t) \sim &
\frac{\Gamma\left(\frac{d}{2\alpha}\right)}{\alpha
\Gamma\left(\frac{d}{2}\right)}\frac{\pi^{\frac{d}{2}}}{\left[\prod_{i=1}^p
a_i\right]}t^{-\frac{d}{2\alpha}}-\frac{\Gamma\left(\frac{q}{2\alpha}\right)}{\alpha
\Gamma\left(\frac{q}{2}\right)}\pi^{\frac{q}{2}}t^{-\frac{q}{2\alpha}}
+O(t).
\end{align*}
Consequently, as $t\rightarrow 0$,
\begin{align*}
 K(t)\sim B(t)+O(t),
\end{align*}where
\begin{align*}B(t)=
\frac{\Gamma\left(\frac{d}{2\alpha}\right)}{\alpha
\Gamma\left(\frac{d}{2}\right)}\frac{\pi^{\frac{d}{2}}}{\left[\prod_{i=1}^p
a_i\right]}t^{-\frac{d}{2\alpha}}e^{-tb^2}.
\end{align*}Proceeding
as in the massive case, we find that $\zeta(s)$ has an analytic
continuation to $\text{Re}\;s>-1$ given by $ \zeta_1(s)+\zeta_2(s)$,
where
\begin{align*}
\zeta_1(s)=\frac{\Gamma\left(\frac{d}{2\alpha}\right)}{\alpha
\Gamma\left(\frac{d}{2}\right)}\frac{\pi^{\frac{d}{2}}}{\left[\prod_{i=1}^p
a_i\right]}\frac{\Gamma\left(s-\frac{d}{2\alpha}\right)}{\Gamma(s)}b^{\frac{d}{\alpha}-2s};
\end{align*}and\begin{align*}
\zeta_2(s)=\frac{1}{\Gamma(s)}\int_0^{\infty}t^{s-1}\left(K(t)-B(t)\right)dt.
\end{align*}This gives
\begin{align}\label{eq1_8_12}
\zeta(0)=\omega_{\alpha,d}\frac{\pi^{\frac{d}{2}}}{\left[\prod_{i=1}^p
a_i\right]}\frac{(-1)^{\frac{d}{2\alpha}}}{
\Gamma\left(\frac{d}{2}+1\right)}b^{\frac{d}{\alpha}},
\end{align}and\begin{align}\label{eq1_8_13}
\zeta'(0)=&\omega_{\alpha,d}\frac{\pi^{\frac{d}{2}}}{\left[\prod_{i=1}^p
a_i\right]}\frac{(-1)^{\frac{d}{2\alpha}}}{
\Gamma\left(\frac{d}{2}+1\right)}b^{\frac{d}{\alpha}}\left\{\psi\left(\frac{d}{2\alpha}+1\right)-\psi(1)-\log
b^2\right\}\\+&(1-\omega_{\alpha,d})\frac{\Gamma\left(\frac{d}{2\alpha}\right)}{\alpha
\Gamma\left(\frac{d}{2}\right)}\frac{\pi^{\frac{d}{2}}}{\left[\prod_{i=1}^p
a_i\right]}\Gamma\left(-\frac{d}{2\alpha}\right)b^{\frac{d}{\alpha}}+\int_0^{\infty}
t^{-1}\left(K(t)-B(t)\right)dt.\nonumber
\end{align}Here
\begin{align*}
\omega_{\alpha,d}=\begin{cases} 1,\hspace{1cm} \text{if}\;\;
\frac{d}{2\alpha}\in\mathbb{N},\\
0,\hspace{1cm}\text{otherwise}.\end{cases}
\end{align*}The quantum potential $V_Q$ can be determined by substituting $\zeta(0)$
and $\zeta'(0)$ from \eqref{eq1_8_12} and \eqref{eq1_8_13} into
\eqref{eq830_3}, which gives us \begin{align}\label{eq1_9_1}
V_Q=&\frac{\omega_{\alpha,d}}{2^{d+1}\pi^{\frac{d}{2}}}\frac{(-1)^{\frac{d}{2\alpha}}}{
\Gamma\left(\frac{d}{2}+1\right)}\left(\frac{\lambda\widehat{\varphi}^2}{2}\right)^{\frac{d}{2\alpha}}
\left\{\log\frac{\lambda\left[\widehat{\varphi}\mu\right]^2}{2}-\psi\left(\frac{d}{2\alpha}+1\right)+\psi(1)
\right\}\\
\nonumber-&\frac{(1-\omega_{\alpha,d})}{2^{d+1}\pi^{\frac{d}{2}}}
\frac{\Gamma\left(\frac{d}{2\alpha}\right)}{\alpha
\Gamma\left(\frac{d}{2}\right)}\Gamma\left(-\frac{d}{2\alpha}\right)\left(\frac{\lambda\widehat{\varphi}^2}{2}\right)^{\frac{d}{2\alpha}}
-\frac{1}{2(2\pi)^q\prod_{i=1}^p L_i}\int_0^{\infty}
t^{-1}\left(K(t)-B(t)\right)dt.
\end{align} From \eqref{eq1_8_12}, we find that $V_Q$ is independent of the
normalization constant $\mu$ if and only if $\alpha\notin
\mathcal{E}_{d}$, where
\begin{align}\label{eq830_5}\mathcal{E}_d=\left\{ d/(2j)\,:\,
j\in\mathbb{N}\right\}.\end{align}

To find the small $b$ expansion of $V_Q$, we note that when
$b<\min\{a_1, \ldots, a_p\}$, we can expand \eqref{eq723_1} as
\begin{align}\label{eq25_1}
\zeta(s) =&\frac{1}{\Gamma(s)}\int_0^{\infty} t^{s-1}
A_0(t)e^{-tb^2}dt
+\frac{1}{\Gamma(s)}\sum_{j=0}^{\infty}\frac{(-1)^jb^{2j}}{j!}\int_0^{\infty}
t^{s+j-1} A(t)
dt\\=&\frac{\pi^{\frac{q}{2}}}{\alpha}\frac{\Gamma\left(\frac{q}{2\alpha}\right)
\Gamma\left(s-\frac{q}{2\alpha}\right)}{\Gamma\left(\frac{q}{2}\right)\Gamma(s)}
b^{\frac{q}{\alpha}-2s}\nonumber\\&+\pi^{\frac{q}{2}}\sum_{j=0}^{\infty}
\frac{(-1)^j}{j!}b^{2j}\frac{\Gamma(s+j)\Gamma\left(
\alpha(s+j)-\frac{q}{2}\right)}
{\Gamma(s)\Gamma(\alpha(s+j))}Z_{E,p}\left(\alpha(s+j)-\frac{q}{2};a_1,\ldots,
a_p\right).\nonumber
\end{align} This gives  a meromorphic
continuation  of $\zeta(s)$  to $\C$, with
\begin{align}\label{eq830_1}
\zeta(0)=&\frac{\pi^{\frac{q}{2}}}{\alpha}\frac{\Gamma\left(\frac{q}{2\alpha}\right)
}{\Gamma\left(\frac{q}{2}\right)}
b^{\frac{q}{\alpha}}\Res_{s=-\frac{q}{2\alpha}}\Gamma(s)+\pi^{\frac{q}{2}}\sum_{j=0}^{\infty}
\frac{(-1)^jb^{2j}}{\Gamma(\alpha j+1)} \Res_{s=\alpha
j-\frac{q}{2}}\Bigl\{\Gamma(s)Z_{E,p}(s;a_1,\ldots,
a_p)\Bigr\}\\
=&\omega_{\alpha,d}\frac{\pi^{\frac{d}{2}}}{\left[\prod_{i=1}^p
a_i\right]}\frac{(-1)^{\frac{d}{2\alpha}}}{
\Gamma\left(\frac{d}{2}+1\right)}b^{\frac{d}{\alpha}},\nonumber
\end{align}and\begin{align}\label{eq830_2}
\zeta'(0)=&\frac{\pi^{\frac{q}{2}}}{\alpha}\frac{\Gamma\left(\frac{q}{2\alpha}\right)
}{\Gamma\left(\frac{q}{2}\right)}
b^{\frac{q}{\alpha}}\left(\PP_{s=-\frac{q}{2\alpha}}\Gamma(s)
-\left(2\log b+\psi(1)\right)\Res_{s=-\frac{q}{2\alpha}}\Gamma(s)
\right)\\&+\pi^{\frac{q}{2}}\sum_{j=0}^{\infty}
\frac{(-1)^jb^{2j}}{\Gamma(\alpha j+1)} \Biggl( \alpha \PP_{s=\alpha
j-\frac{q}{2}}\Bigl\{\Gamma(s)Z_{E,p}(s;a_1,\ldots,
a_p)\Bigr\}\nonumber\\+&\left(\psi(j+1) -\psi(1)-\alpha\psi(\alpha
j+1)\right)\Res_{s=\alpha
j-\frac{q}{2}}\Bigl\{\Gamma(s)Z_{E,p}(s;a_1,\ldots,
a_p)\Bigr\}\Biggr).\nonumber
\end{align}

\vspace{0.5cm} Combining the results above for the massive case and
massless case, we find that when $\lambda$ is small enough, the
quantum potential can be written as a power series $V_{Q,r}$ in
$\lambda\widehat{\varphi}^2$ plus a term $A_Q$, i.e.
\begin{align}\label{eq910_1} V_Q= V_{Q,r}+A_Q,\end{align}where the
term $A_Q$ originates from the $\mathbf{k}=\mathbf{0}$ mode in the
massless case, and is given by
\begin{align}\label{eq903_1}
A_Q=\frac{\left(\frac{\lambda\widehat{\varphi}^2}{2}\right)^{\frac{q}{2\alpha}}}{2^{q+1}\pi^{\frac{q}{2}}\alpha
\left[\prod_{i=1}^p
L_i\right]}\frac{\Gamma\left(\frac{q}{2\alpha}\right)
}{\Gamma\left(\frac{q}{2}\right)} \left( \left(\log
\frac{\lambda[\widehat{\varphi}\mu]^2}{2}
+\psi(1)\right)\Res_{s=-\frac{q}{2\alpha}}\Gamma(s)
-\PP_{s=-\frac{q}{2\alpha}}\Gamma(s)\right).
\end{align}In general, the power of $\widehat{\varphi}^2$ in \eqref{eq903_1} is non-integer. In the massive case,
we do not have such a term and $A_Q=0$. The term $V_{Q,r}$ in  both
the massive and massless cases is equal
to\begin{align}\label{eq830_4}
V_{Q,r}=&\frac{-1}{2^{q+1}\pi^{\frac{q}{2}}\left[\prod_{i=1}^p
L_i\right]}\sum_{j=0}^{\infty}
\frac{(-1)^j\lambda^j\widehat{\varphi}^{2j}}{2^j\Gamma(\alpha j+1)}
\Biggl( \alpha \PP_{s=\alpha
j-\frac{q}{2}}\left\{\Gamma(s)Z_{E,p}\left(s;\frac{2\pi}{L_1},\ldots,
\frac{2\pi}{L_p};m\right)\right\}\\+&\left(\psi(j+1)
-\psi(1)-\alpha\psi(\alpha j+1)-\log\mu^2\right)\Res_{s=\alpha
j-\frac{q}{2}}\left\{\Gamma(s)Z_{E,p}\left(s;\frac{2\pi}{L_1},\ldots,
\frac{2\pi}{L_p};m\right)\right\}\Biggr).\nonumber
\end{align} When $m=0$,
$Z_{E,p}(s; a_1,\ldots, a_p; m)$ is understood as
$Z_{E,p}(s;a_1,\ldots, a_p)$.

In the case where $p=0$ or equivalently $q=d$, i.e. when the
spacetime is $\R^{d}$, we can write down the quantum potential $V_Q$
more explicitly:

\vspace{0.2cm}\noindent $\bullet$\;\;In the massless case, since
$Z_{E,0}(s)=0$, we have $V_{Q,r}=0$. Therefore,

\vspace{0.2cm} $\bullet$ \;\; if $\alpha\notin \mathcal{E}_d$ (see
\eqref{eq830_5}), then
\begin{align}\label{eq917_1}
V_Q=A_Q=\frac{1}{2^{d+1}\pi^{\frac{d-2}{2}}\Gamma\left(\frac{d+2}{2}\right)\sin
\frac{\pi
d}{2\alpha}}\left(\frac{\lambda\widehat{\varphi}^2}{2}\right)^{\frac{d}{2\alpha}};
\end{align}

$\bullet$\;\; if $\alpha\in \mathcal{E}_d$, then
\begin{align}\label{eq917_2}
V_Q=A_Q=\frac{(-1)^{\frac{d}{2\alpha}}
}{2^{d+1}\pi^{\frac{d}{2}}\Gamma\left(\frac{d+2}{2}\right)}
\left(\frac{\lambda\widehat{\varphi}^2}{2}\right)^{\frac{d}{2\alpha}}
\left(\log \frac{\lambda[\widehat{\varphi}\mu]^2}{2}
+\psi(1)-\psi\left(\frac{d}{2\alpha}+1\right)\right).
\end{align}

\vspace{0.2cm}\noindent $\bullet$\;\; In the massive case, using the
fact that $Z_{E,0}(s;m)=m^{-2s}$, we have

\begin{align}\label{eq917_5}
V_Q=V_{Q,r}=&-\frac{\alpha
m^{d}}{2^{d+1}\pi^{\frac{d}{2}}}\sum_{j\in
\mathbb{N}\cup\{0\}\setminus\Xi_{\alpha;d}}
 (-1)^j \frac{\Gamma\left(\alpha
j-\frac{d}{2}\right)}{\Gamma(\alpha
j+1)}\left(\frac{\lambda\widehat{\varphi}^2}{2m^{2\alpha}}\right)^j\\
&-\frac{  m^{d}}{2^{d+1}\pi^{\frac{d}{2}}}\sum_{j\in
 \Xi_{\alpha;d}}
  \frac{(-1)^{\frac{d}{2}-(\alpha+1)j}}{\Gamma(\alpha
j+1)\left(\frac{d}{2}-\alpha j\right)!}\left(\frac{\lambda\widehat{\varphi}^2}{2m^{2\alpha}}\right)^j\times\nonumber\\
&\left(\alpha \left[\psi\left(\frac{d+2}{2}-\alpha j\right)-
\psi(\alpha j+1)\right]
+\psi(j+1)-\psi(1)-\log\left[m^{2\alpha}\mu^2\right]\right);\nonumber
\end{align}
where the set $\Xi_{\alpha;d}$ is given by\begin{align*} \Xi_{
\alpha;d}=\left\{j\in \mathbb{N}\cup\{0\}\,:\, \frac{d}{2}-\alpha j
\in \mathbb{N}\cup\{0\}\right\}.
\end{align*}

\section{Renormalization of the theory}In order to eliminate the
dependence of the effective potential on the arbitrary scaling
length $\mu$, we need to   renormalize the theory. For given $d$ and
$\alpha$, we notice that the term $\log\mu^2$ would  only appear in
the coefficients of $\widehat{\varphi}^{2j}$ for $j \leq
 d/(2\alpha)$. Therefore, we propose to add counterterms $\delta
C_0,\delta C_1,
 \ldots$ of order $\widehat{\varphi}^0, \widehat{\varphi}^2, \ldots$
, up to order $\widehat{\varphi}^{2d_{\alpha}}$, where
$$d_{\alpha}:=\left[ \frac{d}{2\alpha}\right],$$ so that the renormalized
effective potential
$V_{\text{eff}}^{(\text{ren})}(\widehat{\varphi})$ becomes
\begin{align}\label{eq904_1}
V_{\text{eff}}^{(\text{ren})}(\widehat{\varphi})= \frac{1}{2}
m^{2\alpha} \widehat{\varphi}^2+\frac{1}{4!} \lambda
\widehat{\varphi}^4+V_Q +\sum_{j=0}^{d_{\alpha}}\frac{\delta
C_j}{(2j)!} \widehat{\varphi}^{2j}.
\end{align}Upon a  closer inspection of the expressions for $V_Q$ in section 2, we find
that the coefficients of $\log\mu^2$ in $V_Q$ are independent of the
compactification lengths. Therefore we can determine the
counterterms $\delta C_j$, $0\leq j\leq d_{\alpha}$, by the
following conditions
\begin{align}\label{eq26_2}
\left.V_{\text{eff}}^{\text{(ren)}}(\widehat{\varphi})\right|_{\widehat{\varphi}=0,
L_i\rightarrow \infty}=&0,\\
\left.\frac{\pa^2V_{\text{eff}}^{\text{(ren)}}(\widehat{\varphi})}{\pa\widehat{\varphi}^2}\right|_{\widehat{\varphi}=0,
L_i\rightarrow \infty}=&m^{2\alpha},\nonumber\\
\left.\frac{\pa^4V_{\text{eff}}^{\text{(ren)}}(\widehat{\varphi})}{\pa\widehat{\varphi}^4}\right|_{
\widehat{\varphi}= \widehat{\varphi}_2, L_i\rightarrow
\infty}=&\lambda,\nonumber\\
\left.\frac{\pa^{2j}V_{\text{eff}}^{\text{(ren)}}(\widehat{\varphi})}{\pa\widehat{\varphi}^{2j}}\right|_{
\widehat{\varphi}= \widehat{\varphi}_j, L_i\rightarrow
\infty}=&0,\hspace{1cm} 3\leq j\leq d_{\alpha}.\nonumber
\end{align}Here
$\widehat{\varphi}_j$, $2\leq j\leq d_{\alpha}$ are different
renormalization scales. Note that sometimes the notations $\delta
m^{2\alpha}$ and $\delta\lambda$  are used instead of $\delta C_1$
and $\delta C_2$. We would like to emphasize that  for $j\geq 1$,
$\delta C_j$ is defined by the condition above only when $\alpha\leq
\frac{d}{2j}$. When $\alpha>\frac{d}{2j}$, we take $\delta C_j=0$ as
a convention. When $p=0$, the $L_i\rightarrow \infty$ limits in the
definition of the counterterms $\delta C_j$ in \eqref{eq26_2} become
vacuous.

\vspace{0.2cm} To have a unified treatment, we define
$\widehat{\varphi}_0=\widehat{\varphi}_1=0$. Then the conditions
\eqref{eq26_2} that define the counterterms $\delta C_j$, $0\leq
j\leq d_{\alpha}$ can be equivalently expressed as
\begin{align}\label{eq904_2}
\sum_{k=j}^{d_{\alpha}}\frac{\delta C_k
}{(2k-2j)!}\widehat{\varphi}_j^{2(k-j)}=-\left.\frac{\pa^{2j}
V_Q}{\pa
\widehat{\varphi}^{2j}}\right|_{\widehat{\varphi}=\widehat{\varphi}_j,
L_i\rightarrow \infty}.
\end{align}

In the following, we proceed to determine the counterterms for
massive case and massless case separately. We will consider the
massive case first, where we determine the counterterms by
\eqref{eq904_2} and use the formula \eqref{eq830_4} for $V_Q$. The
massless case is more difficult since in this case, the power series
expression of $V_{Q,r}$ (Eq. \eqref{eq830_4}) is only valid when
$\widehat{\varphi} < \sqrt{2/\lambda} \min\left\{
(2\pi)/L_j\right\}_{j=1,\ldots,p}$. Therefore,   the limit
$L_i\rightarrow \infty$, $1\leq i\leq p$ cannot be taken directly on
this formula.

\vspace{0.2cm}

\subsection{The massive case}~

\vspace{0.2cm}
   Using \eqref{eq904_2}, \eqref{eq830_4},
\eqref{eq26_11}, \eqref{eq26_21} and \eqref{eq26_22}, we find that
the counterterms $\delta C_j$, $0\leq j\leq d_{\alpha}$ are
determined by the following linear system
\begin{align}\label{eq905_5}
\Pi\begin{pmatrix}\delta C_0\\\delta C_1\\ \delta C_2\\
\delta C_3\\   \vdots \\ \delta
C_{d_{\alpha}}\end{pmatrix}=\begin{pmatrix} 1 & 0 & 0 & 0 & 0 &\ldots & 0 \\0 & 1 & 0 & 0 & 0 &\ldots & 0 \\
0 & 0&1 & \frac{\widehat{\varphi}^2_2}{2!} &
\frac{\widehat{\varphi}^4_2}{4!} &\ldots &
\frac{\widehat{\varphi}_2^{2(d_{\alpha}-2)}}{[2(d_{\alpha}-2)]!}\\
0&0&0 & 1 & \frac{\widehat{\varphi}_3^2}{2!} & \ldots &
\frac{\widehat{\varphi}_3^{2(d_{\alpha}-3)}}{[2(d_{\alpha}-3)]!}\\
\vdots&\vdots&\vdots & \vdots & \vdots & & \vdots\\
0&0&0 & 0 & 0 & \ldots & 1\end{pmatrix}\begin{pmatrix}\delta C_0\\\delta C_1\\ \delta C_2\\
\delta C_3\\ \vdots \\ \delta
C_{d_{\alpha}}\end{pmatrix} =\begin{pmatrix}T_0\\T_1\\
T_2\\T_3 \\\vdots\\T_{d_{\alpha}}\end{pmatrix},
\end{align}where
\begin{align}\label{eq903_5} T_j=&\frac{(-1)^j  \lambda^j m^{d-2\alpha
j}}{2^{d+j+1}\pi^{\frac{d}{2}}}\Biggl\{\alpha\sum_{ k\in
\mathbb{N}\cup\{0\}\setminus
\mathcal{H}_{\alpha;d,j}}(-1)^k\frac{[2(k+j) ]!\Gamma\left( \alpha
(k+j)-\frac{d}{2}\right)}{[2k]!\Gamma(\alpha(k+j)+1)}\left(\frac{\lambda\widehat{\varphi}^2_j}{2m^{2\alpha}}\right)^k\\
&\nonumber+\sum_{ k\in
\mathcal{H}_{\alpha;d,j}}\frac{(-1)^{k+\frac{d}{2}-\alpha(k+j)}[2(k+j)]!}{[2k]!
\left[\frac{d}{2}-\alpha(k+j)\right]!\Gamma(\alpha(k+j)+1)}
\left(\frac{\lambda\widehat{\varphi}^2_j}{2m^{2\alpha}}\right)^k\times\\&\Biggl[
\alpha\left(\psi\left(\frac{d+2}{2}-\alpha(k+j)\right)-\psi(\alpha(k+j)+1)\right)
+\psi(k+j+1)-\psi(1)-\log\left[m^{2\alpha}\mu^2\right]\Biggr]\Biggr\}.\nonumber
\end{align}
Here $$\mathcal{H}_{\alpha; d,j}=\left\{
k\in\mathbb{N}\cup\{0\}\,:\, \frac{d}{2}-\alpha(k+j)\in
\mathbb{N}\cup\{0\}\right\}.$$ Notice that the matrix $\Pi$ defined
in \eqref{eq905_5} is of the form $\Pi = I + \Pi_0,$  where $I$ is
an $(d_{\alpha}+1)\times (d_{\alpha}+1)$ identity matrix and $\Pi_0$
is a nilpotent matrix with $\Pi_0^{d_{\alpha}+1}=0$. Therefore
\begin{align}\label{eq103_1}\Pi^{-1}= I -\Pi_0+\Pi_0^2-\ldots
+(-1)^{d_{\alpha}}\Pi_0^{d_{\alpha}},\end{align} and one can solve
for the counterterms $\delta C_j$ by multiplying $\Pi^{-1}$
\eqref{eq103_1} on both sides of \eqref{eq905_5}. In particular, by
recalling that $\widehat{\varphi}_0=\widehat{\varphi}_1=0$, we can
easily find that

\vspace{0.2cm}\noindent  $\bullet$\;\; For $\delta C_0$,

\vspace{0.2cm}  $\bullet$\;\; if $d$ is odd, then
\begin{align}\label{eq903_4}
\delta
C_0=\frac{\alpha}{2^{d+1}\pi^{\frac{d}{2}}}\Gamma\left(-\frac{d}{2}\right)
m^{d};
\end{align}

\vspace{0.2cm} $\bullet$\;\; if $d$ is even, then
\begin{align}\label{eq26_28}
\delta C_0=\frac{(-1)^{\frac{d}{2}}m^{d}}{2^{d+1}\pi^{\frac{d}{2}}
\left(\frac{d}{2}\right)!}
\left(\alpha\left[\psi\left(\frac{d+2}{2}\right)-\psi(1)\right]
-\log \left[m^{2\alpha}\mu^2\right]\right) .
\end{align}

\vspace{0.5cm}\noindent $\bullet$\;\; For $\delta m^{2\alpha}$, if
$\alpha\leq \frac{d}{2}$ and

\vspace{0.2cm} $\bullet$\;\; if $\frac{d}{2}-\alpha$ is not a
nonnegative integer, then
\begin{align}\label{eq27_1} \delta
m^{2\alpha}=-\frac{\lambda}{2^{d+1}\pi^{\frac{d}{2}}}\frac{\Gamma\left(\alpha-\frac{d}{2}\right)}{\Gamma(\alpha)}
m^{d-2\alpha};
\end{align}

\vspace{0.2cm} $\bullet$\;\; if $\frac{d}{2}-\alpha\in
\mathbb{N}\cup\{0\}$, then
\begin{align}\label{eq26_29}
\delta
m^{2\alpha}=\frac{(-1)^{\frac{d-2}{2}-\alpha}}{2^{d+1}\pi^{\frac{d}{2}}}
\frac{\lambda
m^{d-2\alpha}}{\left[\frac{d}{2}-\alpha\right]!\Gamma(\alpha+1)}\left(\alpha\left[\psi\left(\frac{d+2}{2}-\alpha
\right)-\psi(\alpha)\right] -\log
\left[m^{2\alpha}\mu^2\right]\right) .
\end{align}

\vspace{0.5cm}\noindent In the case of ordinary scalar field (i.e.
$\alpha=1$) in $d=4$ dimensional spacetime, the formulas for $\delta
C_0$ \eqref{eq26_28} and $\delta m^2$ \eqref{eq26_29} obtained above
are in agreement with the corresponding results in \cite{EK1}. On
the other hand, since $d_{\alpha}=2$ in this case, $\Pi$ is just the
$3\times 3$ identity matrix.  Therefore it is easy to find that
\begin{align*}
\delta\lambda=\delta C_2=T_2=&\frac{  \lambda^2
}{64\pi^{2}}\Biggl\{\sum_{
k=1}^{\infty}(-1)^k\frac{(2k+3)!\Gamma\left(
k\right)}{(2k)!\Gamma(k+2)}
\left(\frac{\lambda\widehat{\varphi}^2_2}{2m^{2\alpha}}\right)^k-6\log[m\mu]^2\Biggr\}\\
=&\frac{  \lambda^2
}{32\pi^{2}}\Biggl\{\frac{\lambda^2\widehat{\varphi}_2^4}{M_1^4}-
\frac{6\lambda\widehat{\varphi}_2^2}{M_1^2}-3\log[M_1\mu]^2\Biggr\},\nonumber
\end{align*}where
$M_1^2=m^2+\frac{1}{2}\lambda\widehat{\varphi}_2^2$. This again
agrees with the result given in \cite{EK1}.

\vspace{0.2cm}
\subsection{The massless case}~

\vspace{0.2cm} Since the series \eqref{eq830_4} is absolutely
convergent if and only if $\widehat{\varphi}
<\sqrt{2/\lambda}\min\left\{ (2\pi)/L_i\right\}_{i=1}^p$, we cannot
take the limit $L_i\rightarrow \infty$, $1\leq i\leq p$ term by term
on the  $2j$-th order derivatives of  \eqref{eq830_4} with respect
to $\widehat{\varphi} $ to obtain the counterterms $\delta C_j$,
$0\leq j\leq d_{\alpha}$ from the equation \eqref{eq904_2},
otherwise we will obtain infinity for each individual term. As a
result, we have to work directly with the expression \eqref{eq1_9_1}
for $V_Q$.

Using \eqref{eq1_9_1}, we find that
\begin{align*}
-\left.\frac{\pa^{2j}V_Q}{\pa\widehat{\varphi}^{2j}}\right|_{L_i\rightarrow
\infty, \widehat{\varphi}=\widehat{\varphi}_j} =&
T_{j,1}+T_{j,2}+T_{j,3},
\end{align*}where
\begin{align*}
T_{j,1}=&\frac{\omega_{\alpha,d}}{2^{d+1}\pi^{\frac{d}{2}}}
\frac{(-1)^{\frac{d}{2\alpha}}}{\Gamma\left(\frac{d}{2}+1\right)}\frac{\lambda^j}{2^j}
\left(\frac{\lambda\widehat{\varphi}_j^2}{2}\right)^{\frac{d}{2\alpha}-j}
\frac{(\frac{d}{\alpha})!}{\left(\frac{d}{\alpha}-2j\right)!}\times\\&
\left\{\psi\left(\frac{d}{2\alpha}+1\right)-\psi(1)-2\psi\left(\frac{d}{\alpha}+1\right)
+2\psi\left(\frac{d}{\alpha}-2j+1\right)-\log
\frac{\lambda\left[\mu\widehat{\varphi}_j\right]^2}{2}\right\},\end{align*}\begin{align*}
T_{j,2}=&\frac{(1-\omega_{\alpha,d})}{2^{d+1}\pi^{\frac{d}{2}}}\frac{\Gamma\left(\frac{d}{2\alpha}\right)}{\alpha
\Gamma\left(\frac{d}{2}\right)}\frac{\lambda^j}{2^j}\Gamma\left(-\frac{d}{2\alpha}\right)
\frac{\Gamma\left(\frac{d}{\alpha}+1\right)}{
\Gamma\left(\frac{d}{\alpha}-2j+1\right)}
\left(\frac{\lambda\widehat{\varphi}_j^2}{2}\right)^{\frac{d}{2\alpha}-j},
\end{align*} and
\begin{align*}
T_{j,3} = \frac{\lambda^j}{2^j}\left.\lim_{L_i\rightarrow \infty}
\frac{1}{2^{q+1}\pi^q\left[\prod_{i=1}^p L_i\right]}\frac{\pa^{2j}
}{\pa
b^{2j}}\int_0^{\infty}t^{-1}\left(K(t)-B(t)\right)dt\right|_{b^2=\frac{\lambda\widehat{\varphi}_j^2}{2}}.
\end{align*}
Notice that
\begin{align*}
\lim_{L_i\rightarrow
\infty}\frac{K(t)-B(t)}{2^{q+1}\pi^q\left[\prod_{i=1}^pL_i\right]}=
&\frac{1}{2^{d}\pi^{\frac{d}{2}}\Gamma\left(\frac{d}{2}\right)}
\int_0^{\infty} w^{d-1} dw
\exp\left\{-t\left(w^{2\alpha}+b^2\right)\right\}\\&-\frac{1}{2^{d+1}
\pi^{\frac{d}{2}}}\frac{\Gamma\left(\frac{d}{2\alpha}\right)}{\alpha\Gamma\left(\frac{d}{2}\right)}
t^{-\frac{d}{2\alpha}} e^{-tb^2}\\
=&0.
\end{align*}
Therefore, $T_{j,3}=0$. Consequently, we find that the counterterms
$\delta C_j$, $0\leq j\leq d_{\alpha}$ are again determined by the
system \eqref{eq905_5}, but with $T_j$ given by
\begin{align}\label{eq913_1}
T_j =&  T_{j,1}+T_{j,2}+T_{j,3} \\
\nonumber=&\frac{\omega_{\alpha,d}}{2^{d+1}\pi^{\frac{d}{2}}}
\frac{(-1)^{\frac{d}{2\alpha}}}{\Gamma\left(\frac{d}{2}+1\right)}\frac{\lambda^j}{2^j}
\left(\frac{\lambda\widehat{\varphi}_j^2}{2}\right)^{\frac{d}{2\alpha}-j}
\frac{(\frac{d}{\alpha})!}{\left(\frac{d}{\alpha}-2j\right)!}\times\\&
\left\{\psi\left(\frac{d}{2\alpha}+1\right)-\psi(1)-2\psi\left(\frac{d}{\alpha}+1\right)
+2\psi\left(\frac{d}{\alpha}-2j+1\right)-\log
\frac{\lambda\left[\mu\widehat{\varphi}_j\right]^2}{2}\right\}\nonumber\\
\nonumber&+\frac{(1-\omega_{\alpha,d})}{2^{d+1}\pi^{\frac{d}{2}}}\frac{\Gamma\left(\frac{d}{2\alpha}\right)}{\alpha
\Gamma\left(\frac{d}{2}\right)}\frac{\lambda^j}{2^j}\Gamma\left(-\frac{d}{2\alpha}\right)
\frac{\Gamma\left(\frac{d}{\alpha}+1\right)}{
\Gamma\left(\frac{d}{\alpha}-2j+1\right)}
\left(\frac{\lambda\widehat{\varphi}_j^2}{2}\right)^{\frac{d}{2\alpha}-j}.
\nonumber
\end{align}
In particular, we find that

\vspace{0.2cm}\noindent $\bullet$\;\;$\delta C_0=0$.

\vspace{0.2cm} \noindent $\bullet$\;\; When $\alpha\leq
\frac{d}{2}$,

\vspace{0.2cm} $\bullet$\;\; if $\alpha <\frac{d}{2}$, $\delta
m^{2\alpha}=0$.

\vspace{0.2cm} $\bullet$\;\; if $\alpha =\frac{d}{2}$, since
$\widehat{\varphi}_1=0$,  the theory is non-renormalizable.

\vspace{0.2cm}\noindent In the  case of ordinary scalar field
($\alpha=1$) in $d=4$ dimensional spacetime, we also have
\begin{align*}\delta \lambda
=-\frac{\lambda^2}{32\pi^2}\left(8+3\log\frac{\lambda[\mu\widehat{\varphi}_2]^2}{2}\right).
\end{align*}

 \vspace{0.5cm}
From the results above, we find that for both the massive case and
the massless case,  the counterterms only depend on the spacetime
dimension $d$ but not on the number of compactified dimensions $p$.
This is expected since in the prescription for the counterterms, we
have taken the limits $L_i\rightarrow \infty, 1\leq i\leq p$.
  By substituting the counterterms obtained above
into \eqref{eq904_1}, together with the explicit formulas for $V_Q$
(Eq. \eqref{eq903_1} and Eq. \eqref{eq830_4}),    the explicit
expression for the renormalized effective potential can be
determined. Since the result is not  illuminating, we omit it here.
In appendix B, we show that with our prescription for the
counterterms, the renormalized effective potential indeed no longer
depends on the parameter $\mu$, but at the expense of introducing
new renormalization scales $\widehat{\varphi}_i$, $2\leq i\leq
d_{\alpha}$.

\section{Renormalized mass and symmetry breaking mechanism}

According to convention, the renormalized topologically generated
mass $m_{T,\text{ren}}^{2\alpha}$ is defined so that for small
$\widehat{\varphi}$, the term of order $\widehat{\varphi}^2$ in
$V_{\text{eff}}^{(\text{ren})}$ is given by
$$\frac{1}{2}m_{T,\text{ren}}^{2\alpha}\widehat{\varphi}^2.$$ In
this section, we derive explicit formulas for
$m_{T,\text{ren}}^{2\alpha}$ and discuss their signs. Symmetry
breaking mechanisms appear when $m_{T, \text{ren}}^{2\alpha}<0$.

\vspace{0.5cm}\noindent $\bullet$\;\;In the massive case, we obtain
from \eqref{eq904_1},  \eqref{eq830_4}, \eqref{eq27_1},
\eqref{eq26_29}, \eqref{eq26_14} and \eqref{eq26_15} that given $d,
p$ and $\alpha$,
\begin{align}\label{eq27_2}
m_{T,\text{ren}}^{2\alpha}=&m^{2\alpha}+\frac{\lambda\pi^{\alpha}m^{\frac{d}{2}-\alpha}}{(2\pi)^{\frac{d}{2}+\alpha}\Gamma(\alpha)}
\times\\ &\sum_{\mathbf{k}\in \Z^p\setminus\{
\mathbf{0}\}}\left(\sum_{i=1}^p
\left[L_ik_i\right]^2\right)^{-\frac{d-2\alpha}{4}}
K_{\frac{d-2\alpha}{2}} \left(m\sqrt{\sum_{i=1}^p \left[L_i
k_i\right]^2}\right).\nonumber
\end{align}When $p=0$, this reduces to $m_{T,\text{ren}}^{2\alpha}=m^{2\alpha}$. Namely, the
renormalized topologically generated mass is equal to the bare mass.
The result \eqref{eq27_2} agrees with the result in \cite{EK1} when
$d=4$ and $\alpha=1$. Since the modified Bessel function
$K_{\nu}(z)$ is positive for any $\nu\in\R$ and $z\in \R^+$, we
conclude immediately from \eqref{eq27_2} one of the main results of
our paper:

\vspace{0.3cm}\noindent $\bullet$\;\; \emph{In the massive case, the
renormalized topologically generated mass
$m_{T,\text{ren}}^{2\alpha}$ is strictly positive for any $d, p$ and
$\alpha$. Hence  quantum fluctuations do not lead to symmetry
breaking in this case}.

\vspace{0.3cm}  The massless case is more interesting. From the
previous section, we find that the mass is non-renormalizable when
$\alpha=\frac{d}{2}$. In fact, for the ordinary interacting scalar
field theory (i.e. $\alpha=1$), it is well known that the theory is
non-renormalizable when $d=2$. For $\alpha \neq \frac{d}{2}$, the
mass counterterm $\delta m^{2\alpha}$ is identically zero.
Therefore, for $\alpha\neq\frac{d}{2}$, we find from \eqref{eq903_1}
and \eqref{eq830_4} that the renormalized topologically generated
mass $m_{T,\text{ren}}^{2\alpha}$ is given by

\vspace{0.2cm}\noindent $\bullet$\;\; If $p=0$, then
$m_{T,\text{ren}}^{2\alpha}=0$;

\vspace{0.2cm}\noindent $\bullet$\;\; If $p\geq 1$, and

 \vspace{0.2cm}
$\bullet$\;\; if $\alpha \neq \frac{q}{2}$, then
\begin{align}\label{eq911_1}
m_{T,\text{ren}}^{2\alpha}=&\frac{\lambda}{\Gamma(\alpha)}\frac{1}{2^{q+1}\pi^{\frac{q}{2}}\left[\prod_{i=1}^p
L_i\right]}\Gamma\left(\alpha-\frac{q}{2}\right)Z_{E,p}\left(\alpha-\frac{q}{2};
\frac{2\pi}{L_1},\ldots,\frac{2\pi}{L_p}\right)\\
=&\frac{\lambda}{\Gamma(\alpha)}\frac{1}{2^{2\alpha+1}\pi^{\frac{d}{2}}}\Gamma\left(\frac{d}{2}-\alpha\right)
Z_{E,p}\left( \frac{d}{2}-\alpha; L_1, \ldots, L_p\right).\nonumber
\end{align}

\vspace{0.2cm} $\bullet$\;\; if $\alpha = \frac{q}{2}$, then
\begin{align}\label{eq913_5}
m_{T,\text{ren}}^{2\alpha}=&\frac{\lambda}{\Gamma(\alpha+1)}\frac{1}{2^{q+1}\pi^{\frac{q}{2}}\left[\prod_{i=1}^p
L_i\right]}\left\{ 1+\alpha\Bigl[\psi(\alpha)-\psi(1)\Bigr]+\alpha
Z_{E,p}'\left( 0 ; \frac{2\pi}{L_1}, \ldots,
\frac{2\pi}{L_p}\right)\right\}.
\end{align}

\vspace{0.2cm}\noindent We would like to point out that when
$\alpha=\frac{q}{2}$, there is a term proportional to
$$\frac{\lambda\widehat{\varphi}^2}{2}\log\frac{\lambda\widehat{\varphi}^2}{2}$$ in the renormalized effective potential.
This may give rise to ambiguity in the definition of $m_{T,
\text{ren}}^{2\alpha}$ in this case.

  When $d=4$ and $\alpha=1$, the results of
\eqref{eq911_1} agree with the corresponding results in \cite{EK1}
for $q=1,3,4$. Note that \eqref{eq911_1} shows that when $\alpha\neq
q/2$, up to the factor
\begin{align}\label{eq104_3}\frac{\lambda}{\Gamma(\alpha)}\frac{1}{2^{2\alpha+1}
\pi^{\frac{d}{2}}},\end{align} the renormalized mass
$m_{T,\text{ren}}^{2\alpha}$ depends on the spacetime dimension $d$
and the fractional order of the Klein Gordon field $\alpha$,  in the
combination $d-2\alpha$. Therefore, the renormalized mass of a
fractional Klein--Gordon field of fractional order $\alpha$ in a
$d$--dimensional spacetime would be essentially the same (up to a
multiplicative factor) as the renormalized mass of an ordinary
Klein-Gordon field ($\alpha=1$) in a spacetime with fractional
dimension $d+2-2\alpha$.

 To study the sign of the renormalized
topologically generated mass $m_{T,\text{ren}}^{2\alpha}$ when
$p\geq 1$, we first note that the function $\Gamma(s)$ is positive
for all $s\in \R^+$, whereas for the Epstein zeta function
$Z_{E,p}\left(s; L_1, \ldots, L_p\right)$, it is obvious from its
definition by infinite series \eqref{eq104_1} that it is positive
for all $(L_1,\ldots, L_p)\in (\R^+)^p$ when $s>\frac{p}{2}$.
Therefore, we can conclude immediately from \eqref{eq911_1} that

\vspace{0.2cm} $\bullet$\;\; \emph{If
$0<\alpha<\frac{q}{2}=\frac{d-p}{2}$ or $\alpha > \frac{d}{2}$,
quantum fluctuations lead to positive $m_{T,\text{ren}}^{2\alpha}$.
Symmetry breaking mechanism does not appear in these cases}.

\vspace{0.3cm}\noindent Now we turn to the case $\frac{d-p}{2}<
\alpha <\frac{d}{2}$. The argument leading to the sign of the
function $\Gamma(s)Z_{E,p}(s; L_1,\ldots, L_p)$ is rather involved
and lengthy, which will be dealt with  in a separate paper
\cite{LT3}. Here we just give the results:

\vspace{0.2cm} \noindent -- if $p=1$, then for all
$0<s<\frac{1}{2}$, $\Gamma(s)Z_{E,p}(s; L)<0$;

\vspace{0.2cm}\noindent -- if $2\leq p\leq 9$, then for any fixed
$s\in (0, p/2)$, there is a nonempty region $\Omega_{s,p}^+$ of
$(L_1,\ldots, L_p)\in (\R^+)^p$ where $\Gamma(s)Z_{E,p}(s;
L_1,\ldots, L_p)>0$ and a nonempty region $\Omega_{s,p}^-$ of
$(L_1,\ldots, L_p)\in (\R^+)^p$ where $\Gamma(s)Z_{E,p}(s;
L_1,\ldots, L_p)<0$.

\vspace{0.2cm}\noindent -- if $p\geq 10$, there exists an odd number
of points $\gamma_{p,1},\ldots, \gamma_{p, 2n_p+1}$ such that
$0<\gamma_{p,1}< \ldots<\gamma_{p, 2n_p+1}<p/4$ and if we let $I_p$
to be the union of the disjoint closed intervals $[\gamma_{p,1},
\gamma_{p,2}]$, $\ldots$, $[\gamma_{p,2n_p-1}, \gamma_{p, 2n_p}]$,
$[\gamma_{p,2n_p+1}, (p/2)-\gamma_{p,2n_p+1}]$, $[(p/2)-\gamma_{p,
2n_p}, (p/2)-\gamma_{p,2n_p-1}]$, $\ldots$, $[(p/2)-\gamma_{p,2},
(p/2)-\gamma_{p,1}]$, then for all $s\in I_p$, $\Gamma(s)Z_{E,p}(s;
L_1,\ldots, L_p)\geq 0$; and for all $s\in (0, p/2)\setminus I_p$,
there is a nonempty region $\Omega_{s,p}^+$ of $(L_1,\ldots, L_p)\in
(\R^+)^p$ where $\Gamma(s)Z_{E,p}(s; L_1,\ldots, L_p)>0$ and a
nonempty region $\Omega_{s,p}^-$ of $(L_1,\ldots, L_p)\in (\R^+)^p$
where $\Gamma(s)Z_{E,p}(s; L_1,\ldots, L_p)<0$.

\vspace{0.3cm}\noindent Applying these results to the renormalized
mass $m_{T, \text{ren}}^{2\alpha}$, we find that

\vspace{0.2cm} $\bullet$\;\; \emph{If $p=1$ and $\frac{d-1}{2}<
\alpha< \frac{d}{2}$, quantum fluctuations lead to  negative $m_{T,
\text{ren}}^{2\alpha}$. Symmetry breaking mechanism appears in this
case, but there is no symmetry restoration}.

\vspace{0.2cm} $\bullet$\;\; \emph{If $2\leq p\leq 9$ and
$\frac{d-p}{2}< \alpha< \frac{d}{2}$, quantum fluctuations lead to
topological mass generation. The sign of the renormalized mass
$m_{T,\text{ren}}^{2\alpha}$ can be positive or negative, depending
on the relative ratios of the compactification lengths. Therefore,
symmetry breaking mechanism appears in this case. Varying the
compactification lengths of the torus  will lead to  symmetry
restoration}.

\vspace{0.2cm} $\bullet$\;\; \emph{If $p\geq 10$ and $\alpha \in
J_p$, where $J_p$ is the union \begin{align*}
J_p=&\bigcup_{i=1}^{n_p}\left[\frac{d-p}{2}+\gamma_{p,2i-1},\frac{d-p}{2}+\gamma_{p,2i}\right]\bigcup\left[\frac{d-p}{2}+
\gamma_{p, 2n_p+1}, \frac{d}{2}-\gamma_{p, 2n_p+1}\right]
\\&\bigcup_{i=1}^{n_p}\left[\frac{d}{2}-\gamma_{p,2i},\frac{d}{2}-\gamma_{p,2i-1}\right]
\end{align*} of disjoint closed intervals, quantum fluctuations lead to positive $m_{T,\text{ren}}^{2\alpha}$.
Symmetry breaking mechanism does not appear in this case.}

\vspace{0.2cm} $\bullet$\;\; \emph{If $p\geq 10$ and $\alpha \in
\left(\frac{d-p}{2},\frac{d}{2}\right)\setminus  J_p$, quantum
fluctuations lead to topological mass generation. The sign of the
renormalized mass $m_{T,\text{ren}}^{2\alpha}$ can be positive or
negative, depending on the relative ratios of the compactification
lengths. Therefore, symmetry breaking mechanism appears in this
case. Varying the compactification lengths of the torus  will lead
to  symmetry restoration}.

\vspace{0.3cm}\noindent Finally, using the formula \eqref{eq914_1},
we find that

\vspace{0.2cm}\noindent --  If $a_1=\ldots=a_p\rightarrow 0$,
$Z_{E,p}'(0; a_1, \ldots, a_p)\rightarrow -\infty$;

\vspace{0.2cm} \noindent --  If $a_1=\ldots=a_p\rightarrow \infty$,
$Z_{E,p}'(0; a_1, \ldots, a_p)\rightarrow \infty$.

\vspace{0.2cm}\noindent Applying these to \eqref{eq913_5} with
$\alpha=q/2$ gives us:

\vspace{0.2cm} $\bullet$\;\; \emph{For any $d$ and $p$, if $\alpha=
\frac{d-p}{2}$, quantum fluctuations lead to topological mass
generation. The sign of the renormalized mass
$m_{T,\text{ren}}^{2\alpha}$ can be positive or negative, depending
on the relative ratios of the compactification lengths. Therefore,
symmetry breaking mechanism appears in this case. Symmetry
restoration can be realized by suitably varying the compactification
lengths}.

\vspace{0.5cm} In the above discussion, we fix the spacetime
dimension $d$ and the number of compactified dimensions $p$, and
study the condition on the order $\alpha$ of fractional
Klein--Gordon field for the presence of symmetry breaking mechanism.
We observe that for $p\leq 9$, there is a simple criterion on
$\alpha$ for the existence of symmetry breaking mechanism. In
contrast, when $p\geq 10$, the criterion on $\alpha$ for the
presence of symmetry breaking mechanism  becomes complicated. It
would be interesting to explore the physical significance of this
dichotomy between $p\leq 9$ and $p\geq 10$. Now, if we assume $d$
and $\alpha$ being fixed, and $d\leq 9$, then we can conclude that
symmetry breaking mechanism exists if and only
 if   the number of compactified dimensions $p$ is $\geq
 d-2\alpha$.  However, when $d\geq 10$, this condition becomes necessary but not
 sufficient. In Table 1, we tabulated the subset of values of
 $d-2\alpha\in (0, p)$ for which symmetry breaking mechanism does not
 appear, when $10\leq p\leq 21$.

\vfill\pagebreak \vspace{0.2cm} \noindent Table 1: The subset
$\mathcal I_p$ of $d-2\alpha\in (0,p)$ where symmetry breaking
mechanism does not appear, when $10\leq p\leq 21$.

\vspace{0.2cm}\noindent
\begin{tabular}{|c|c||c|c|}
\hline
\hspace{0.5cm}$p$\hspace{0.5cm}  & $\mathcal I_p$  & \hspace{0.5cm} $p$ \hspace{0.5cm} & $\mathcal I_p$\\
\hline
 10 & \hspace{0.5cm}[2.1799,    7.8201]\hspace{0.5cm} & 11 & \hspace{0.5cm} [1.2802,    9.7198]\hspace{0.5cm} \\
 12 & [0.7952,   11.2048] & 13 & \hspace{0.5cm}[0.4995,   12.5005]\hspace{0.5cm} \\
 14 & [0.3124,   13.6876] & 15 & \hspace{0.5cm}[0.1928,   14.8072]\hspace{0.5cm} \\
 16 & [0.1170,   15.8830] & 17 & \hspace{0.5cm}[0.0695,   16.9305]\hspace{0.5cm} \\
 18 & [0.0404,   17.9596] & 19 & \hspace{0.5cm}[0.0229,   18.9771]\hspace{0.5cm} \\
 20 & [0.0127,   19.9873] & 21 & \hspace{0.5cm}[0.0069,  20.9931]\hspace{0.5cm} \\
 \hline
\end{tabular}

\vspace{0.5cm} For $10\leq p\leq 21$, we see from this table that
each of the sets $\mathcal I_p$ is of the form $(2\gamma_{p,1}, p
-2\gamma_{p,1})$ with $\mathcal I_p\subseteq \mathcal I_{p+1}$. In
\cite{LT3}, we show that for all $p\geq 10$, $\mathcal I_p$ is
indeed a subset of $\mathcal I_{p+1}$. Since we have shown that for
any $p$, there is no symmetry breaking when $d-2\alpha<0$ or
$d-2\alpha>p$, we can conclude from Table 1 that increasing the
number of compactified dimensions $p$ tends to elude symmetry
breaking. In the case of ordinary Klein--Gordon field (i.e.
$\alpha=1$), it is easy to verify from the data in Table 1 that when
$p\geq 12$, there is no integer value of $d-2$ lying in the set
$(0,p)\setminus I_p$. Therefore, for $p\geq 12$, symmetry breaking
cannot happen in ordinary Klein--Gordon field theory. This is an
interesting fact since compactifying some of the spacetime
dimensions is a mechanism to induce symmetry breaking, but we find
that there is an upper limit to the number of dimensions that can be
compactified such that there still exists symmetry breaking
mechanism.

\begin{figure}\centerline{
             \epsfig{file=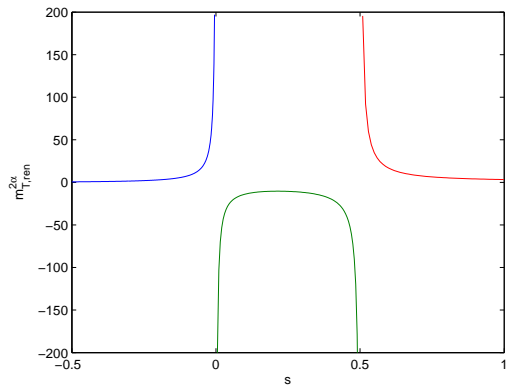, scale=0.5}
             }
 \caption{The   dependence of the renormalized mass
$m_{T,\text{ren}}^{2\alpha}$ (up to the factor \eqref{eq104_3}) on
$s=\frac{d}{2}-\alpha$ when $p=1$ and $V=L_1=1$. }\end{figure}

\begin{figure}\centering \epsfxsize=.49\linewidth \epsffile{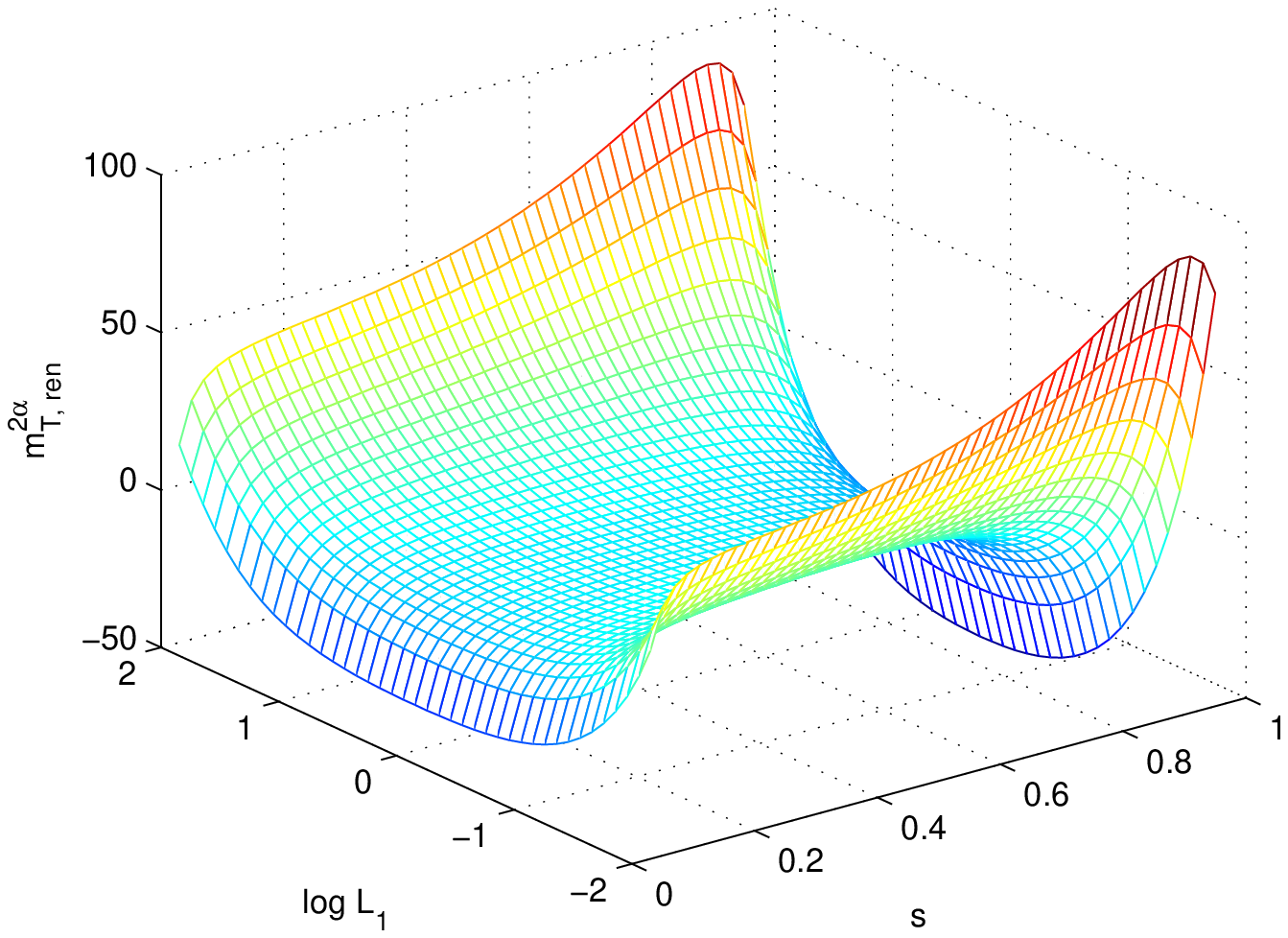}\centering \epsfxsize=.49\linewidth
\epsffile{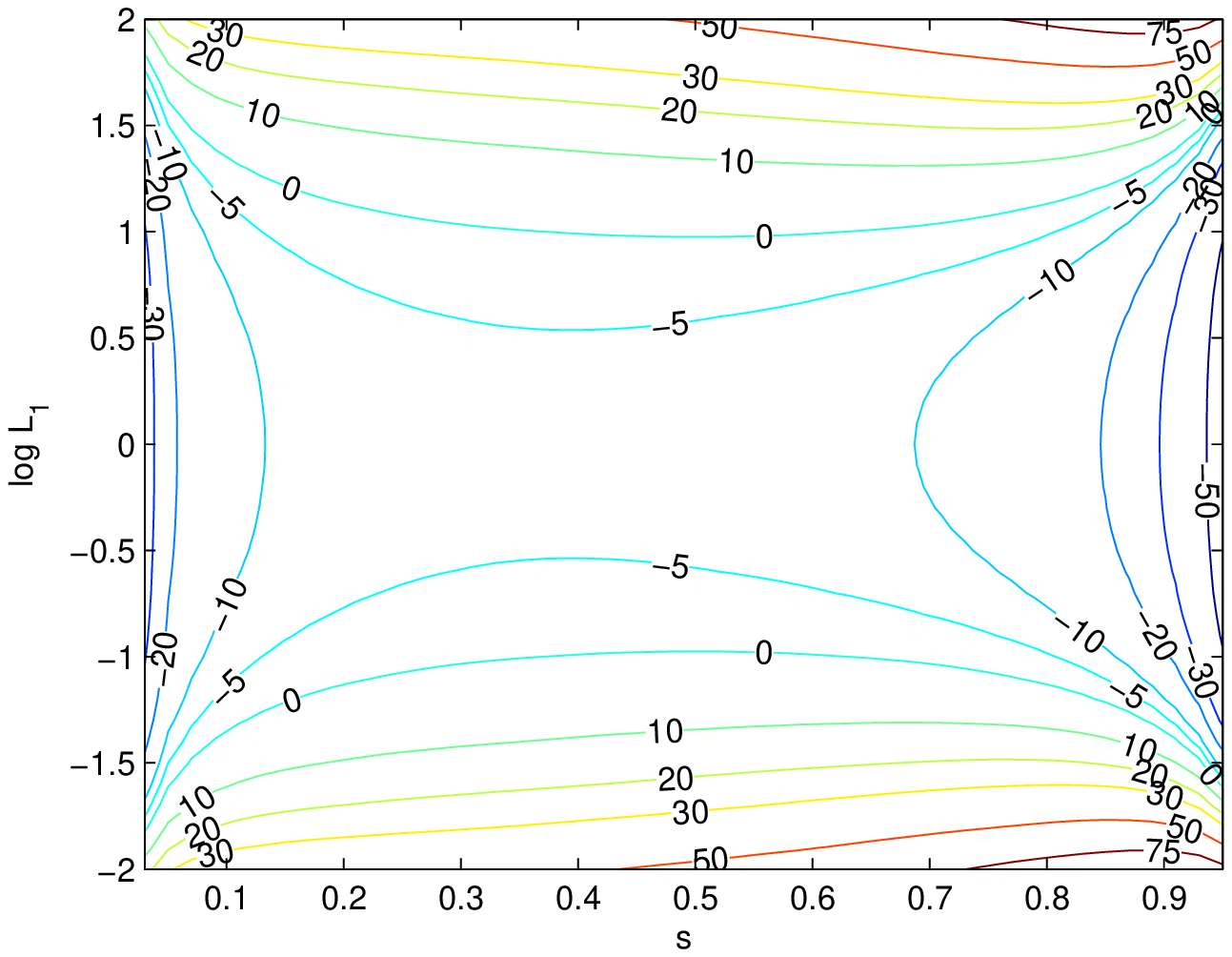} \caption{The graph and the contour lines of
the renormalized mass $m_{T,\text{ren}}^{2\alpha}$ as a function of
$s=\frac{d}{2}-\alpha$ and $\log L_1$ when $p=2$, $V=L_1L_2=1$. Due
to the symmetry with respect to the interchange of  $L_1$ and $L_2$,
these graphs show the symmetry with respect to $\log L_1\mapsto
-\log L_1$. }\end{figure}

\begin{figure}\centering \epsfxsize=.49\linewidth \epsffile{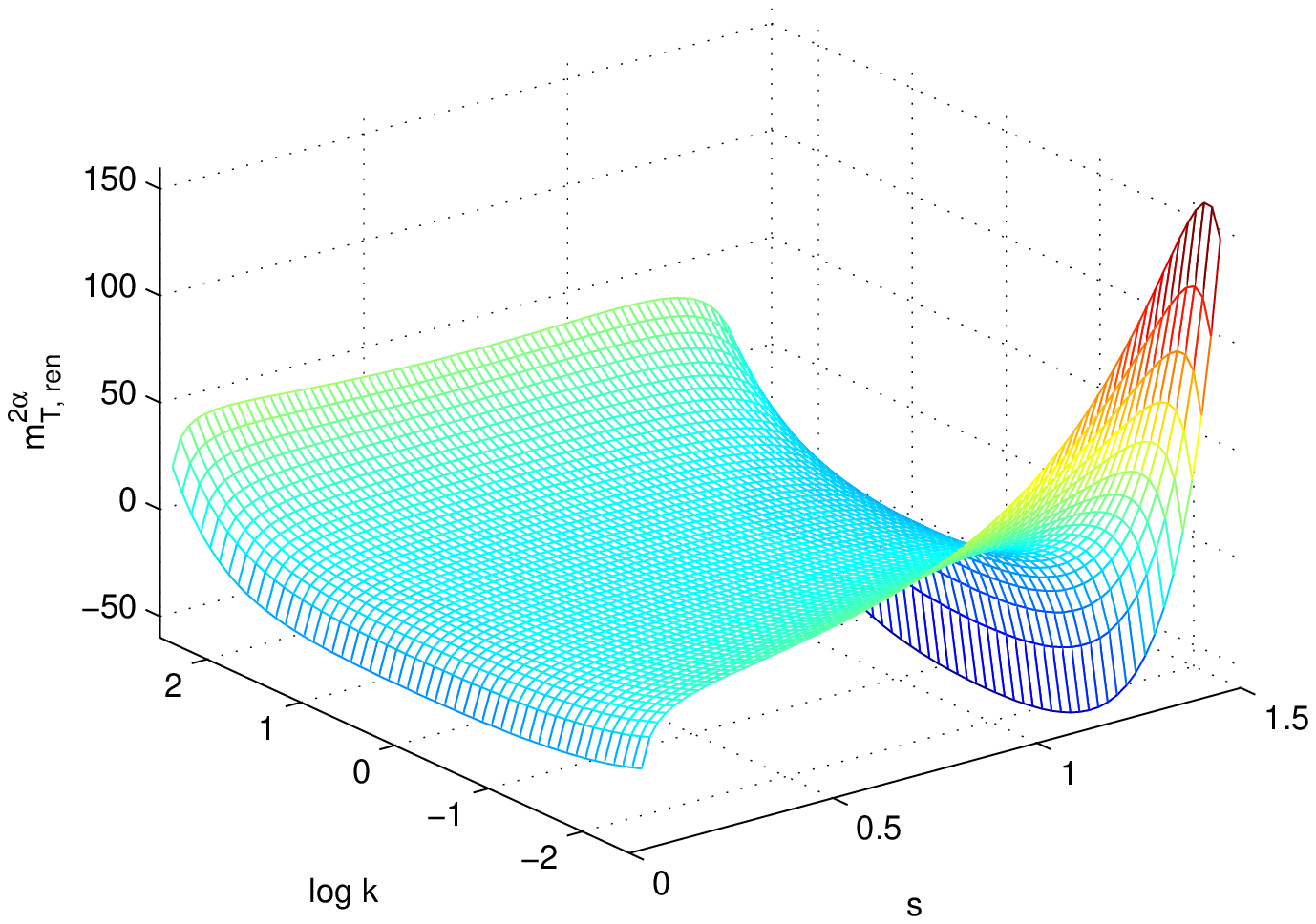}\epsfxsize=.49\linewidth
\epsffile{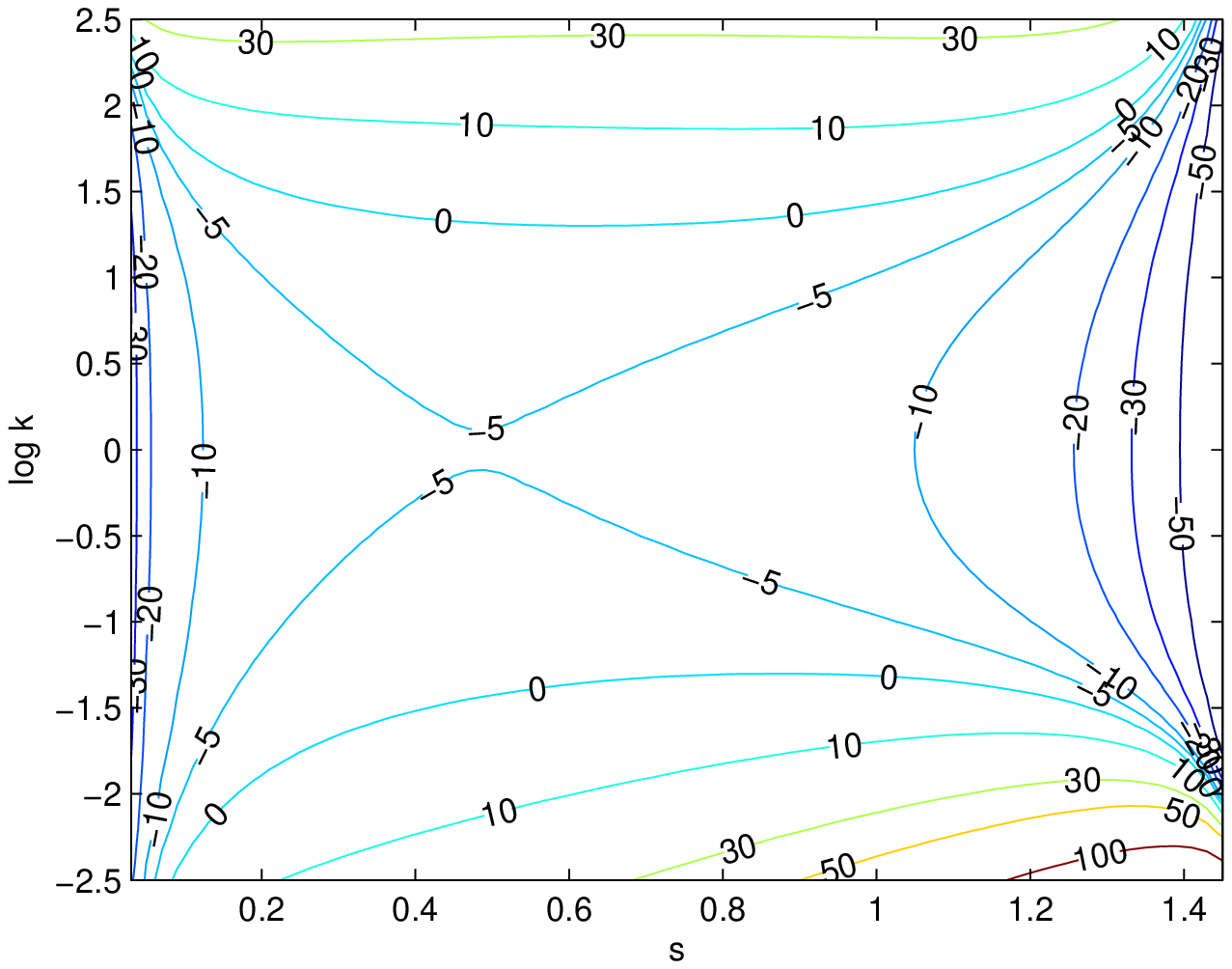} \caption{The graph and the contour lines of
the renormalized mass $m_{T,\text{ren}}^{2\alpha}$ (up to the factor
\eqref{eq104_3}) as a function of $s=\frac{d}{2}-\alpha$ and $\log
k$, when $p=3$, $V=L_1L_2L_3=1$ and $L_1:L_2:L_3=k:1:1$.
}\end{figure}

\begin{figure}\centering \epsfxsize=.49\linewidth
\epsffile{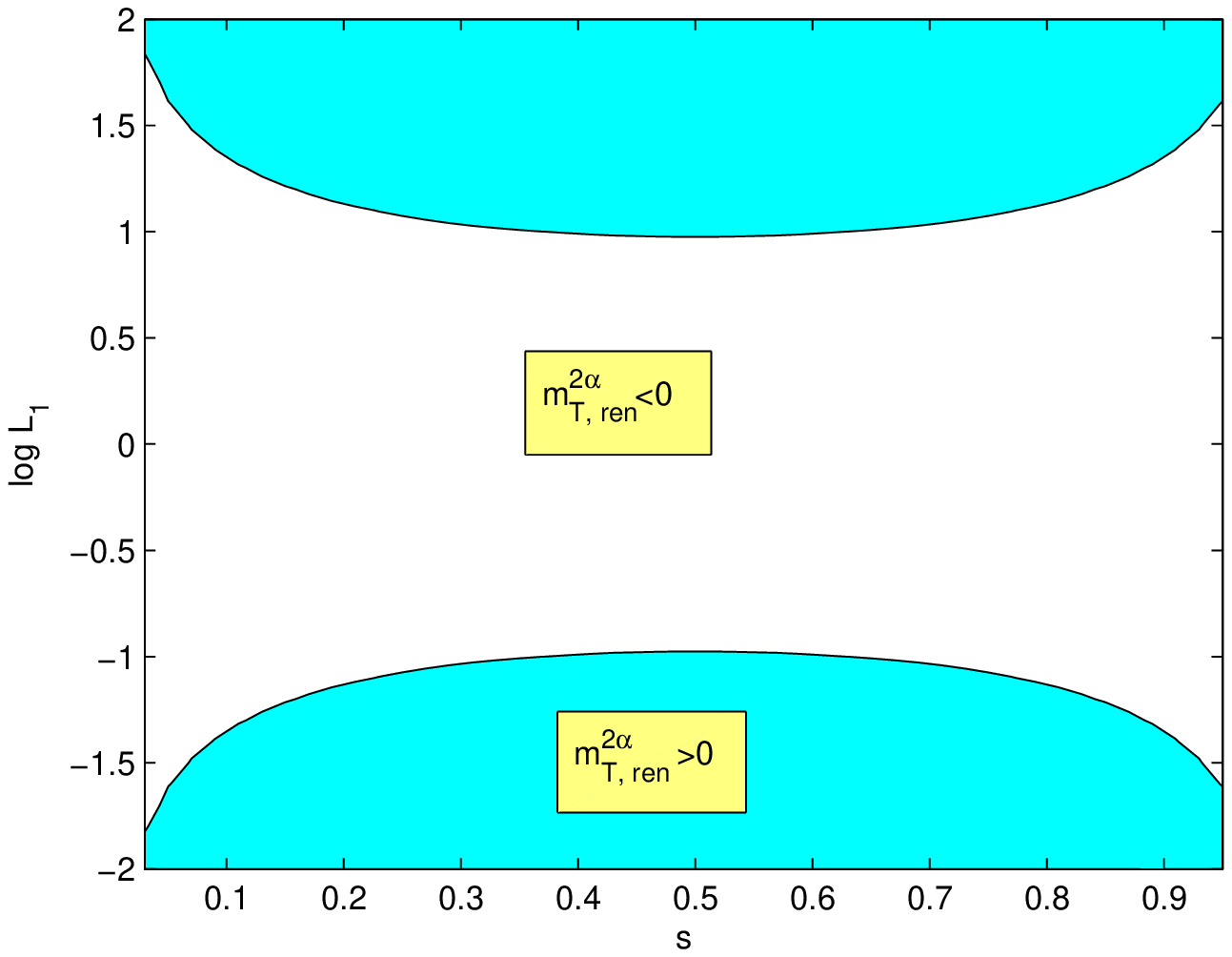}\epsfxsize=.49\linewidth
\epsffile{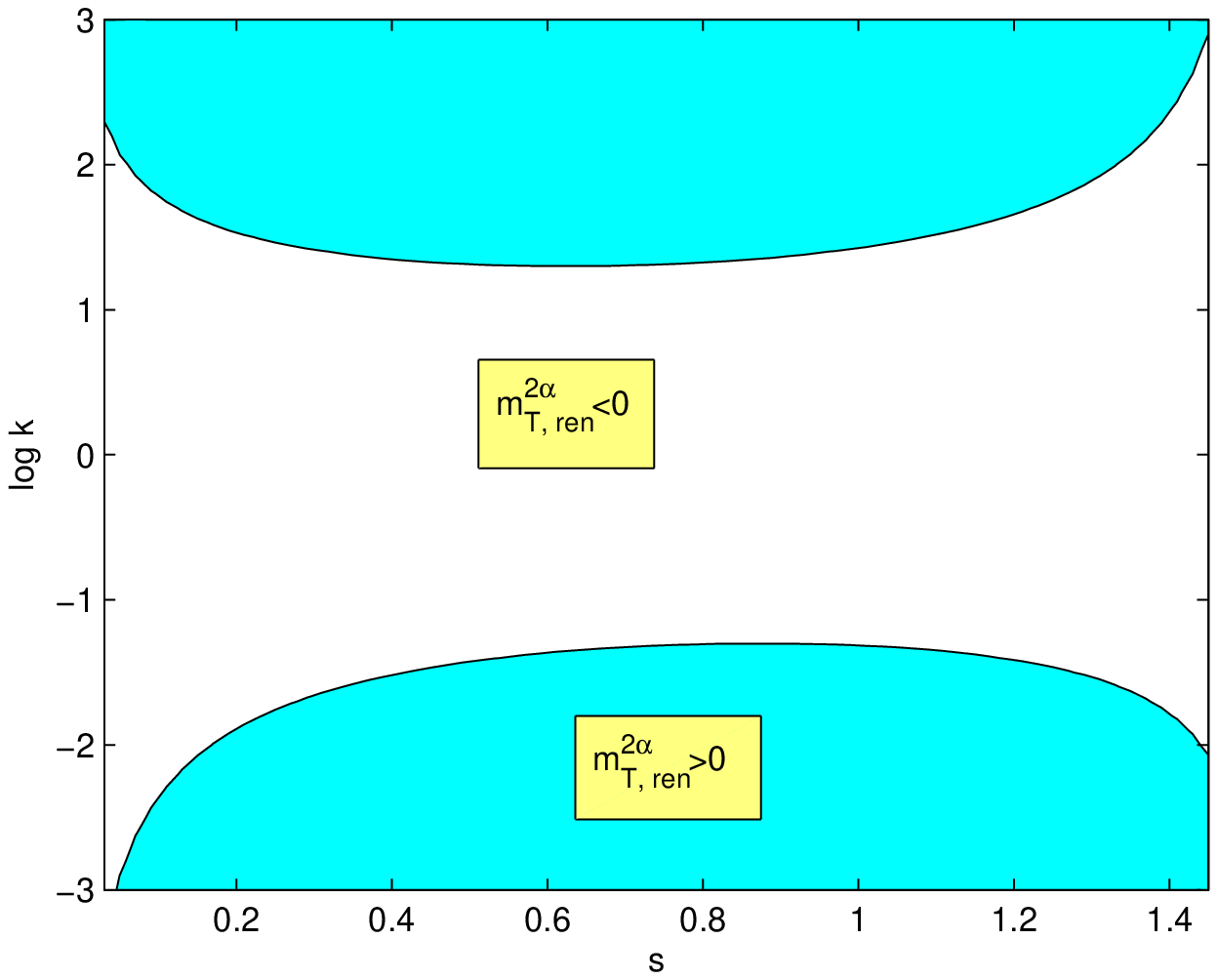}\caption{Left: The   region where
$m_{T,ren}^{2\alpha}>0$ and $m_{T,\text{ren}}^{2\alpha}<0$ for $p=2$
and $V=L_1L_2=1$. Right: The region where
$m_{T,\text{ren}}^{2\alpha}>0$ and $m_{T,\text{ren}}^{2\alpha}<0$
for $p=3$, $V=L_1L_2L_3=1$, $L_1:L_2:L_3=k:1:1$. Here
$s=\frac{d}{2}-\alpha$.}\end{figure}

\begin{figure}\centering \epsfxsize=.49\linewidth \epsffile{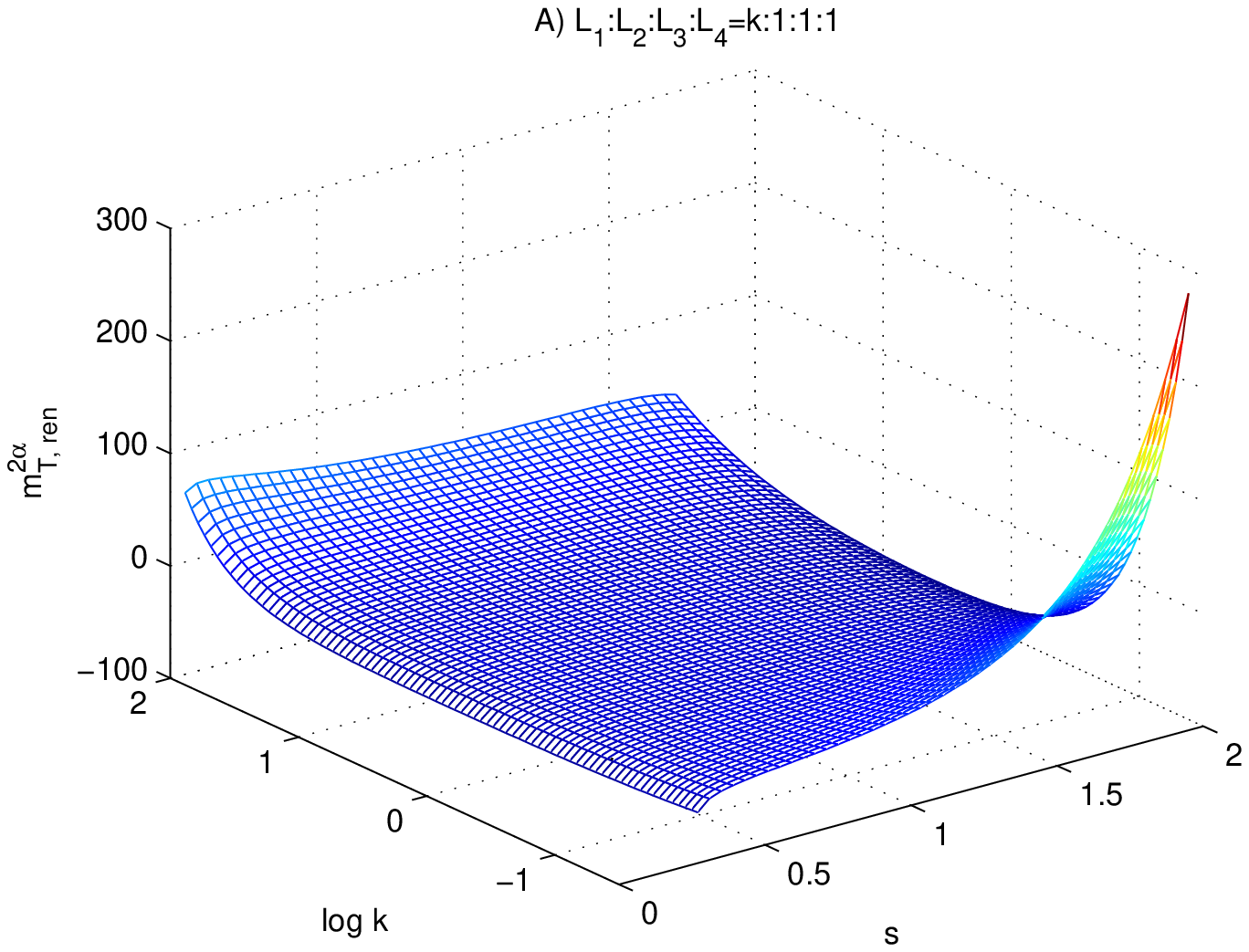}
\epsfxsize=.49\linewidth \epsffile{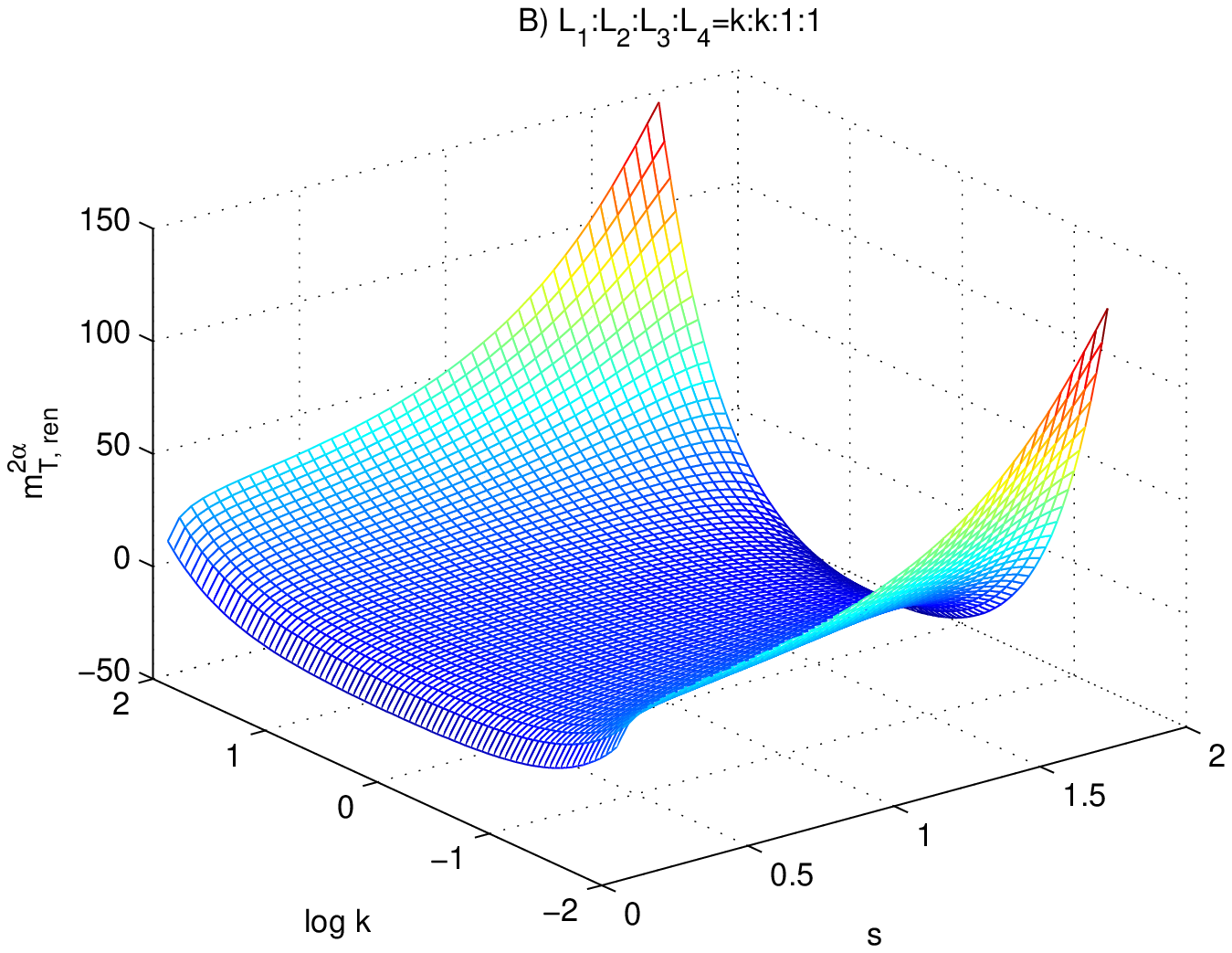}\caption{The graphs
of the renormalized mass  $m_{T,\text{ren}}^{2\alpha}$ (up to the
factor \eqref{eq104_3}) as a function of $s=\frac{d}{2}-\alpha$ and
$\log k$ when $p=4$ and $V=L_1L_2L_3L_4=1$. For A),
$L_1:L_2:L_3:L_4=k:1:1:1$. For B),
$L_1:L_2:L_3:L_4=k:k:1:1$.}\end{figure}

\begin{figure}\centering \epsfxsize=.49\linewidth
\epsffile{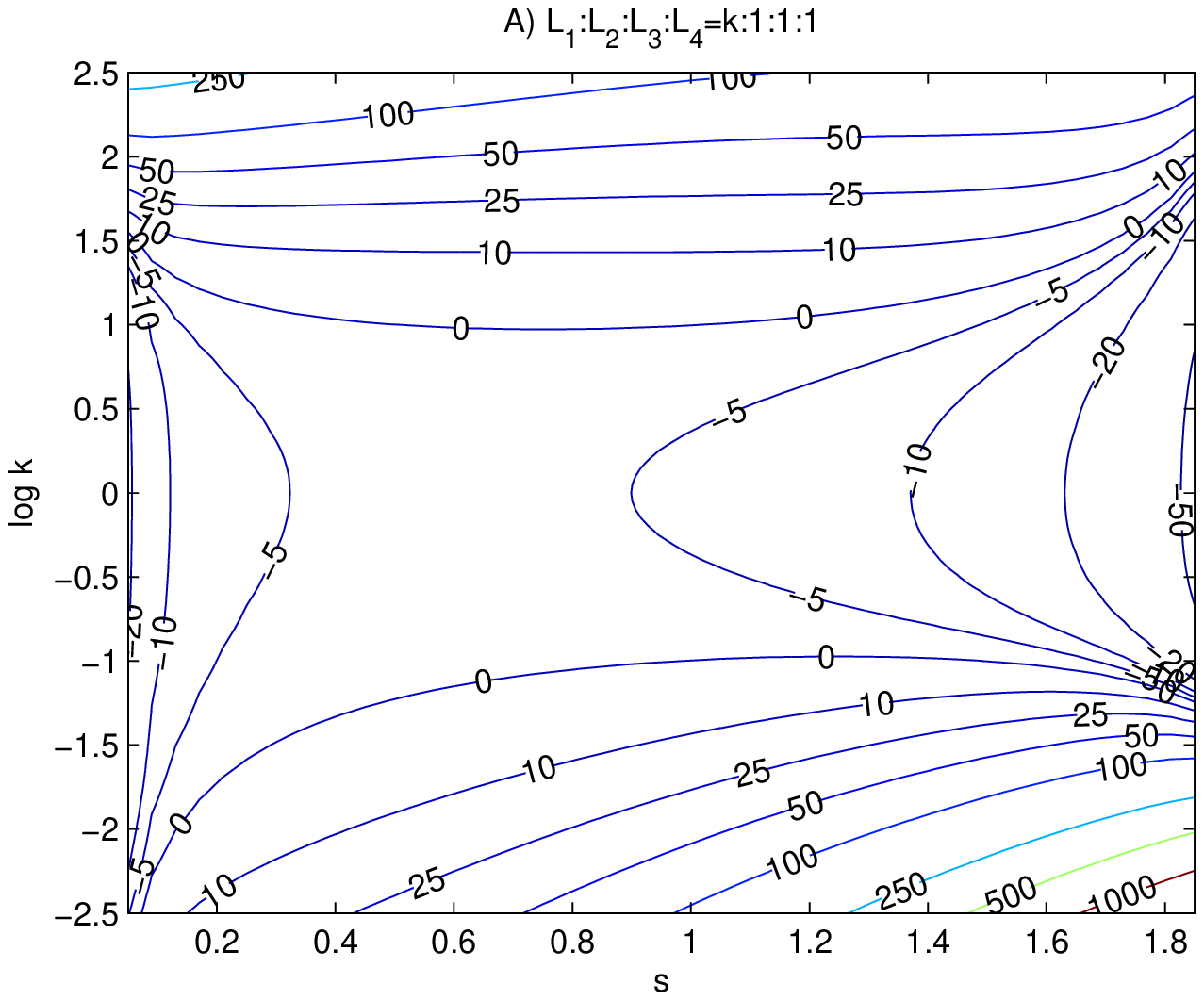} \epsfxsize=.49\linewidth
\epsffile{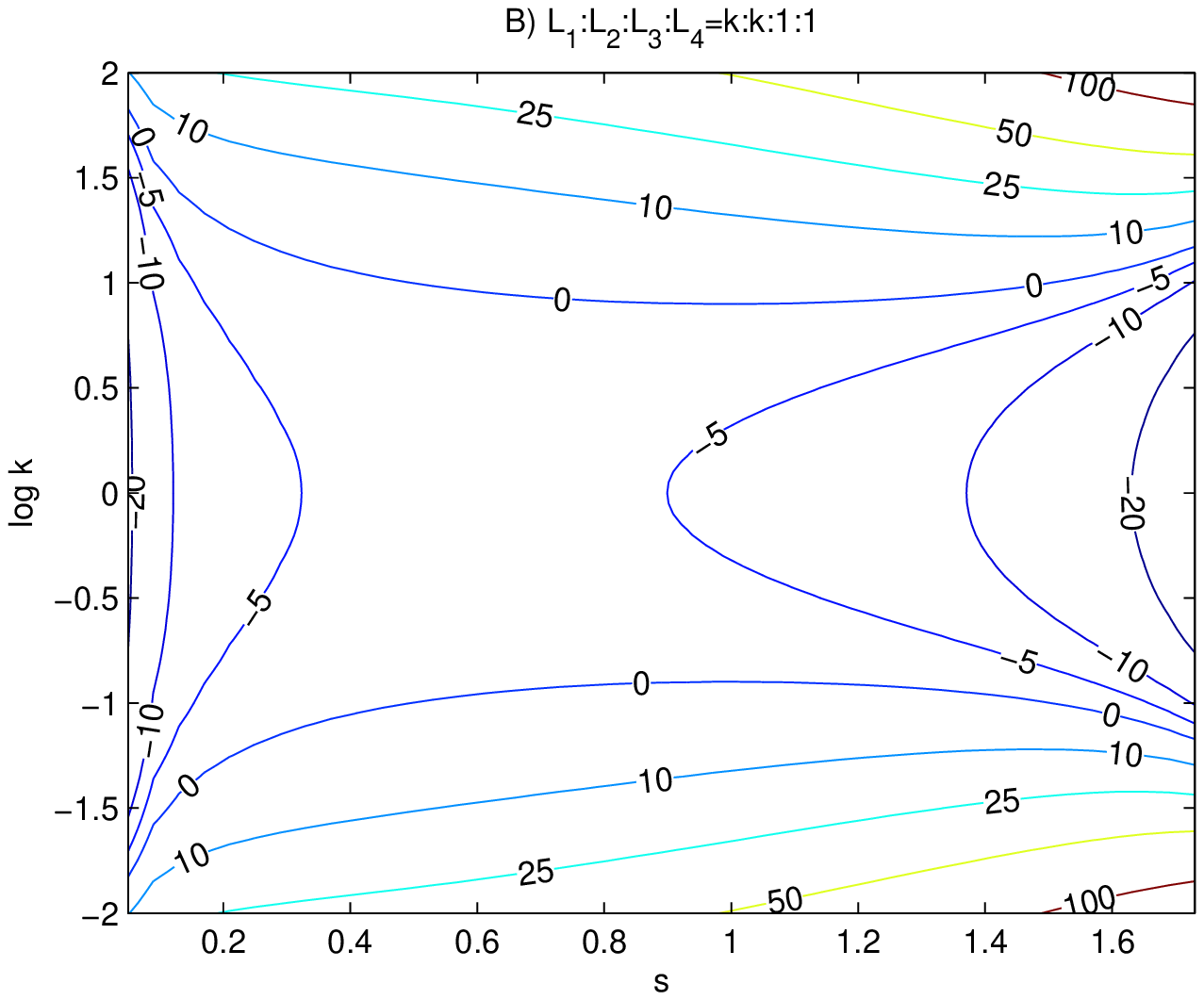} \caption{The contour lines of the graphs in
Figure 5.}\end{figure}

\begin{figure}\centering \epsfxsize=.49\linewidth
\epsffile{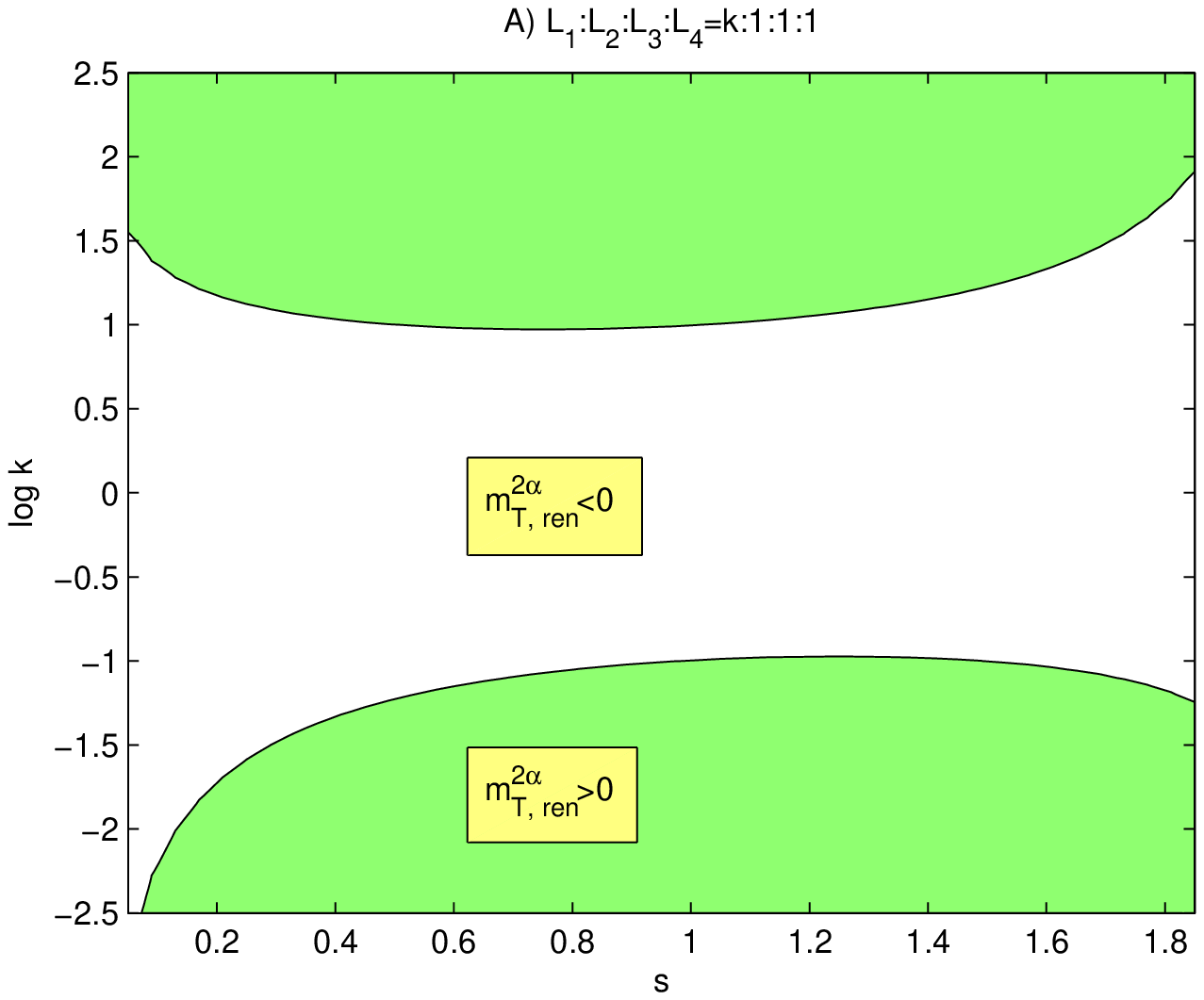}\epsfxsize=.49\linewidth
\epsffile{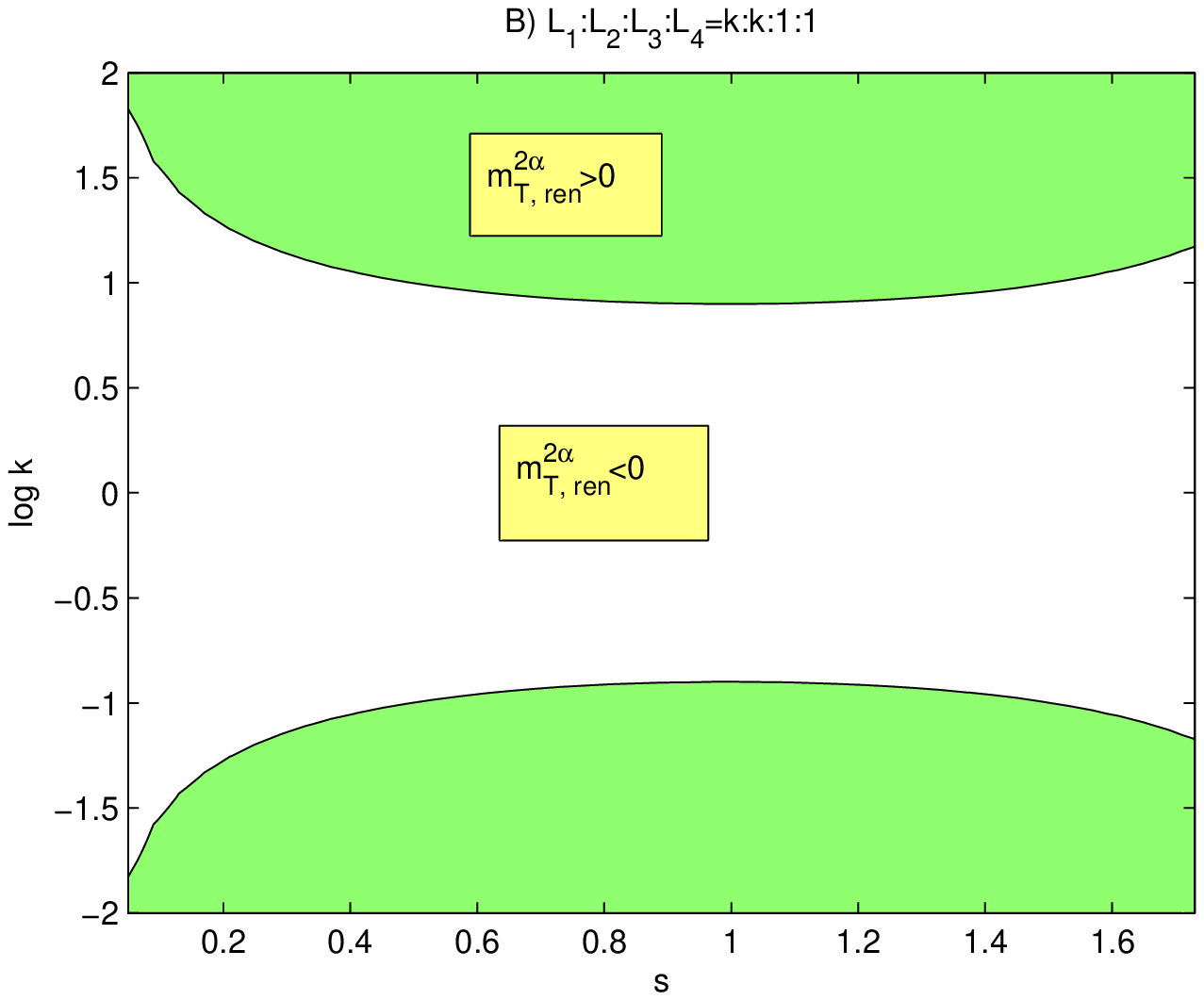}\caption{The  regions where
$m_{T,\text{ren}}^{2\alpha}>0$ and $m_{T,\text{ren}}^{2\alpha}<0$
when $p=4$ and $V=L_1L_2L_3L_4=1$. For A) $L_1:L_2:L_3:L_4=k:1:1:1$.
For B) $L_1:L_2:L_3:L_4=k:k:1:1$. Here $s=\frac{d}{2}-\alpha$.}
\end{figure}

\begin{figure}\centering \epsfxsize=.32\linewidth
\epsffile{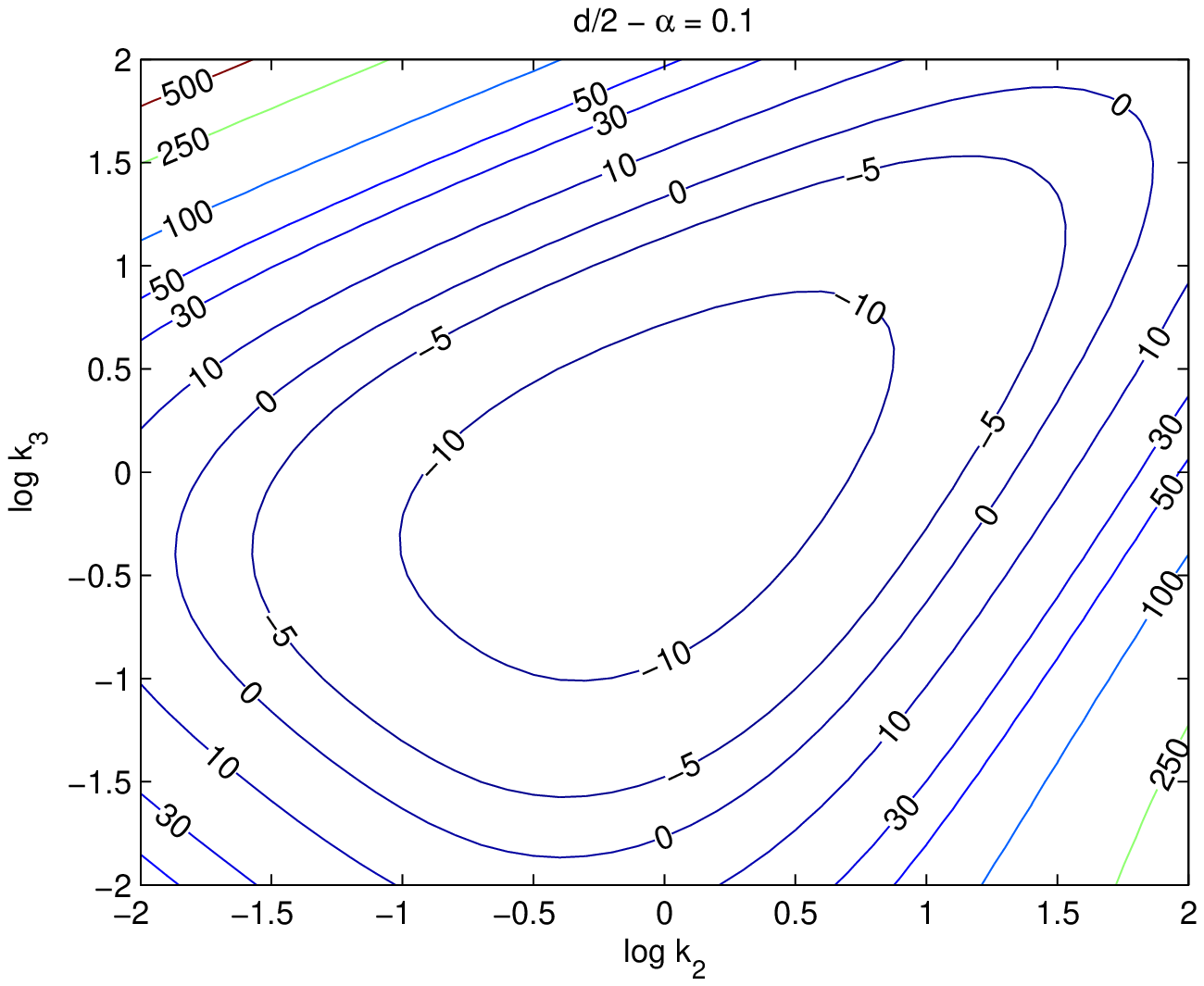}\epsfxsize=.32\linewidth
\epsffile{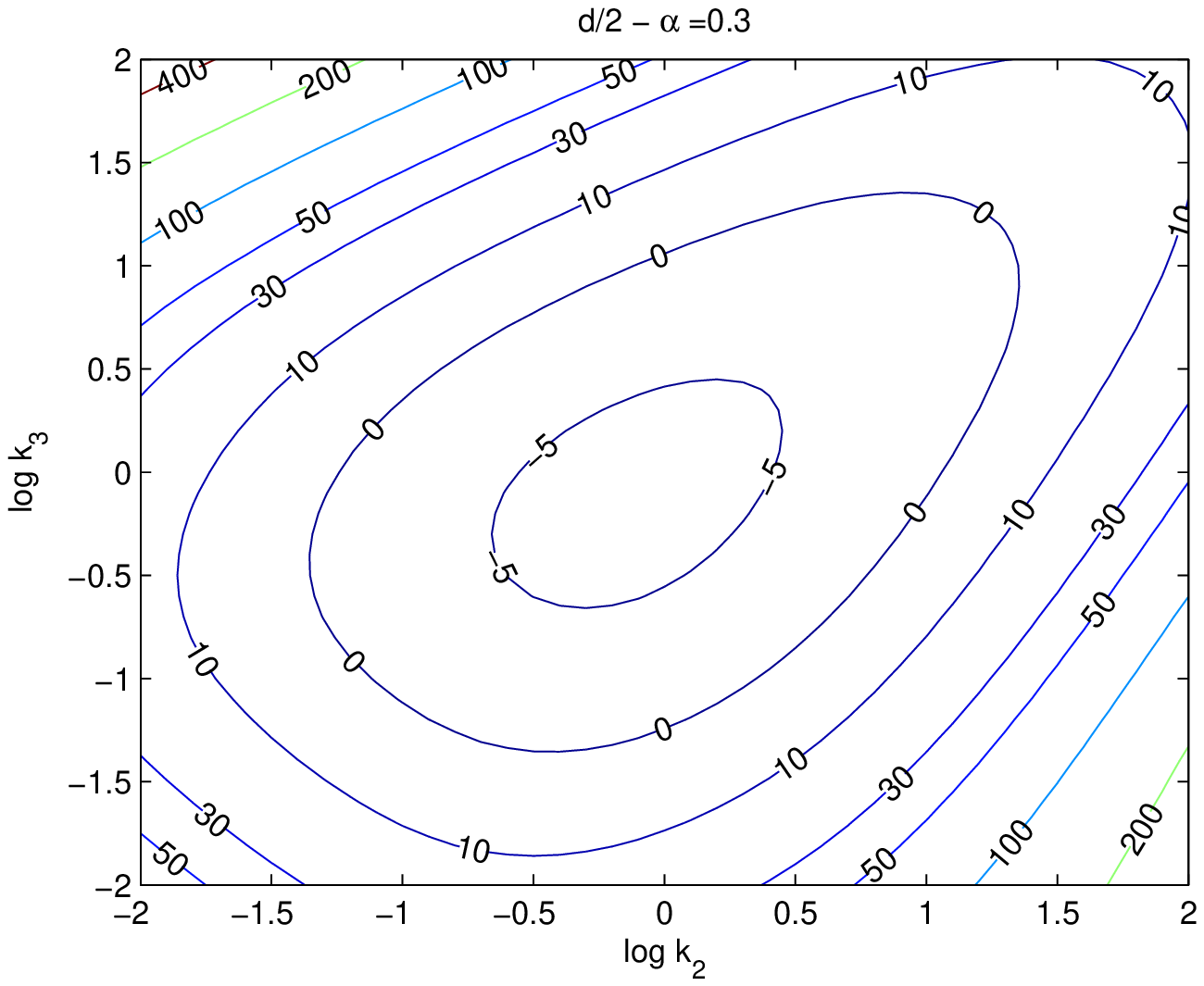}\epsfxsize=.32\linewidth
\epsffile{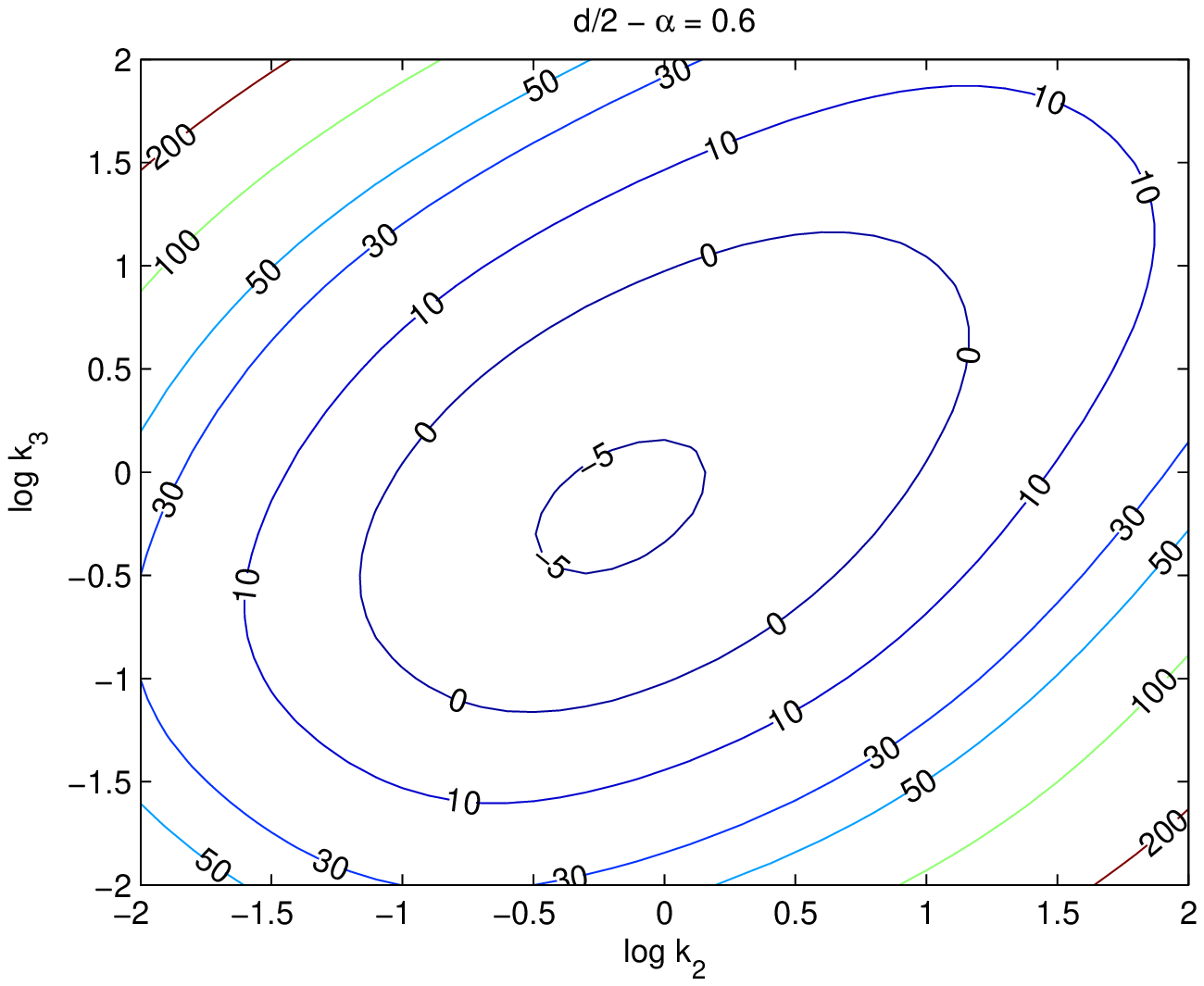} \\ \epsfxsize=.32\linewidth
\epsffile{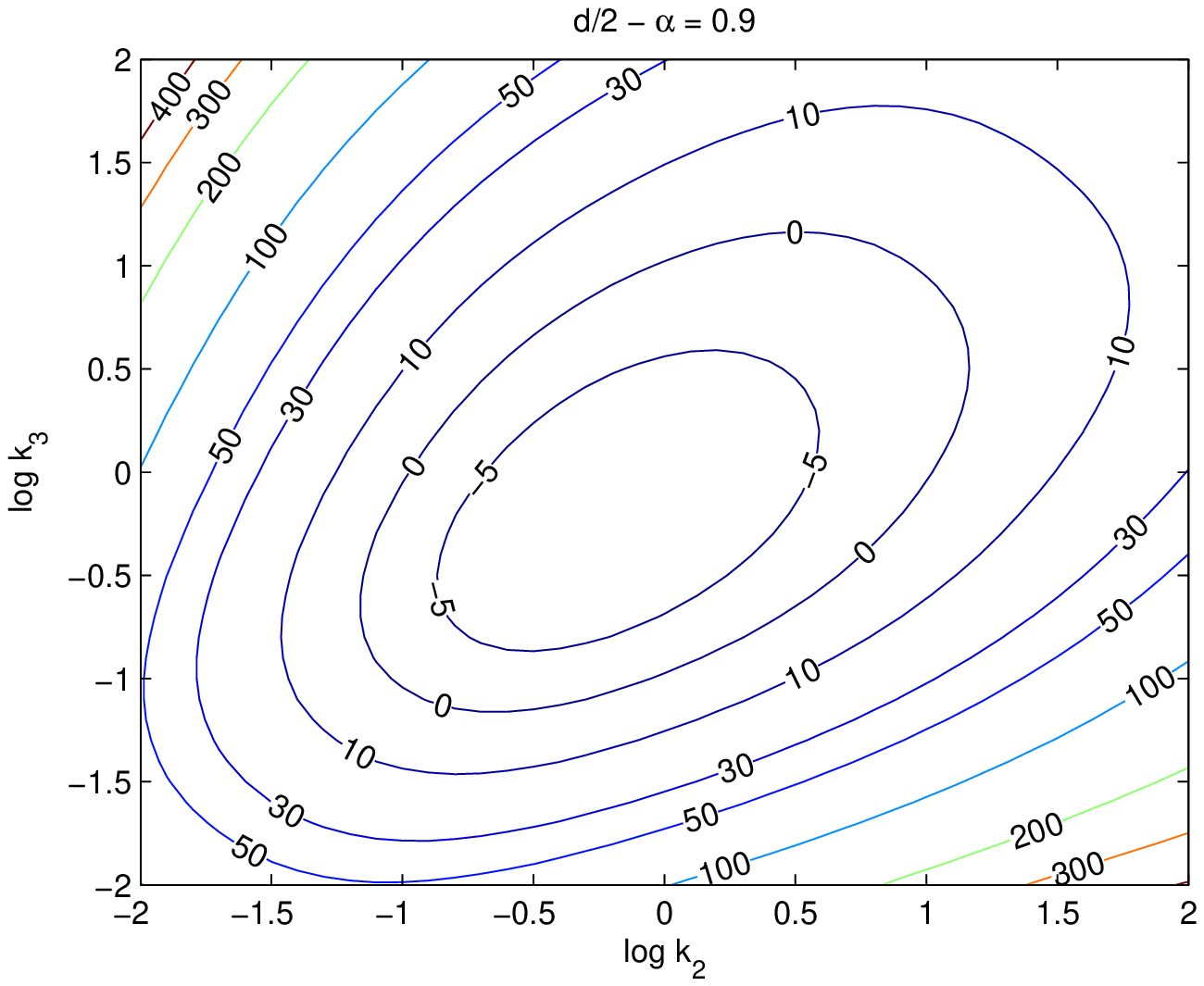} \epsfxsize=.32\linewidth
\epsffile{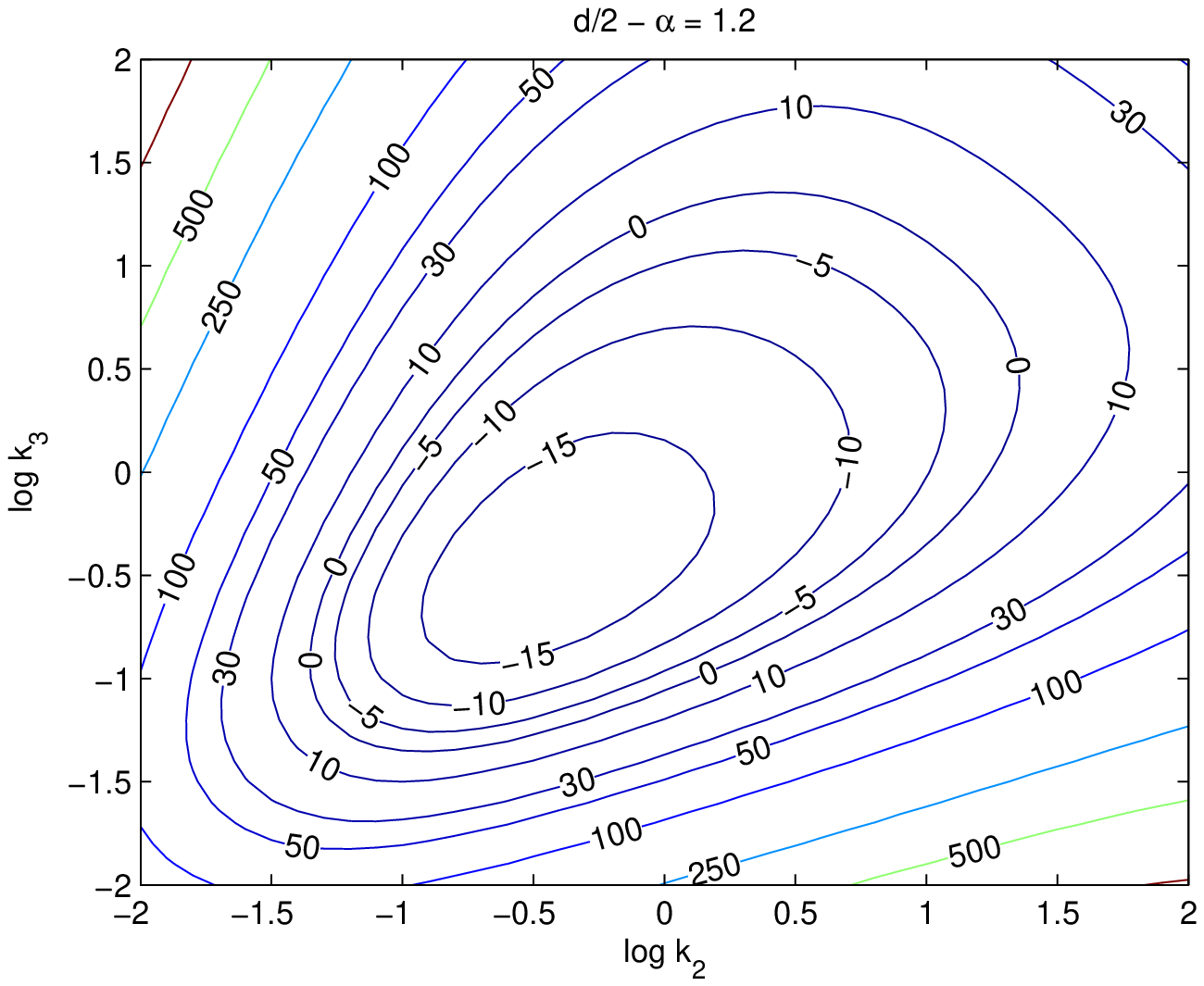} \epsfxsize=.32\linewidth
\epsffile{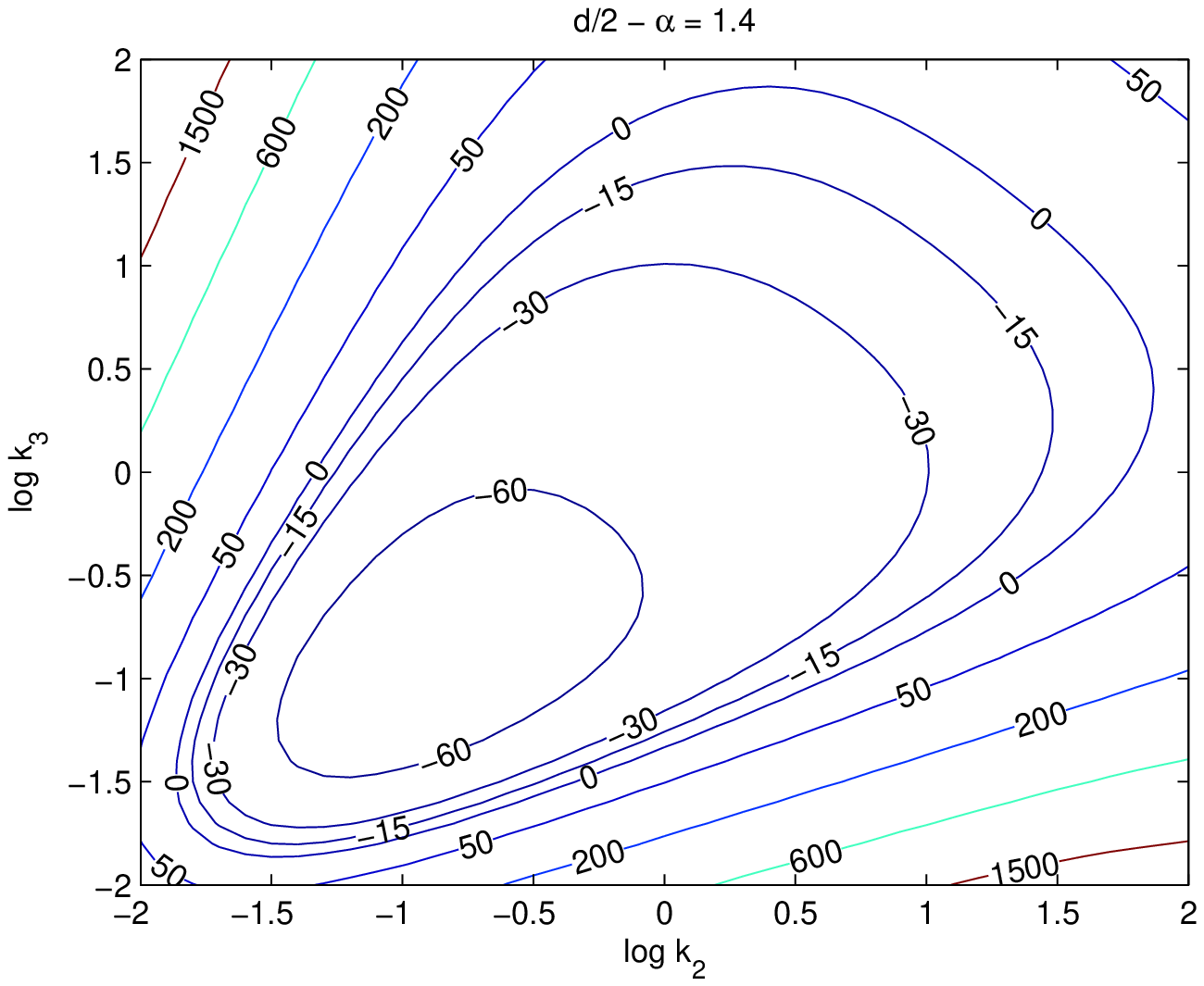} \caption{The contour lines of the
renormalized mass $m_{T, \text{\text{ren}}}^{2\alpha}$ (up to the
factor \eqref{eq104_3}) as a function of $\log k_2$ and $\log k_3$.
Here $p=3$, $V=L_1L_2L_3=1$ and  $L_1:L_2:L_3=1:k_2:k_3$. The values
of $\frac{d}{2}-\alpha$ are $0.3, 0.6, 0.9, 1.2, 1.4$
respectively.}\end{figure}

\begin{figure}\centering \epsfxsize=.32\linewidth
\epsffile{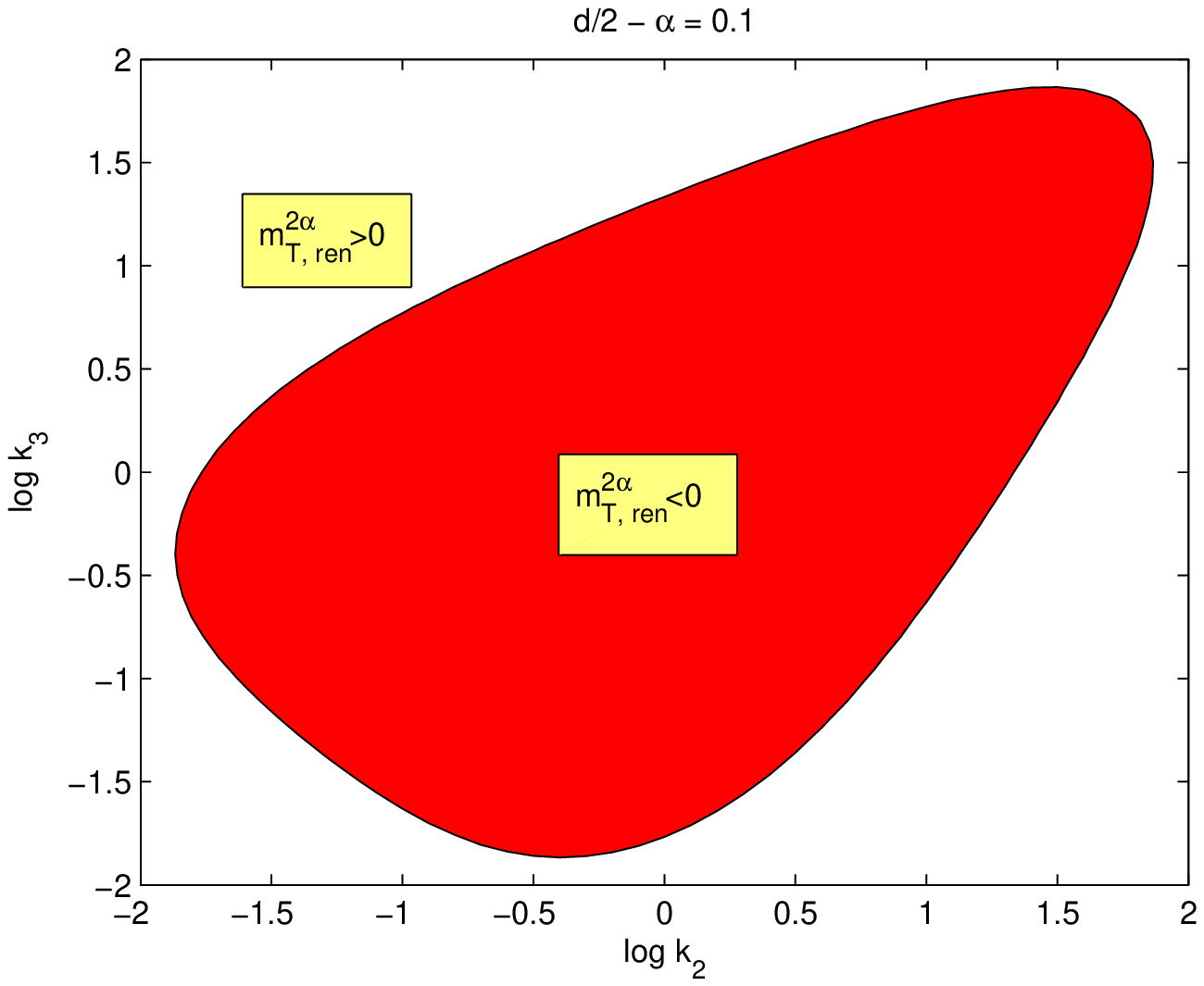}\epsfxsize=.32\linewidth
\epsffile{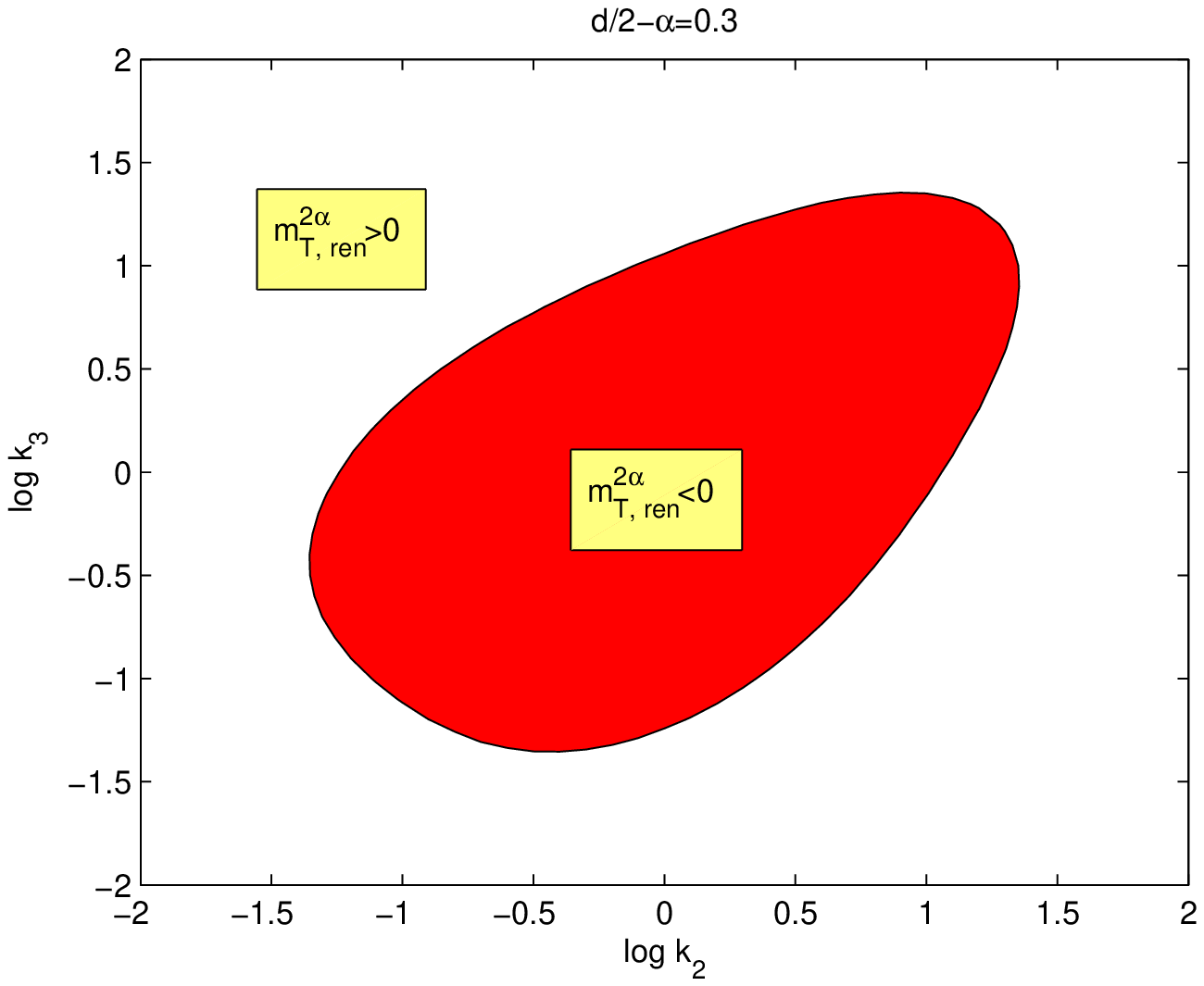} \epsfxsize=.32\linewidth
\epsffile{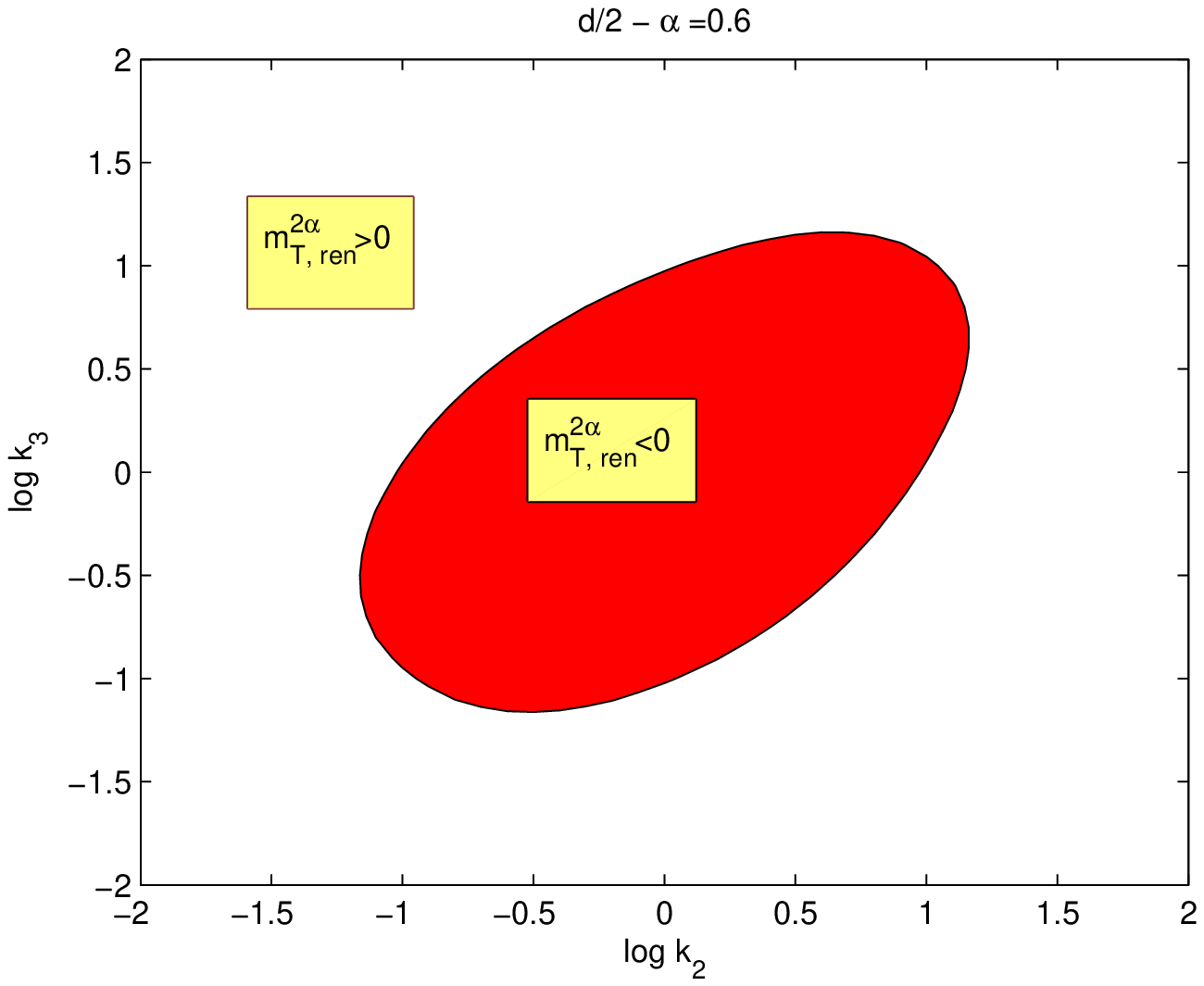}\\
\epsfxsize=.32\linewidth \epsffile{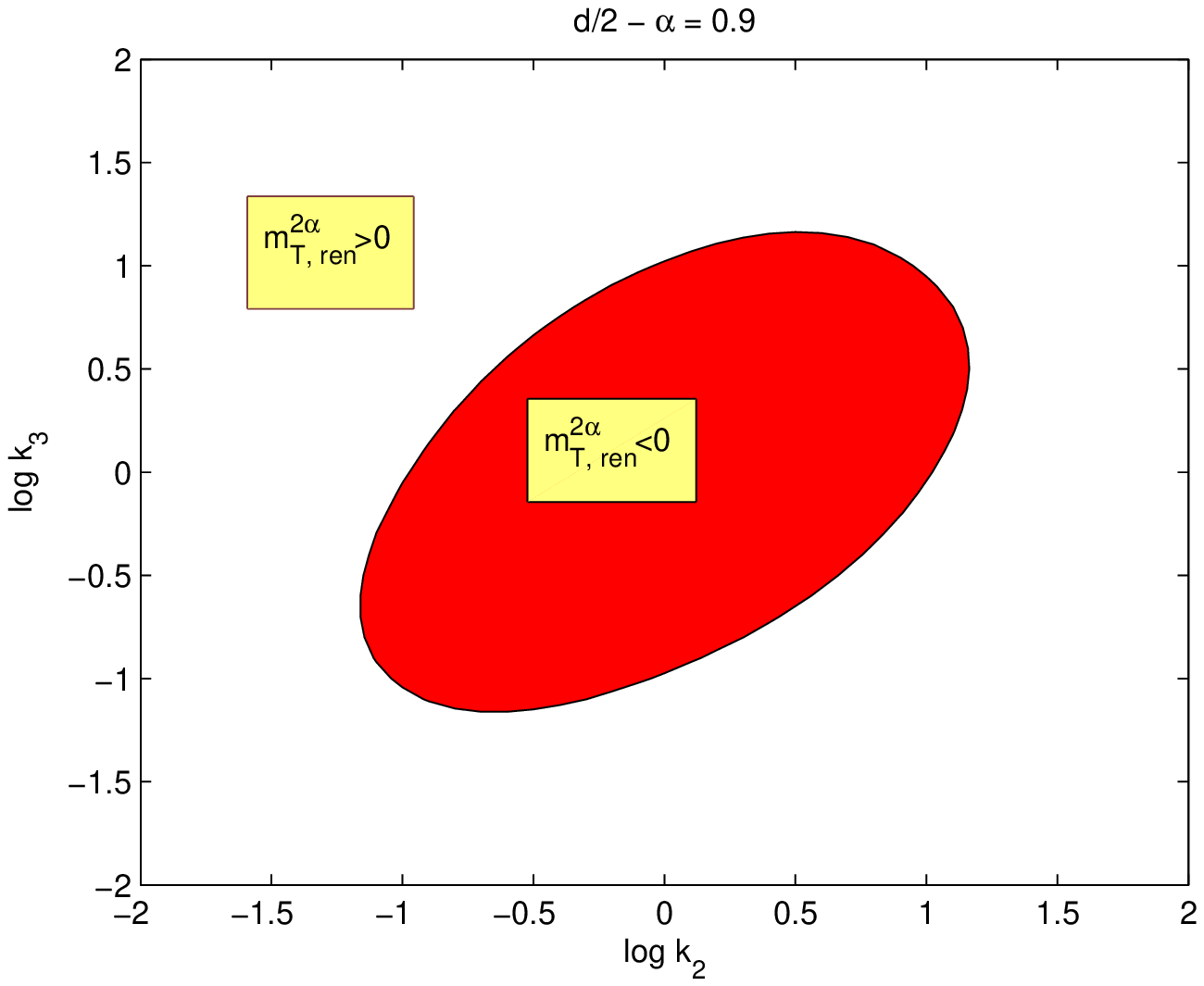}
\epsfxsize=.32\linewidth \epsffile{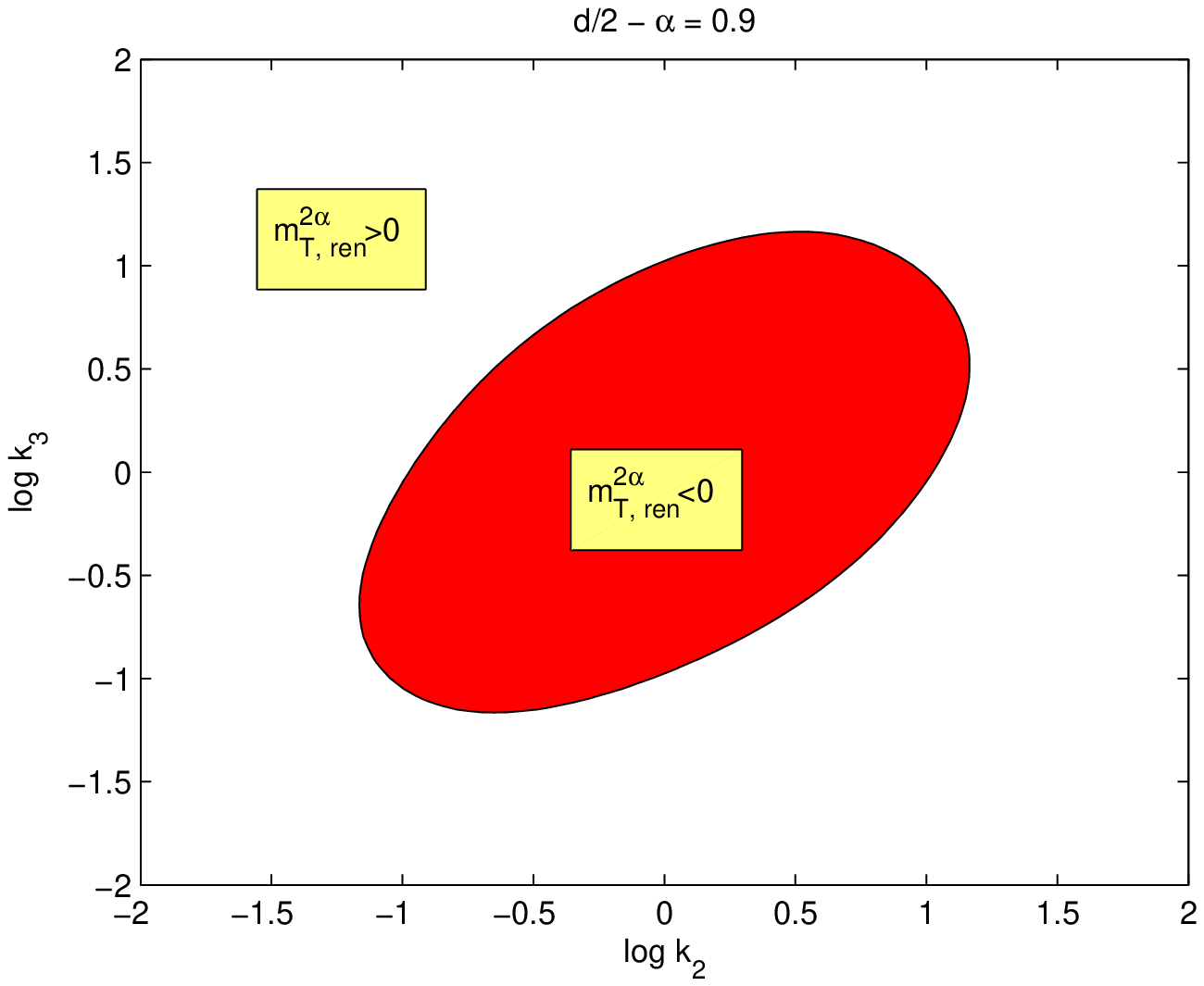}
\epsfxsize=.32\linewidth \epsffile{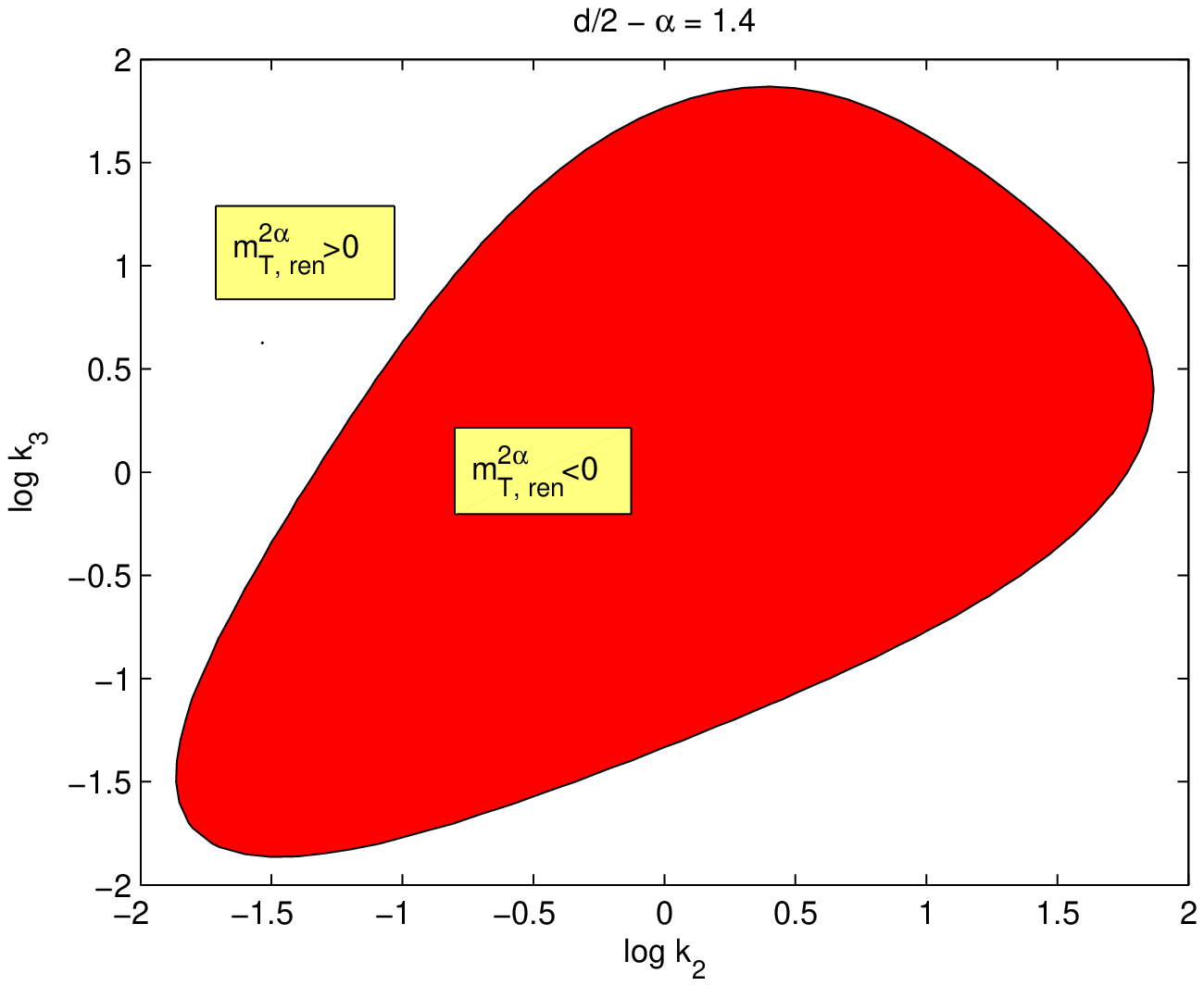}\caption{The regions
where $m_{T,\text{ren}}^{2\alpha}>0$ and
$m_{T,\text{ren}}^{2\alpha}<0$ for $p=3$,
 $V=L_1L_2L_3=1$,
  and
$L_1:L_2:L_3=1:k_2:k_3$. Here the values of $\frac{d}{2}-\alpha$
 are $0.1, 0.3, 0.6,0.9, 1.2, 1.4$. }

\end{figure}

\begin{figure}\centering \epsfxsize=.32\linewidth
\epsffile{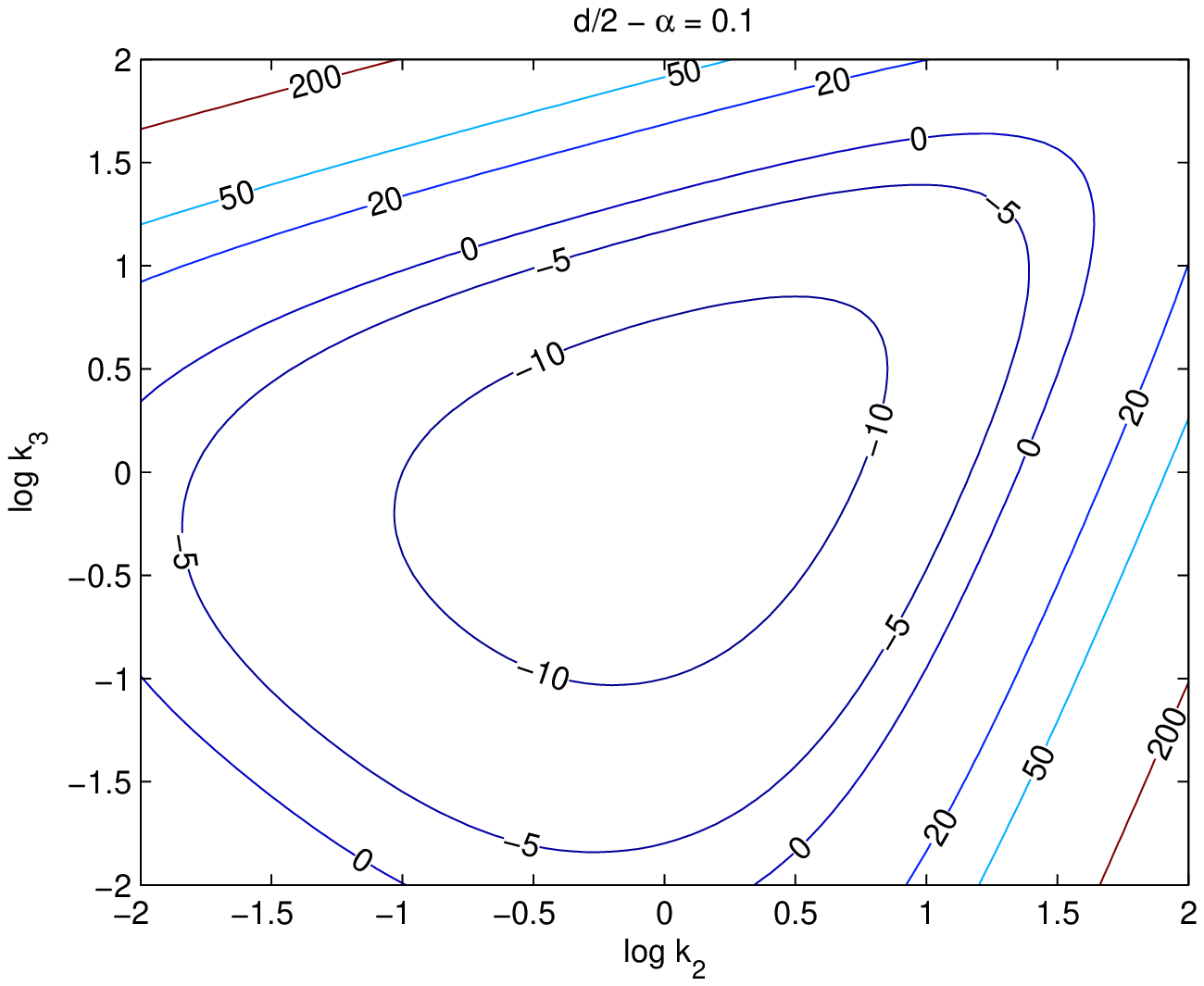}\epsfxsize=.32\linewidth
\epsffile{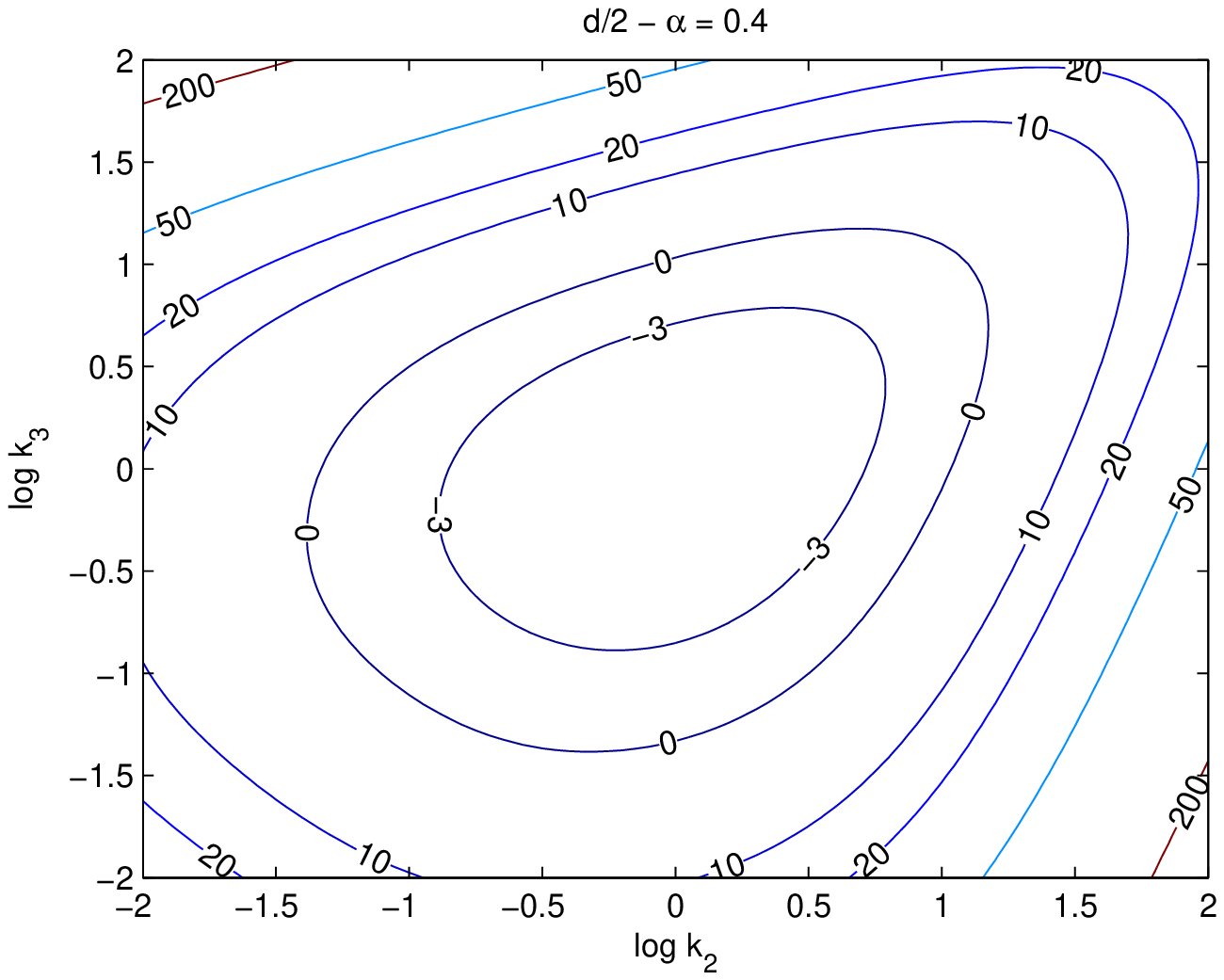}\epsfxsize=.32\linewidth
\epsffile{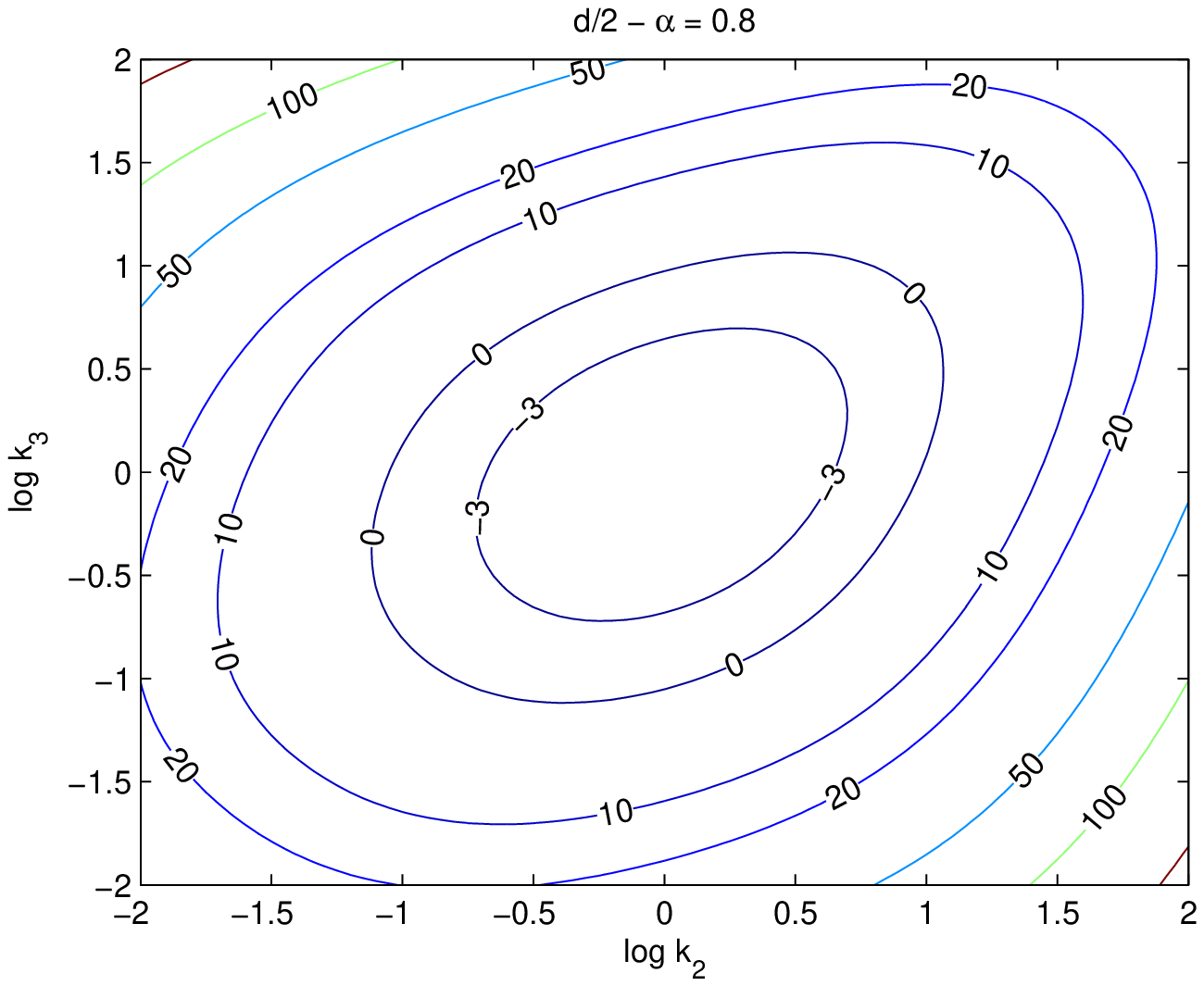}\\ \epsfxsize=.32\linewidth
\epsffile{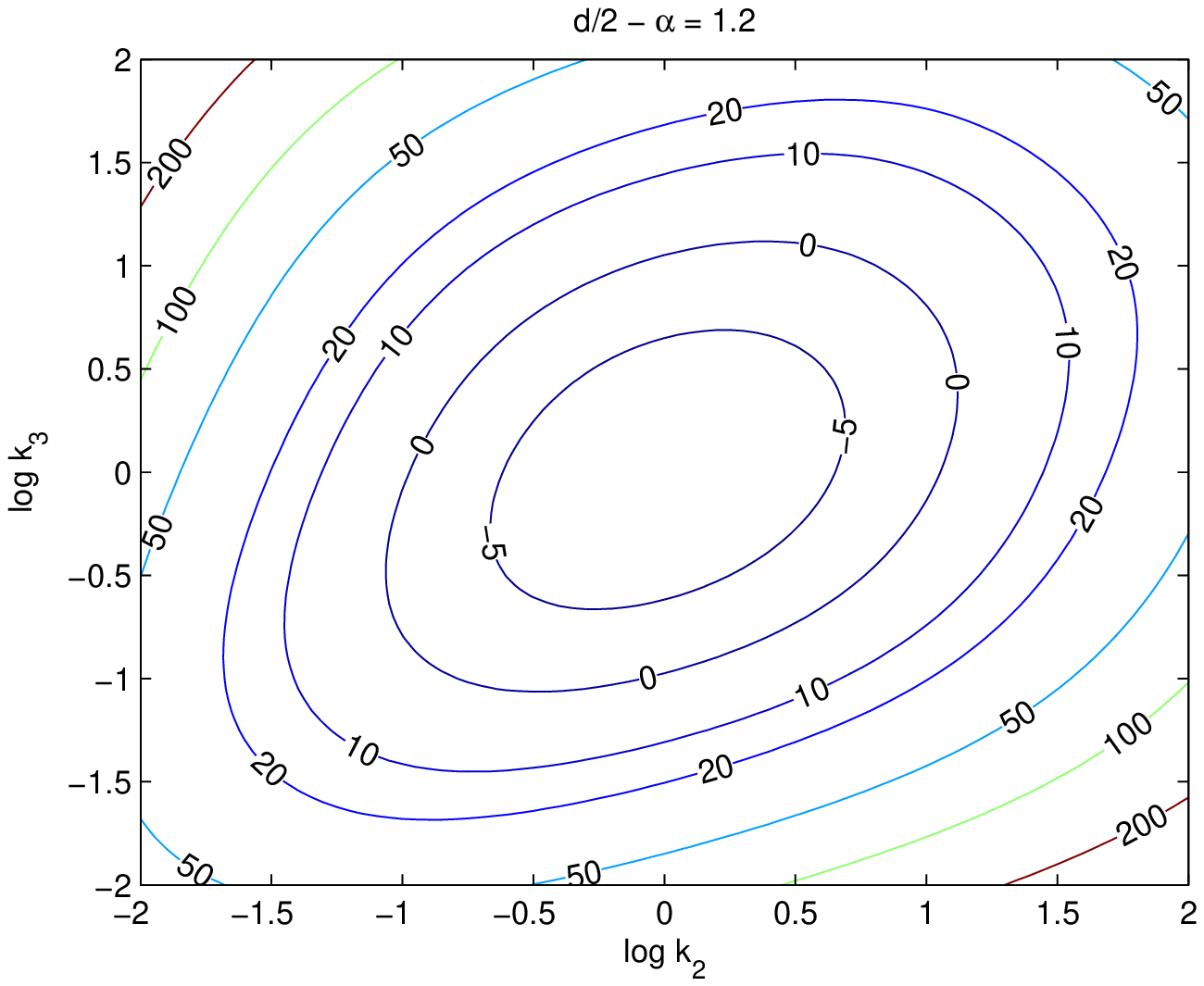} \epsfxsize=.32\linewidth
\epsffile{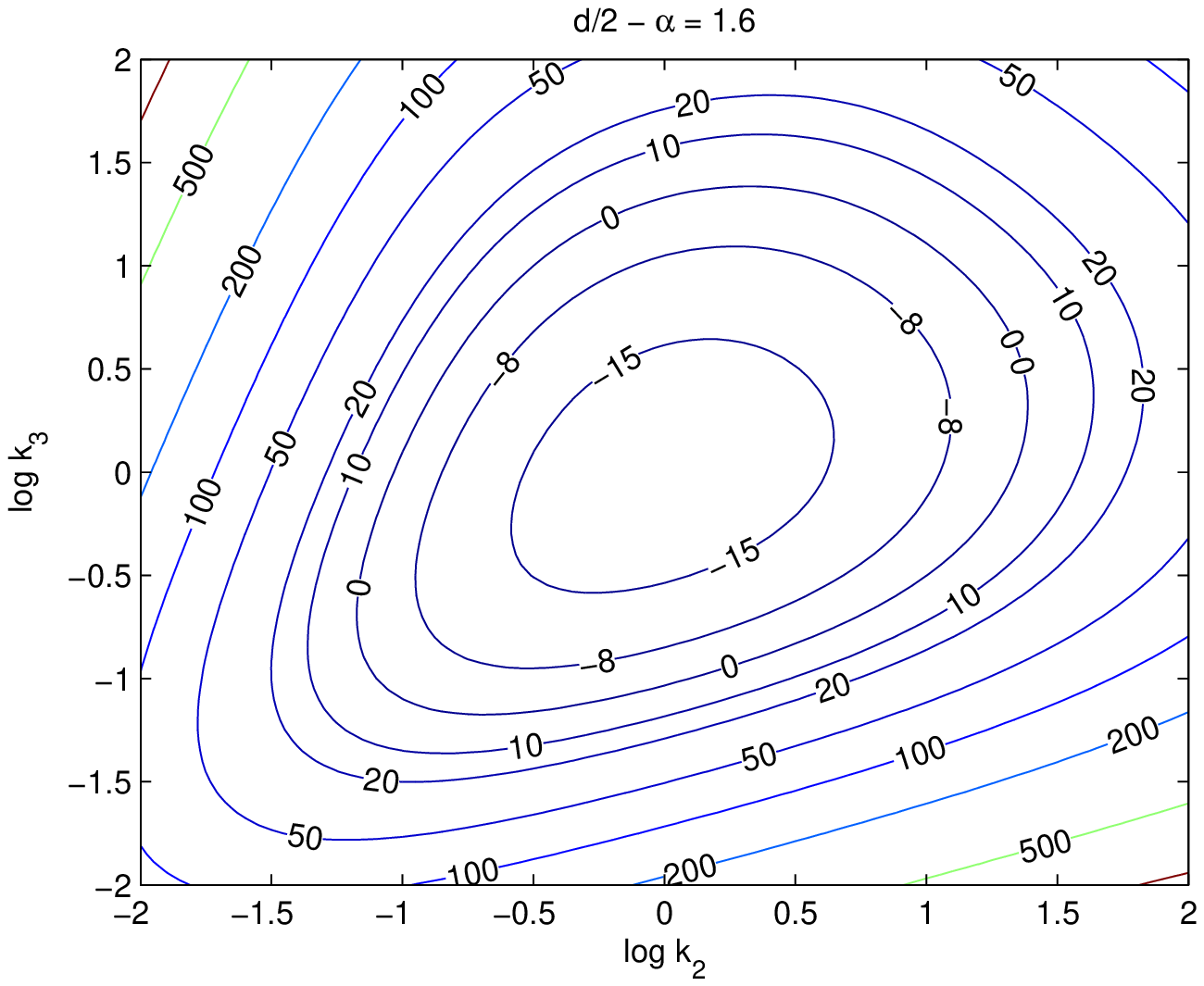} \epsfxsize=.32\linewidth
\epsffile{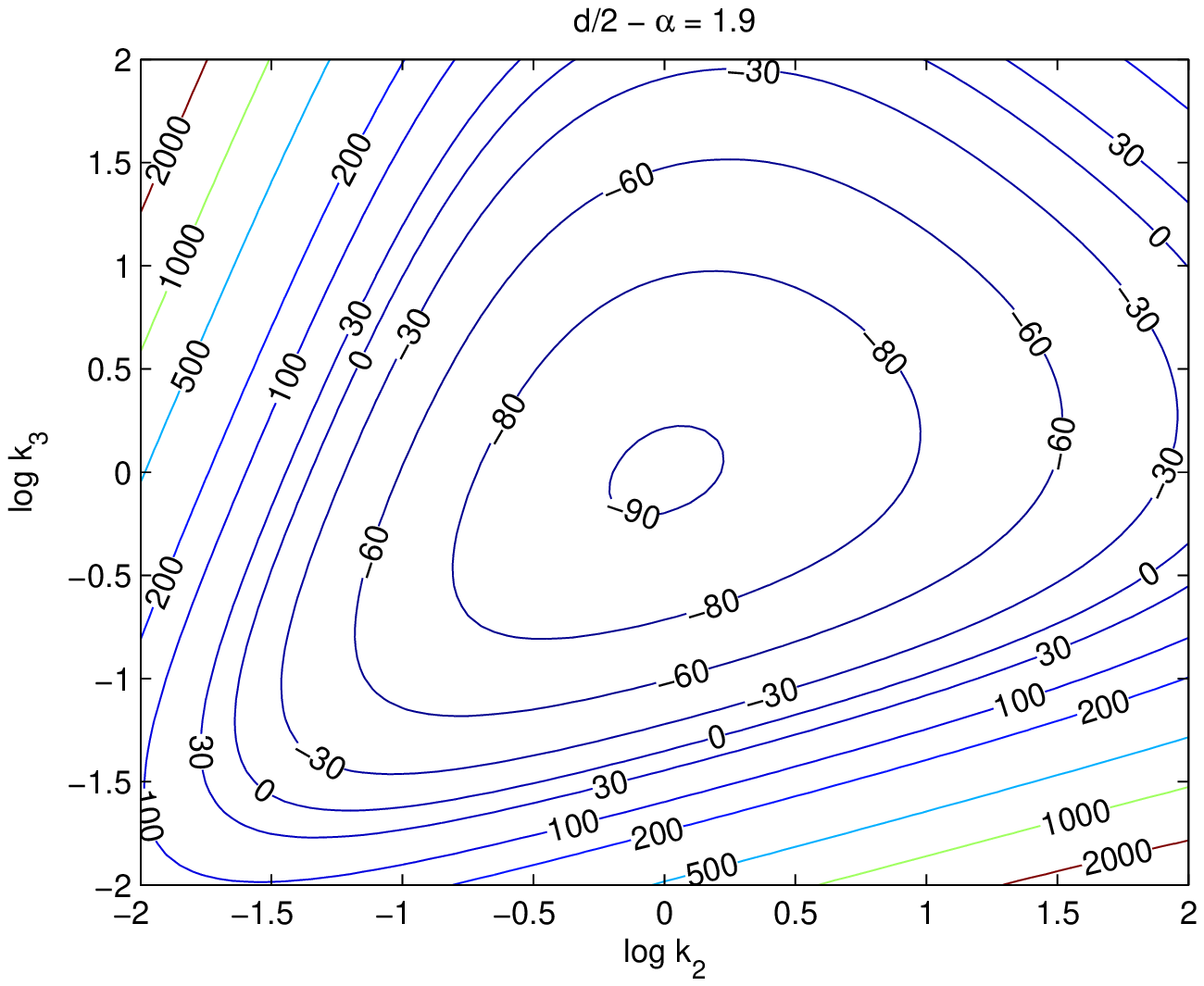} \caption{The contour lines of the
renormalized mass $m_{T,\text{ren}}^{2\alpha}$ (up to the factor
\eqref{eq104_3}) as a function of $\log k_2$ and $\log k_3$, when
$p=4$, $V=L_1L_2L_3L_4=1$, $L_1:L_2:L_3=1:k_2:k_3:k_4$ and $k_4=1$.
Here the values of $\frac{d}{2}-\alpha$ are $0.1, 0.4, 0.8, 1.2,
1.6, 1.9$.}\end{figure}

\begin{figure}\centering \epsfxsize=.32\linewidth
\epsffile{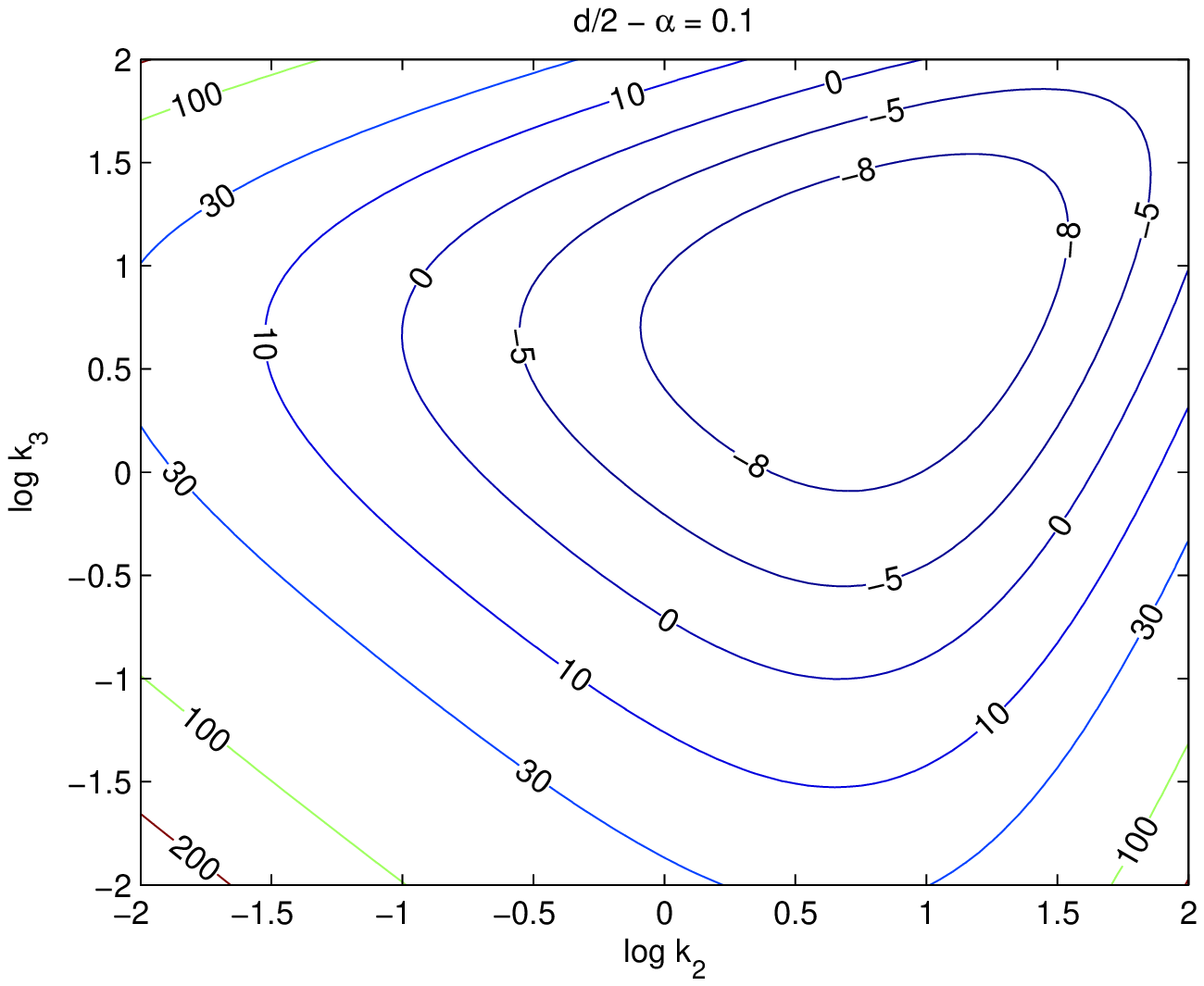}\epsfxsize=.32\linewidth
\epsffile{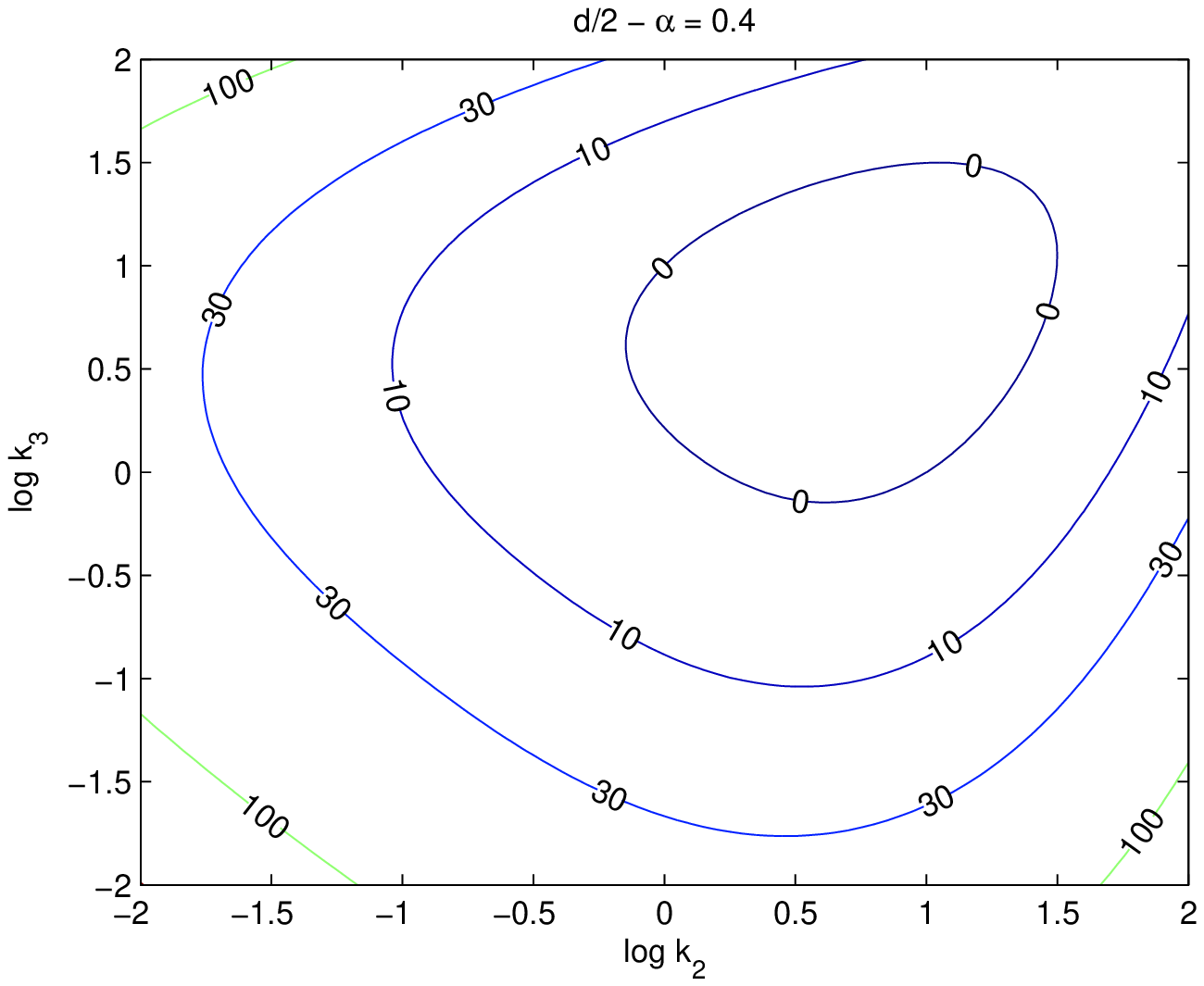}\epsfxsize=.32\linewidth
\epsffile{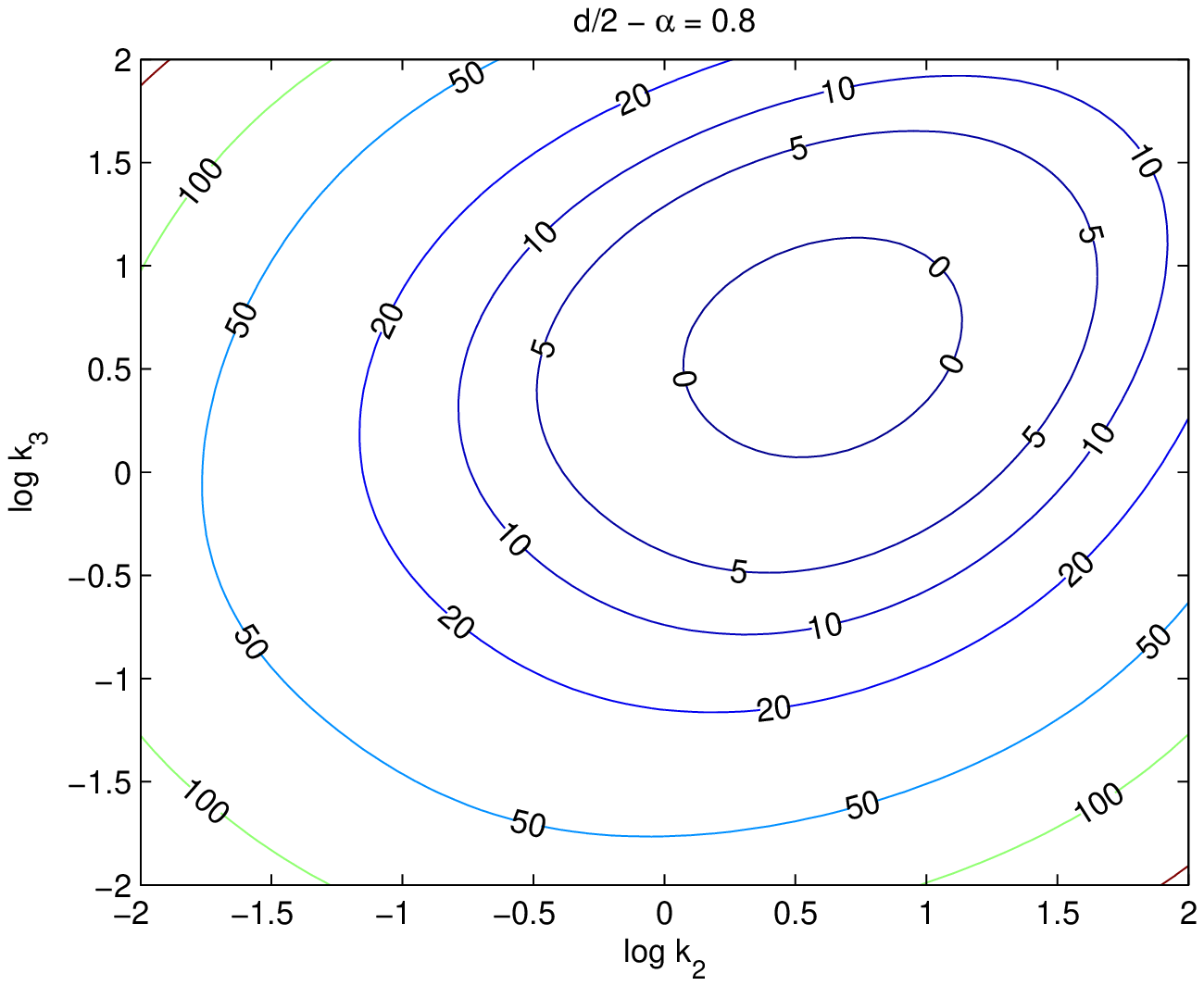} \\\epsfxsize=.32\linewidth
\epsffile{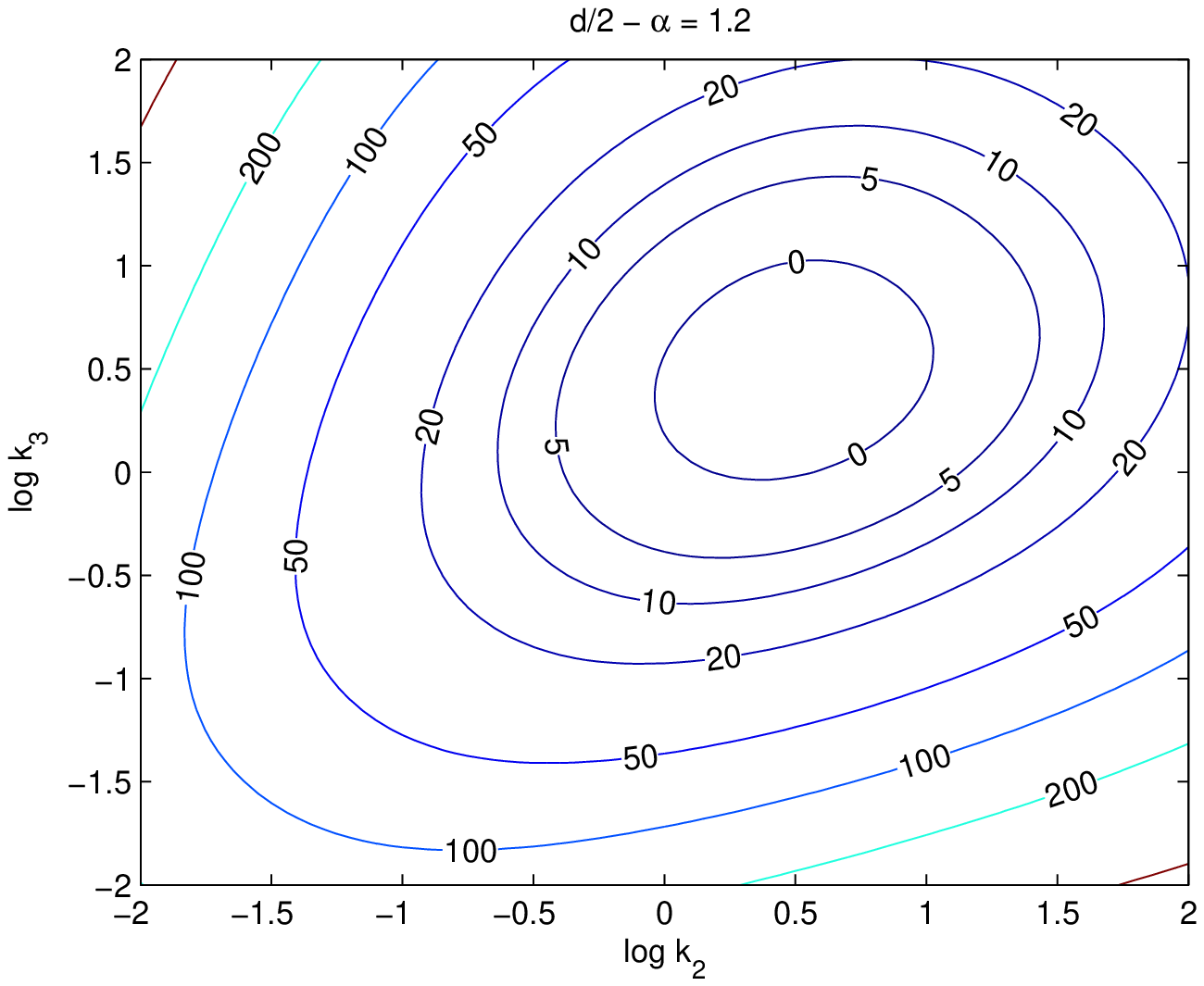} \epsfxsize=.32\linewidth
\epsffile{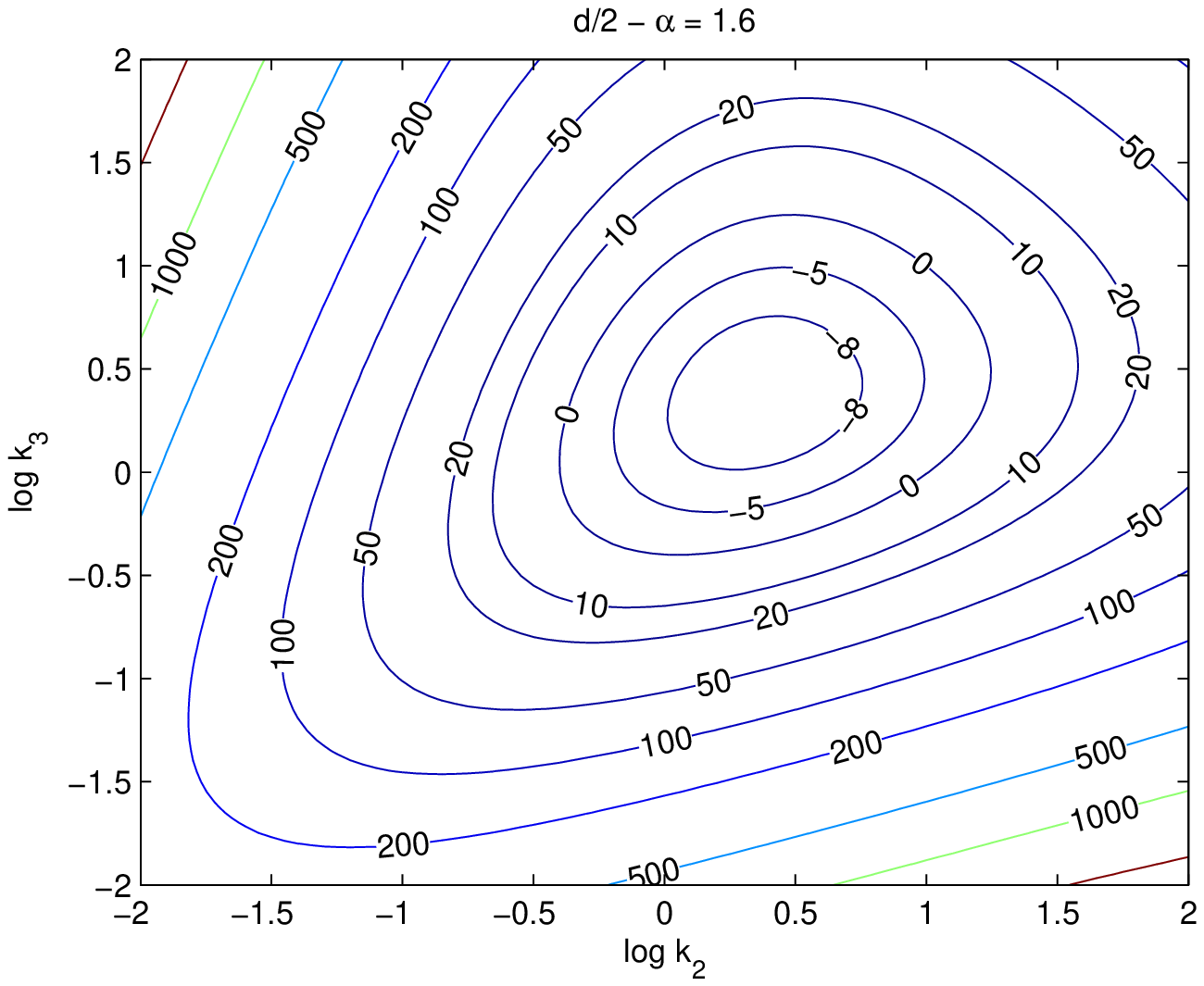} \epsfxsize=.32\linewidth
\epsffile{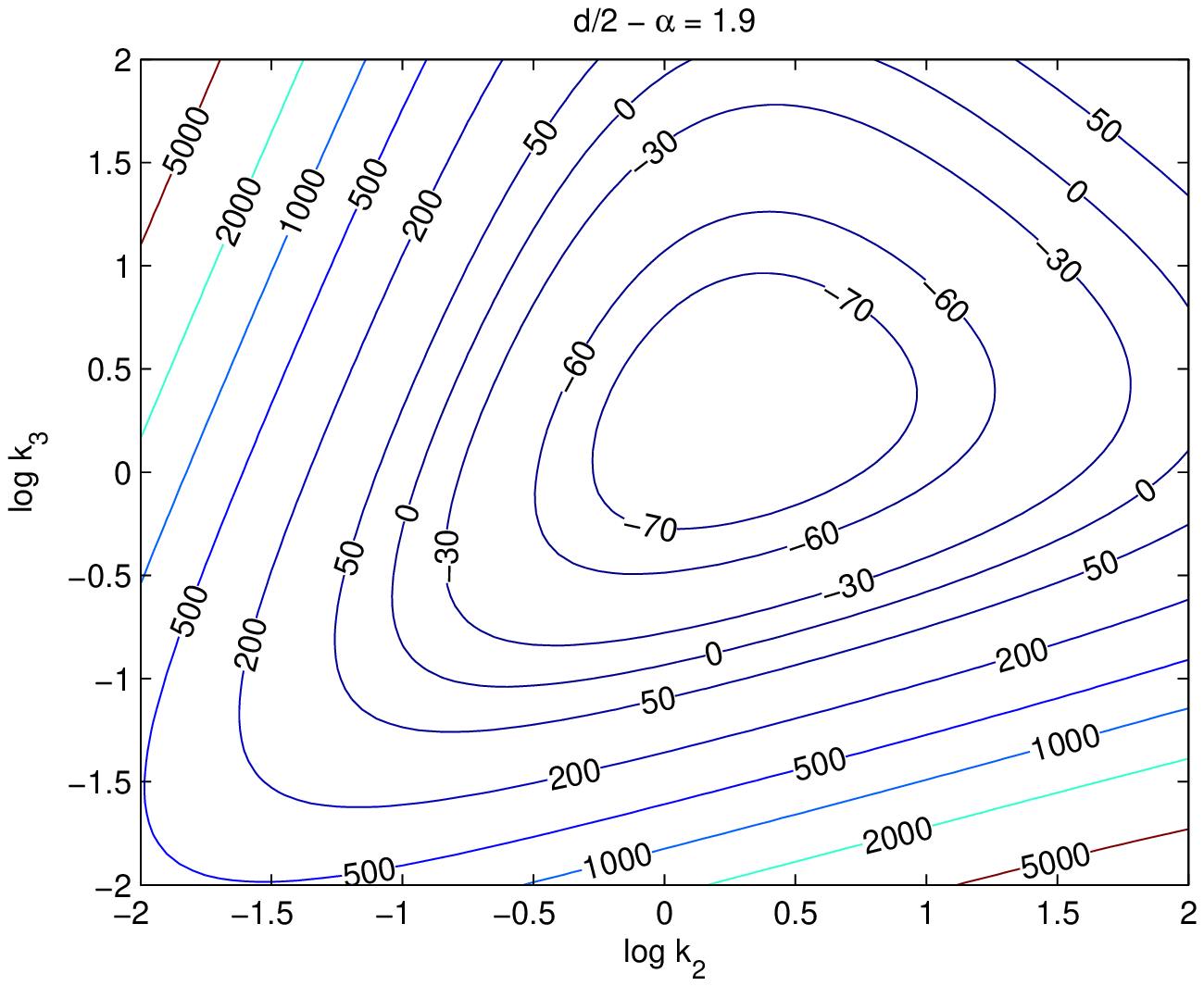} \caption{The contour lines of the
renormalized mass $m_{T,\text{ren}}^{2\alpha}$ (up to the factor
\eqref{eq104_3}) as a function of $\log k_2$ and $\log k_3$, when
$p=4$, $V=L_1L_2L_3L_4=1$, $L_1:L_2:L_3=1:k_2:k_3:k_4$ and $k_4=3$.
Here the values of $\frac{d}{2}-\alpha$ are $0.1, 0.4, 0.8, 1.2,
1.6, 1.9$.}\end{figure}

\begin{figure}\centering \epsfxsize=.32\linewidth
\epsffile{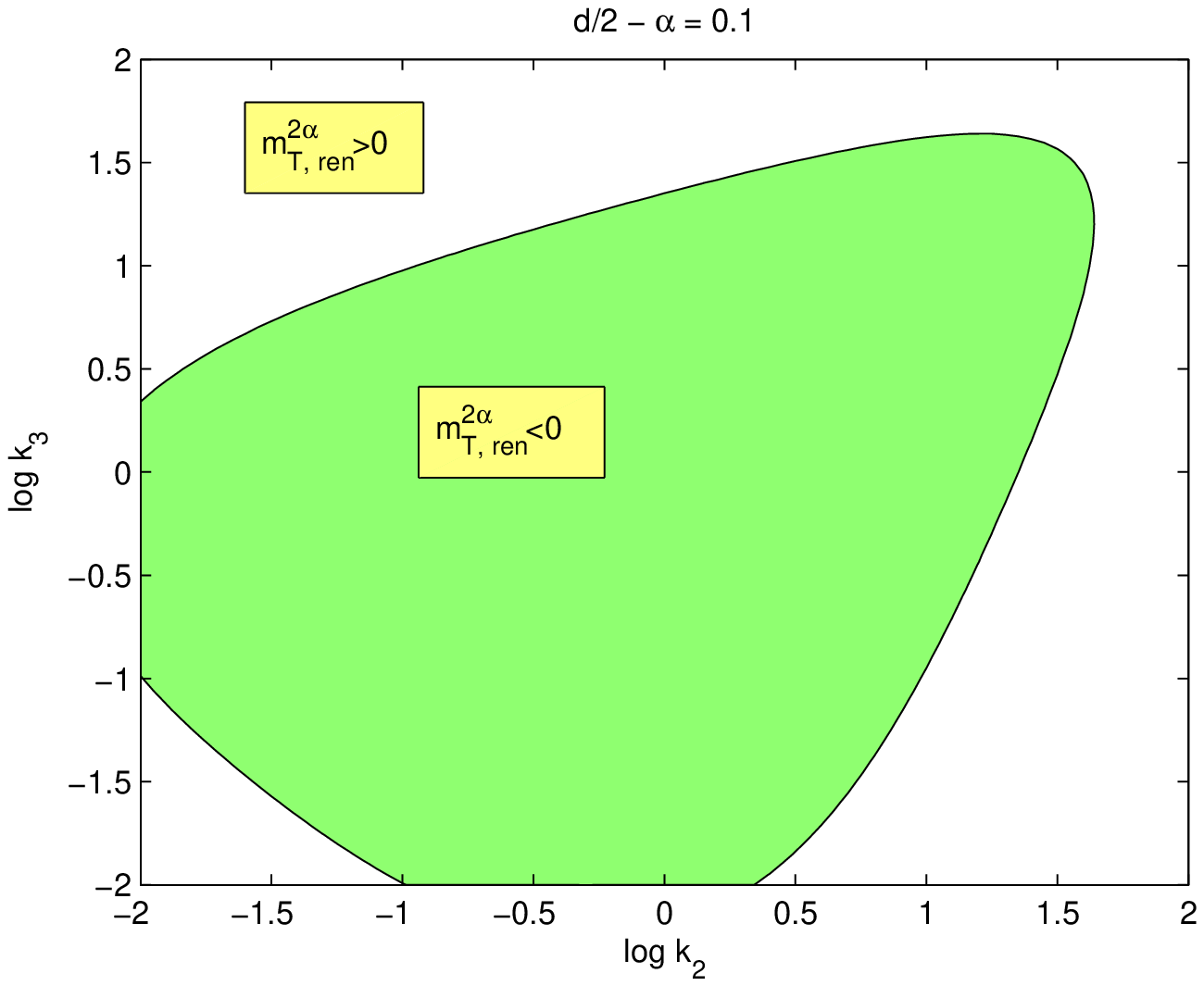}\epsfxsize=.32\linewidth
\epsffile{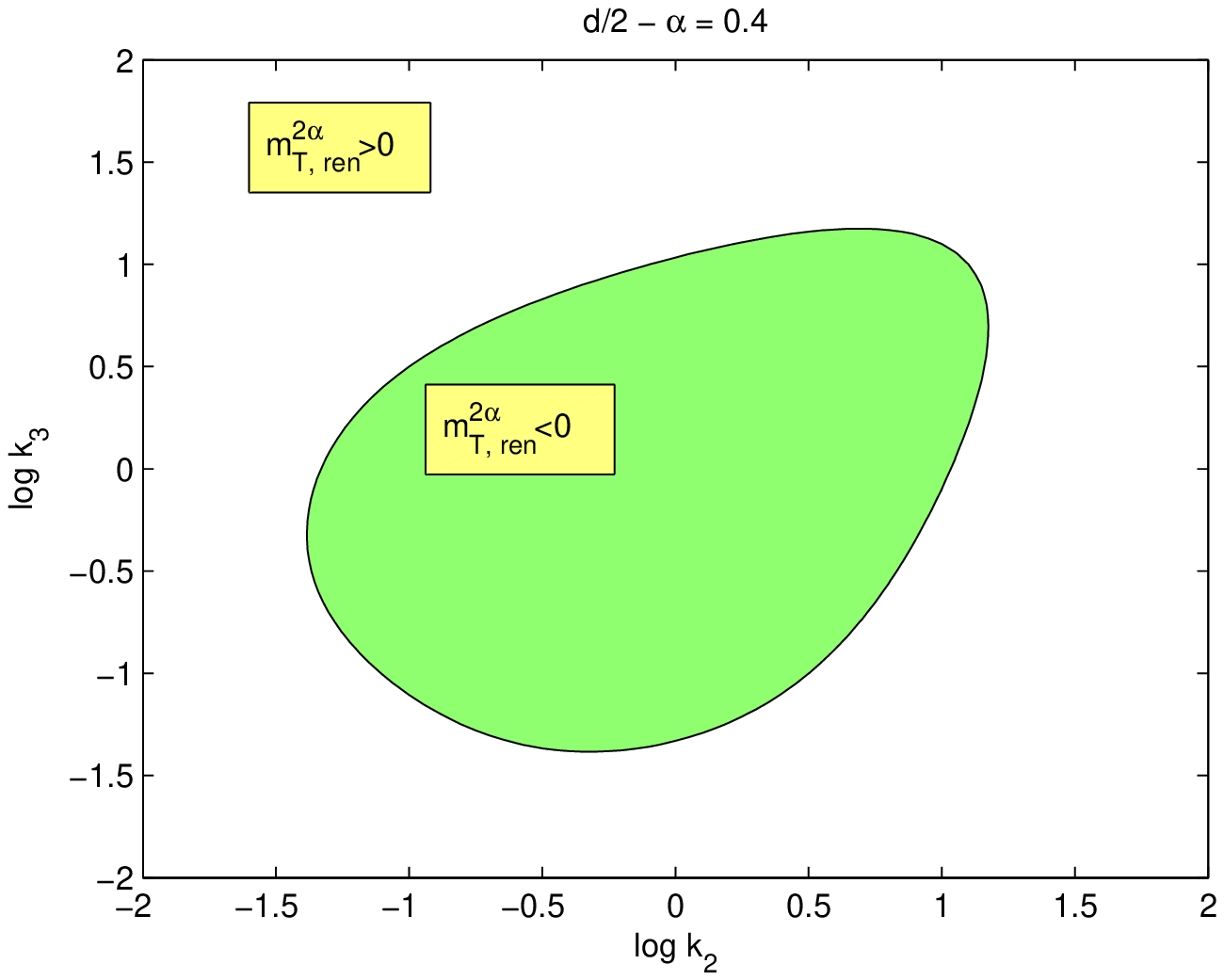}
\epsfxsize=.32\linewidth \epsffile{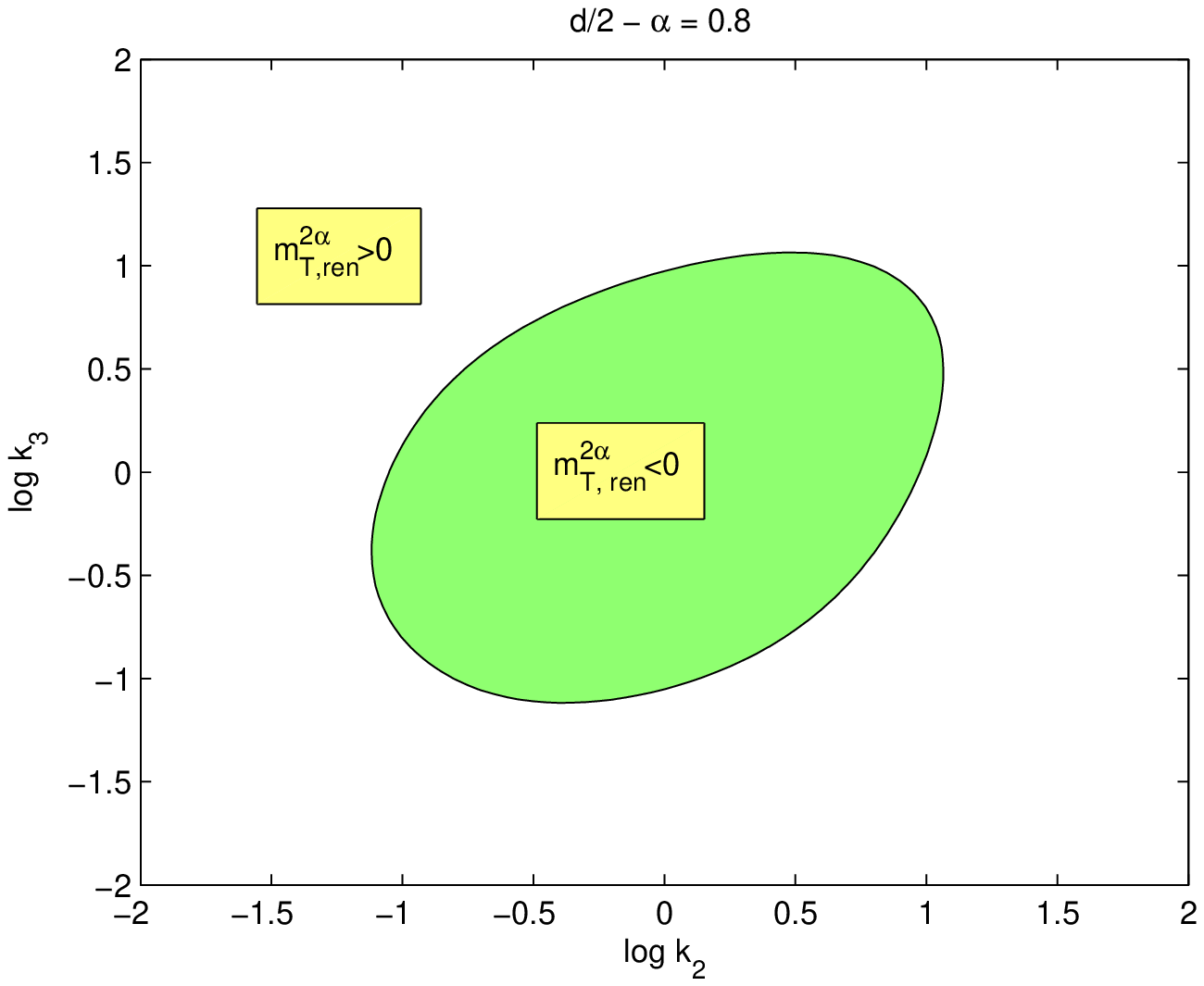}\\
\epsfxsize=.32\linewidth \epsffile{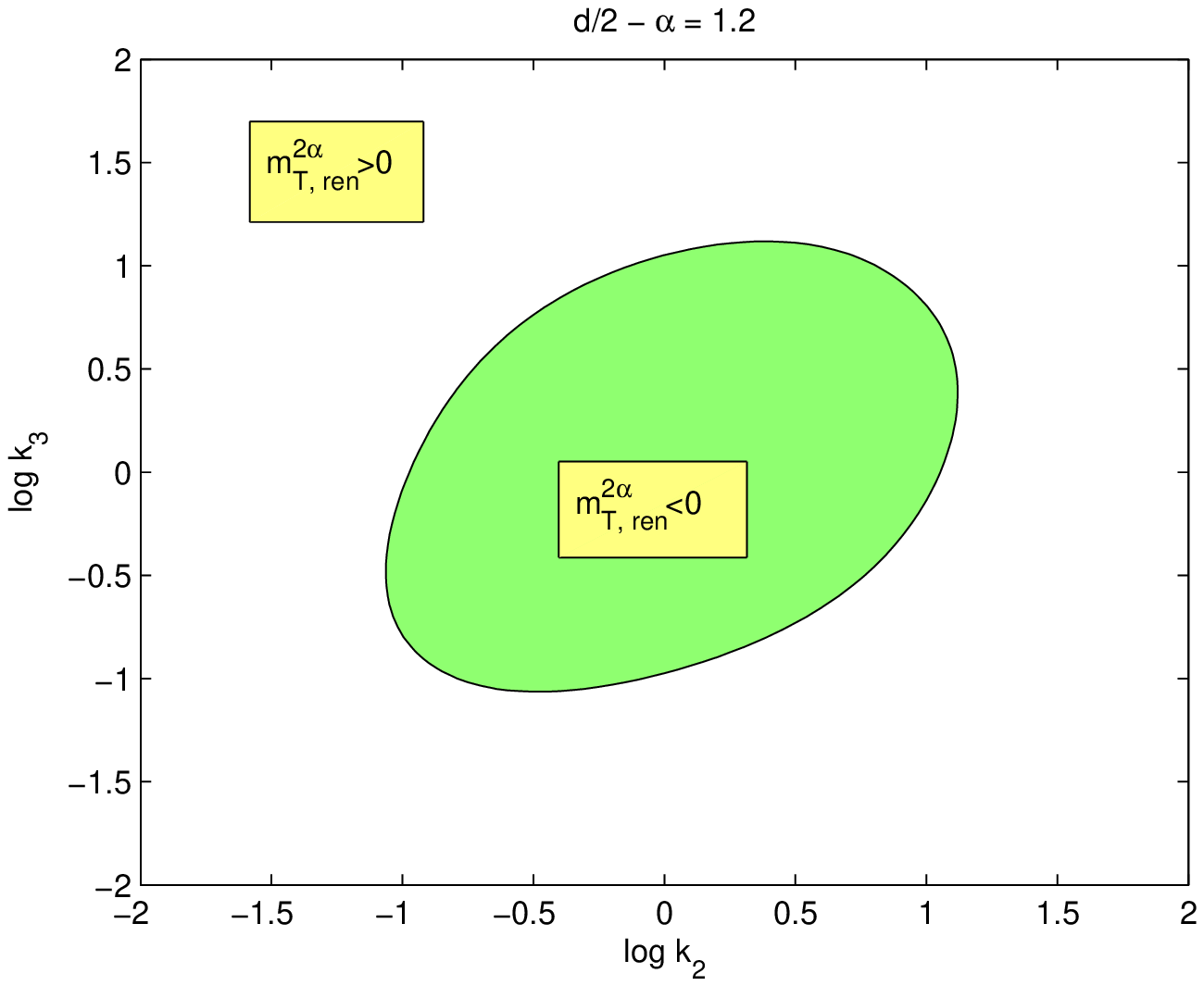}
\epsfxsize=.32\linewidth \epsffile{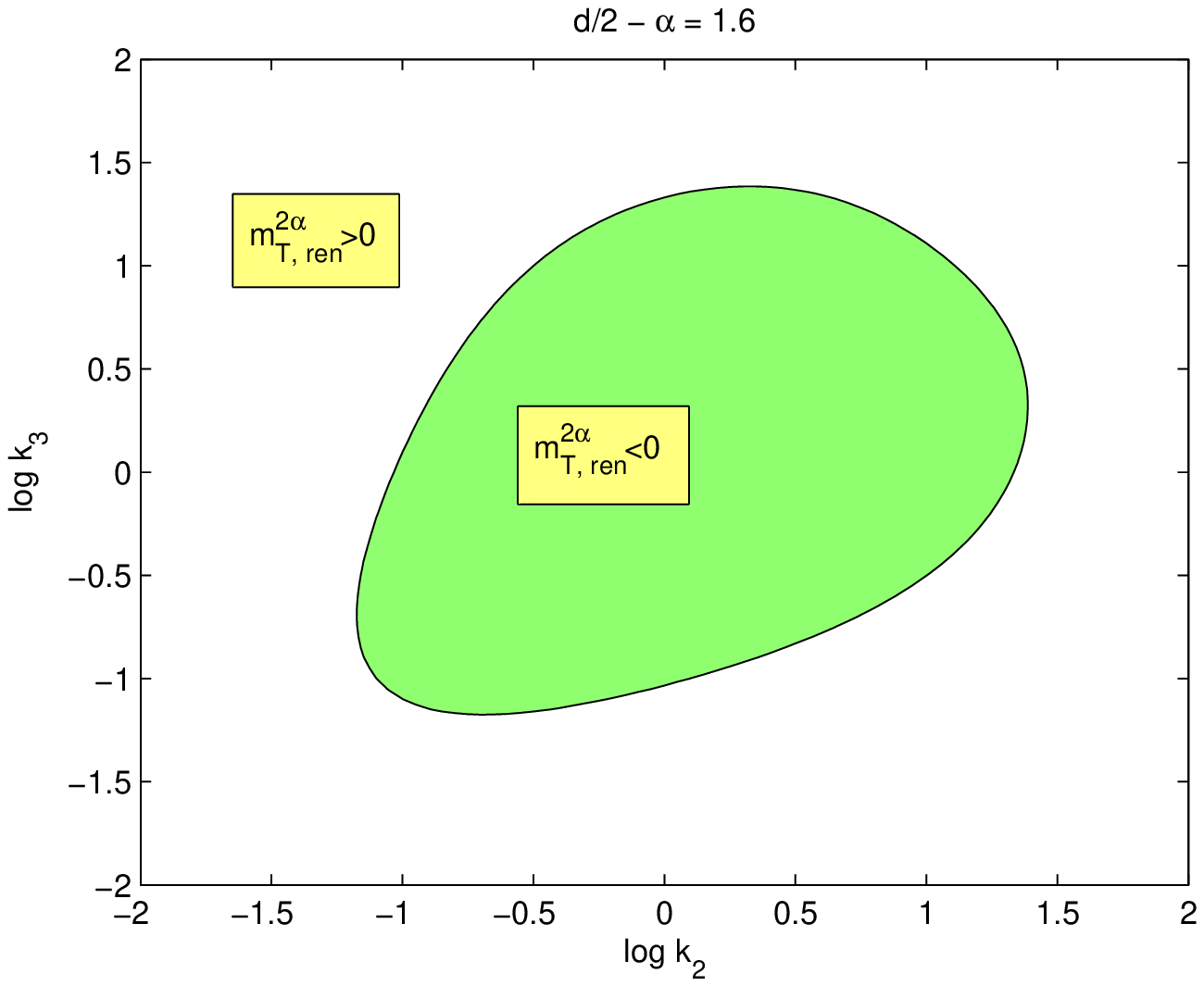}
\epsfxsize=.32\linewidth \epsffile{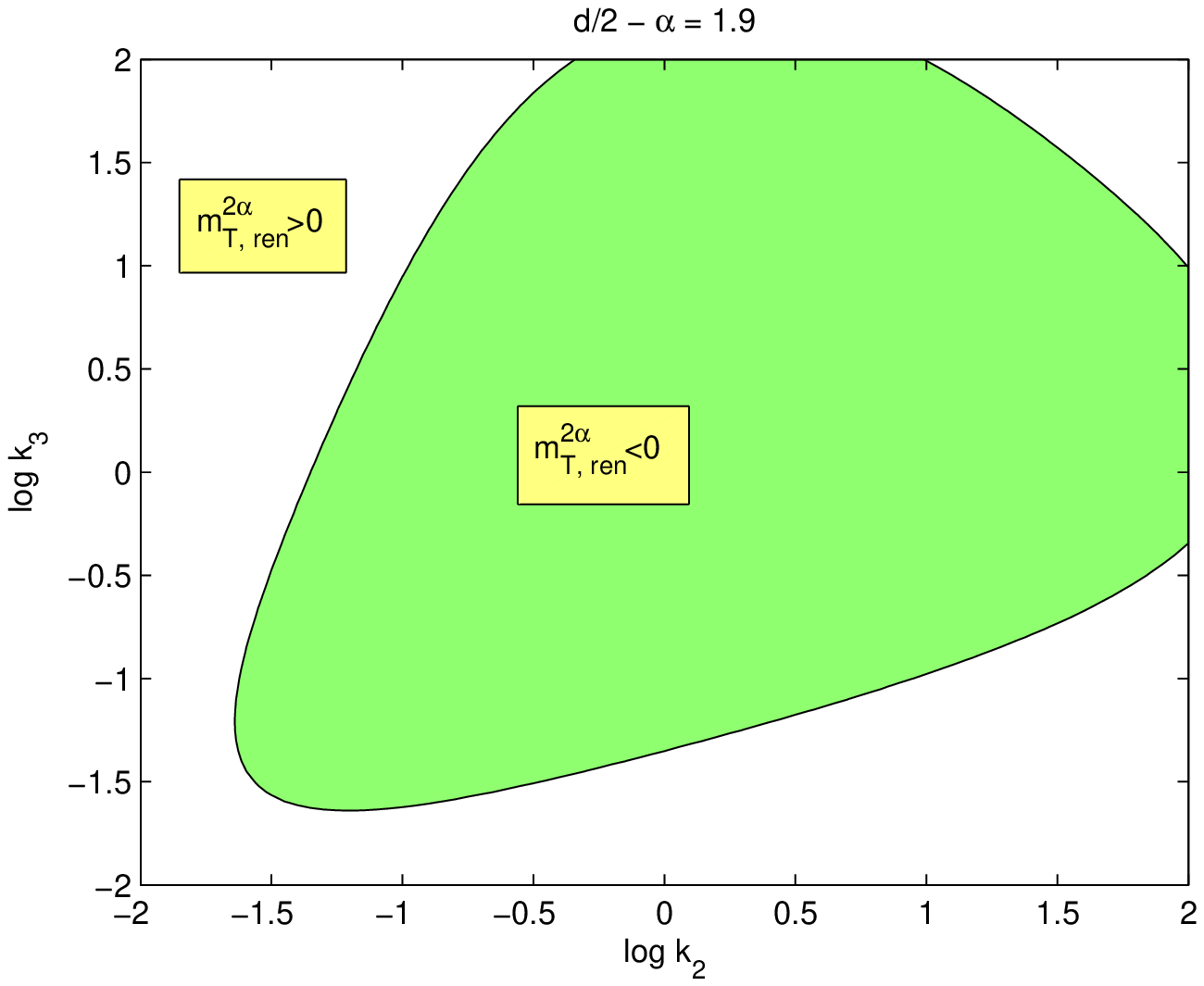}\caption{The regions
where $m_{T,\text{ren}}^{2\alpha}>0$ and
$m_{T,\text{ren}}^{2\alpha}<0$ for $p=4$, $V=L_1L_2L_3L_4=1$,
$L_1:L_2:L_3=1:k_2:k_3:k_4$ and $k_4=1$. The values of
$\frac{d}{2}-\alpha$ are $0.1, 0.4, 0.8, 1.2, 1.6, 1.9$. }
\end{figure}

\begin{figure}\centering \epsfxsize=.32\linewidth
\epsffile{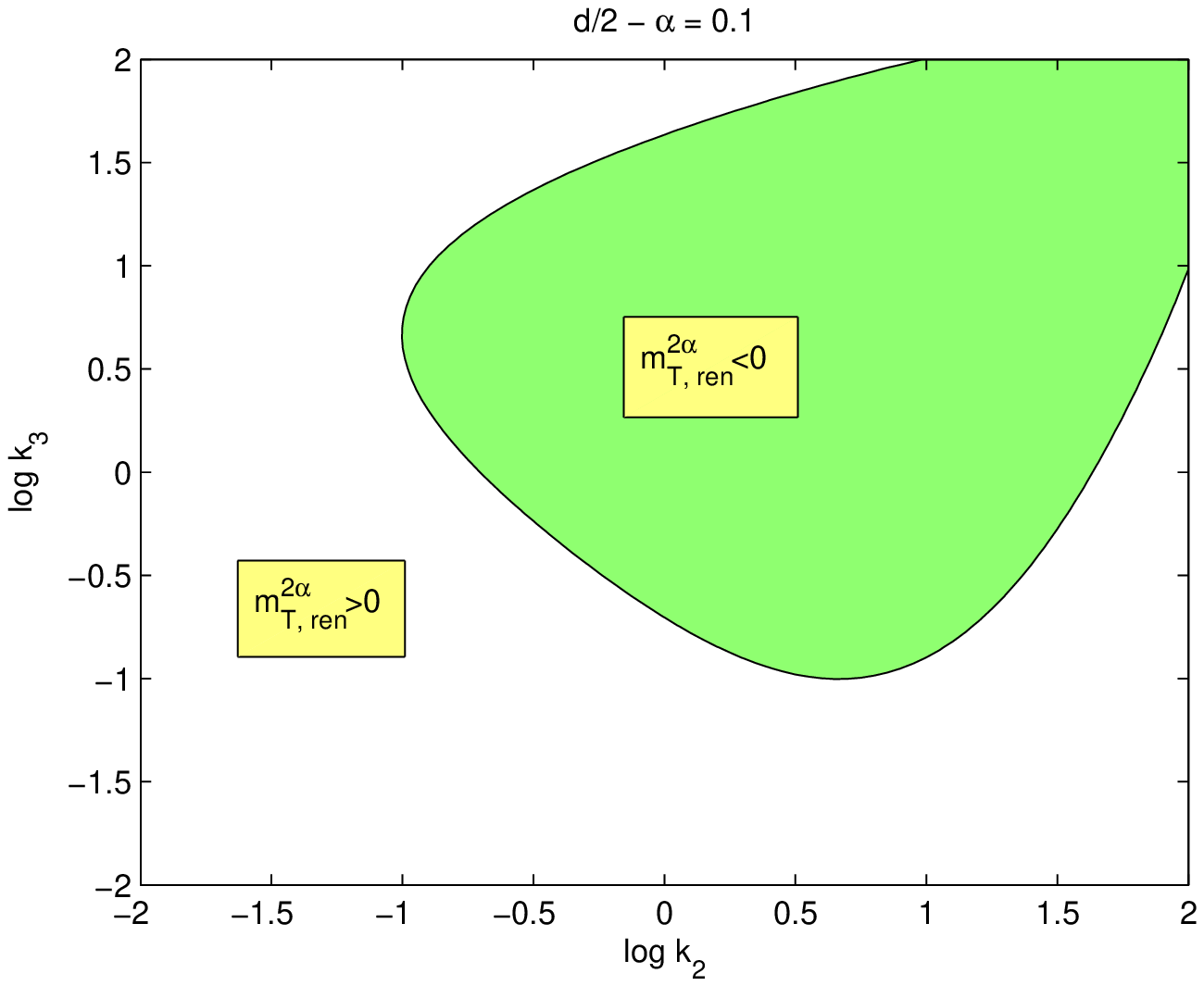}\epsfxsize=.32\linewidth
\epsffile{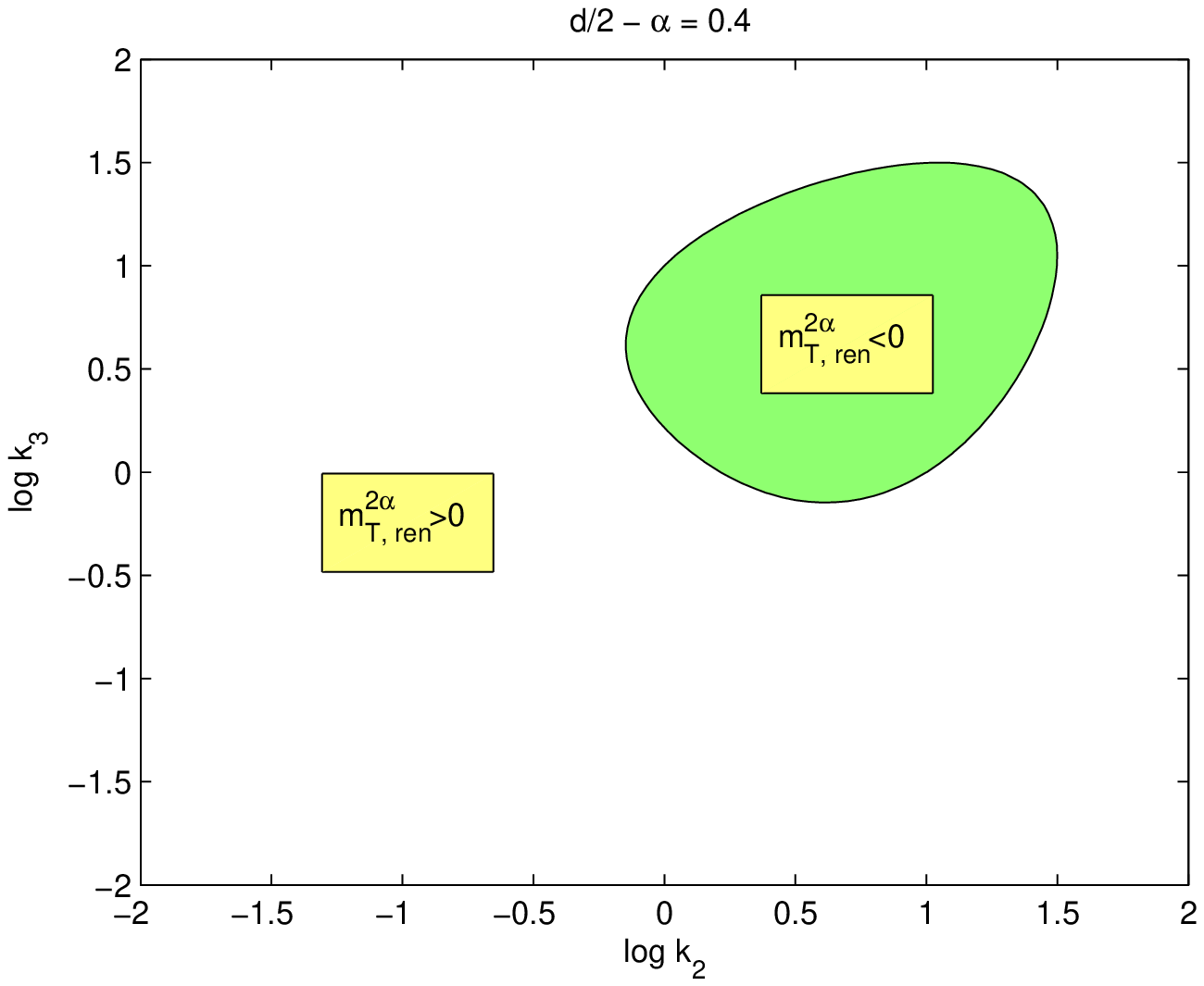}
\epsfxsize=.32\linewidth \epsffile{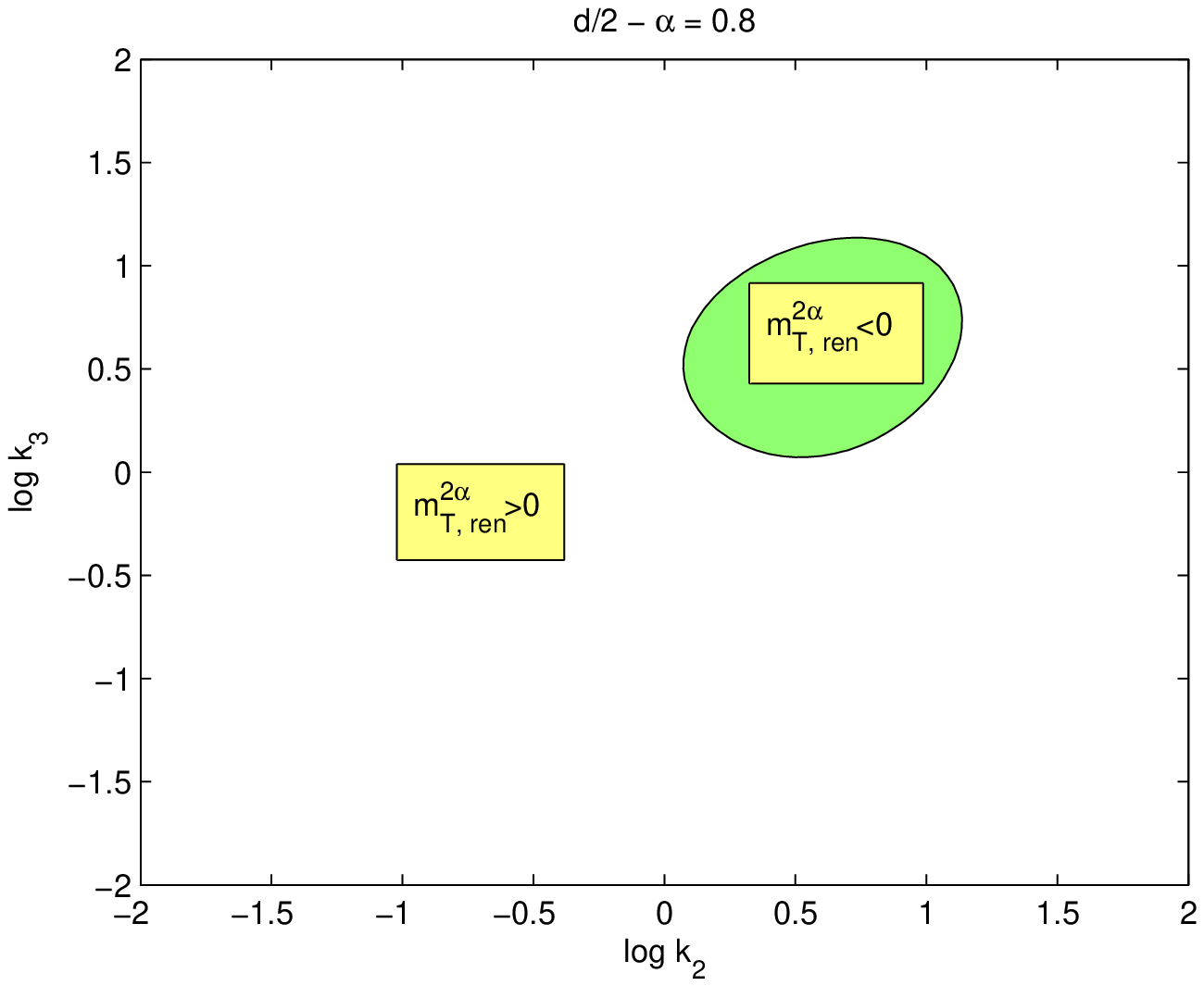}\\
\epsfxsize=.32\linewidth \epsffile{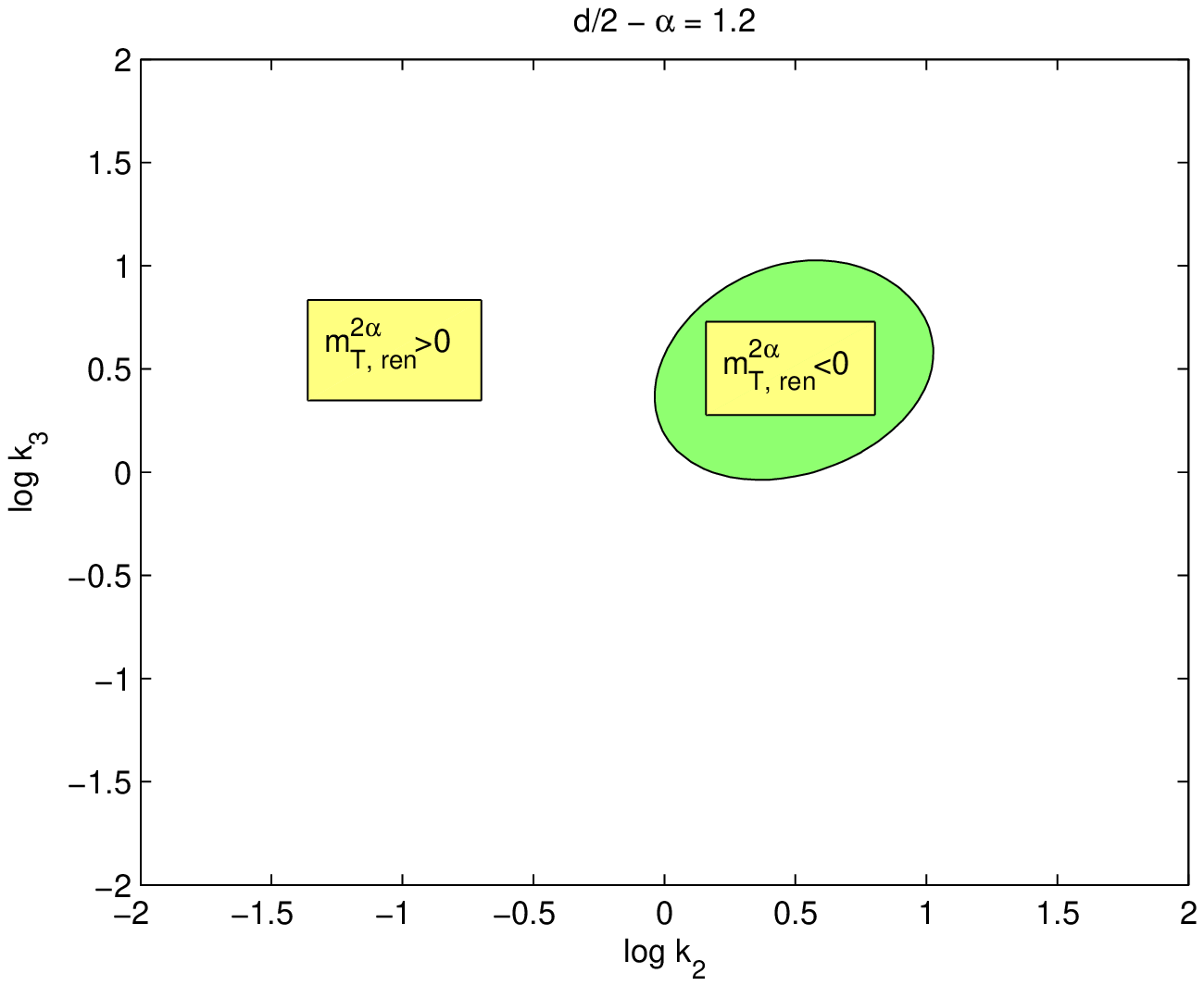}
\epsfxsize=.32\linewidth \epsffile{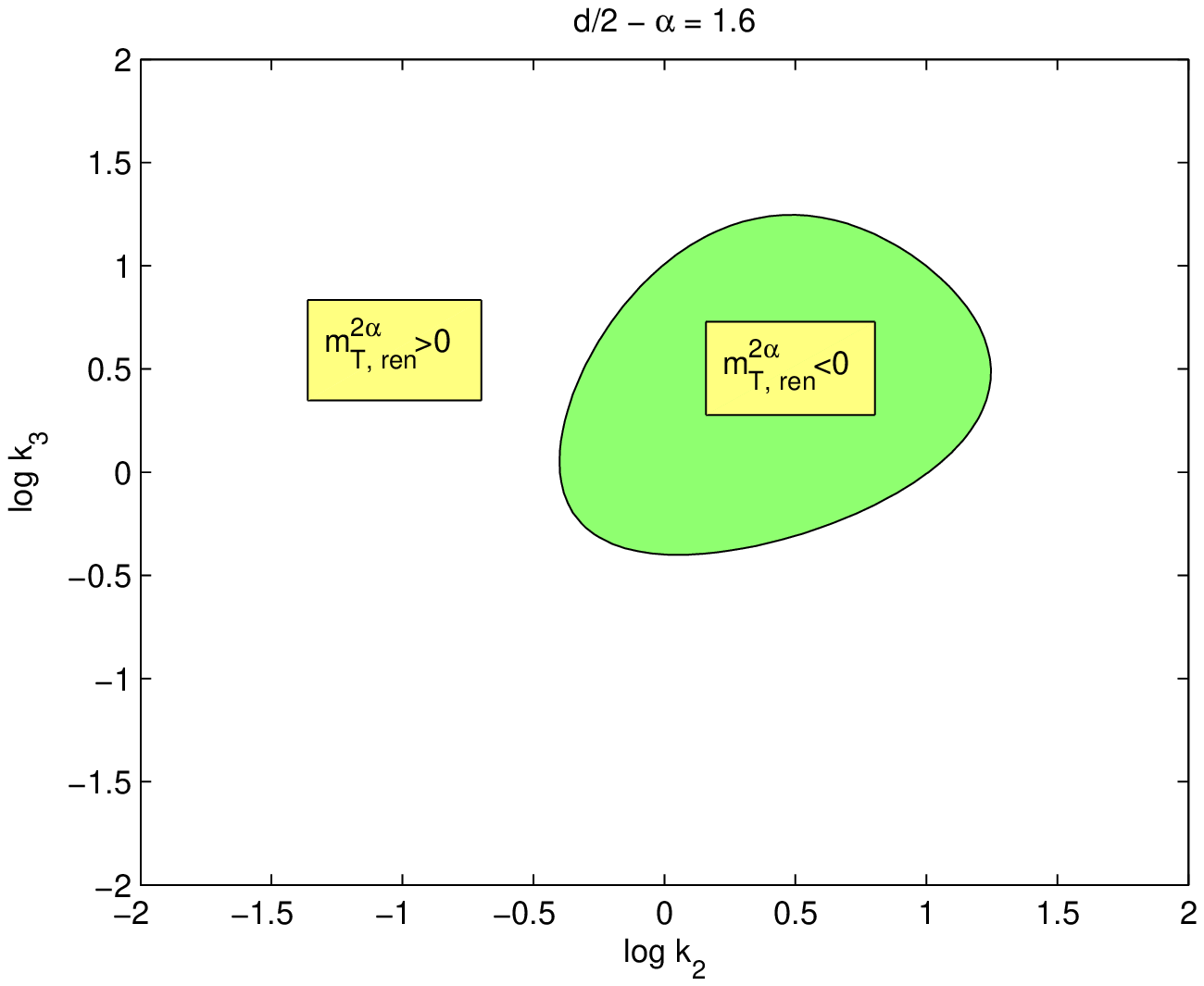}
\epsfxsize=.32\linewidth \epsffile{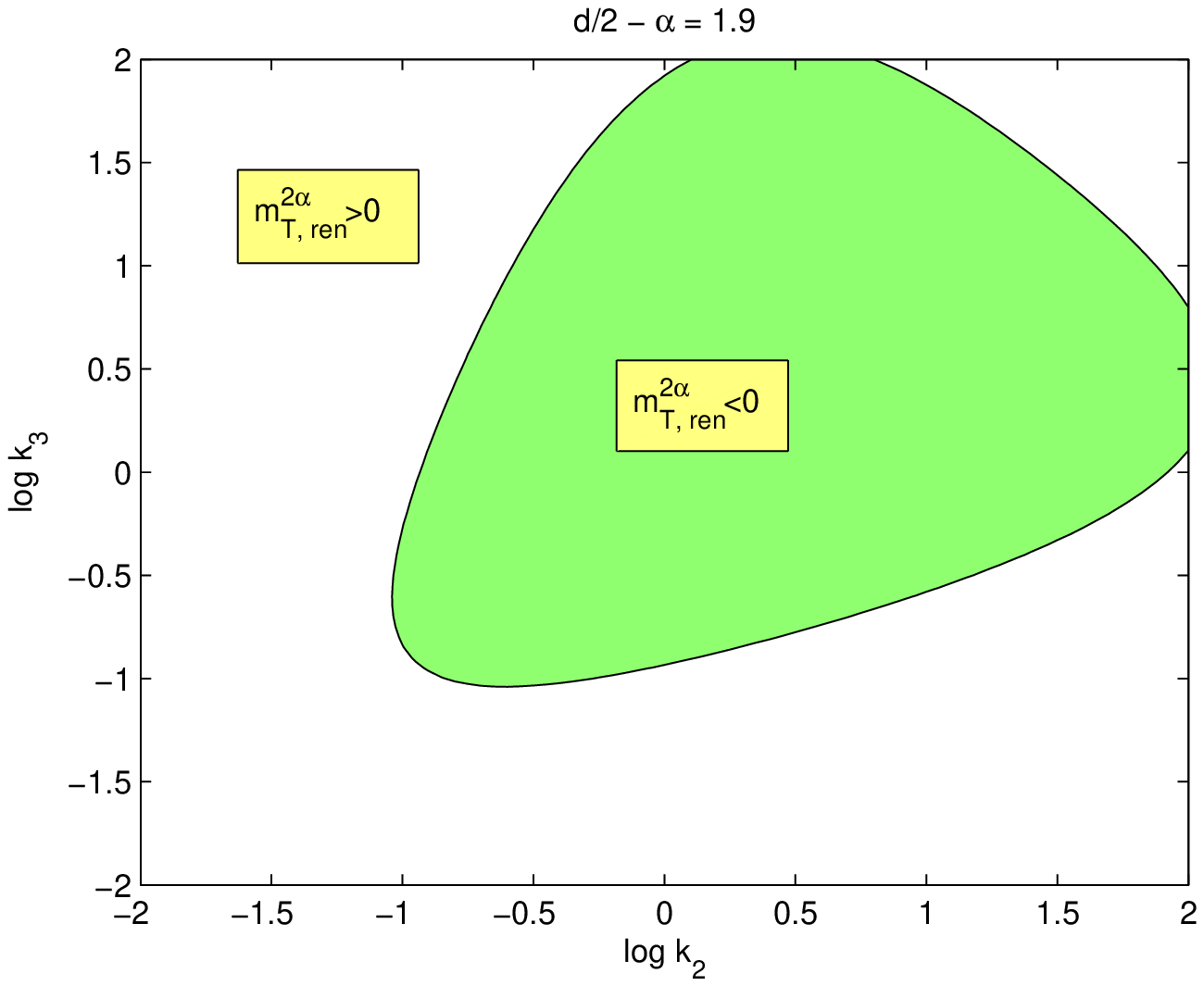}\caption{The
regions where $m_{T,\text{ren}}^{2\alpha}>0$ and
$m_{T,\text{ren}}^{2\alpha}<0$ for $p=4$, $V=L_1L_2L_3L_4=1$,
$L_1:L_2:L_3=1:k_2:k_3:k_4$ and $k_4=3$. The values of
$\frac{d}{2}-\alpha$ are $0.1, 0.4, 0.8, 1.2, 1.6, 1.9$.}

\end{figure}

The investigation of the sign and magnitude of the renormalized mass
$m_{T, \text{ren}}^{2\alpha}$ when $p=1,2,3,4$ is carried out
graphically. In Figures 1--13, we show   the dependence of the
renormalized mass $m_{T, \text{ren}}^{2\alpha}$ (up to the
multiplicative factor \eqref{eq104_3}) on the compactification
lengths $L_1,\ldots, L_p$ and the regions where symmetry breaking
mechanism appears. Using \eqref{eq917_10}, it is found  that if
$\alpha \neq q/2$, then under the simultaneous scaling $L_i\mapsto r
L_i$, the renormalized topologically generated mass
$m_{T,\text{ren}}^{2\alpha}$ transforms according to
$$m_{T,\text{ren}}^{2\alpha}\mapsto r^{2\alpha- d}m_{T,\text{ren}}^{2\alpha}.$$
Therefore, when we study the dependence of $m_{T,
\text{ren}}^{2\alpha}$ on the variables $(L_1, \ldots, L_p)$, we fix
a degree of freedom. This can be done by letting
$\mathcal{V}_p=\prod_{i=1}^pL_i=1$.

In Figure 1, we plot the renormalized mass $m_{T,
\text{ren}}^{2\alpha}$ as a function of $\frac{d}{2}-\alpha$ when
$p=1$ . The graph shows clearly that symmetry breaking appears only
when $0<\frac{d}{2}-\alpha<\frac{1}{2}$.  Figures 2,   3,  5 and
  6 demonstrate  the cases with $p=2$, $p=3$ and $p=4$.
These graphs show that for suitable choices of the compactification
lengths, symmetry breaking appears for all values of
$\frac{d}{2}-\alpha$ lying in the range $(0, \frac{p}{2})$. In
Figures 4 and
  7, the unshaded regions are the regions where
$m_{T,\text{ren}}^{2\alpha}<0$ and symmetry breaking exists. In all
the cases considered, i.e., $p=1, 2, 3, 4$, these regions  contain
the   point $L_1=\ldots=L_p$ where all compactification lengths are
equal, which corresponds to the lines $\log L_1=0$ or $\log k=0$ in
the graphs.

When $p\geq 3$, after fixing one degree of freedom in the variables
$(L_1,\ldots, L_p)$ by setting $\mathcal{V}_p=1$, we still have at
least another two degrees of freedom. Figures 8, 10 and 11 are
contour plots that show the dependence of the renormalized mass
$m_{T, \text{ren}}^{2\alpha}$ on the other two degrees of freedom of
the compactification lengths, for some specific values of
$d-2\alpha$. In Figures 9, 12 and 13, the corresponding regions
where $m_{T, \text{ren}}^{2\alpha}<0$ are shaded. From these graphs,
we find that the $m_{T, \text{ren}}^{2\alpha}<0$ regions that lead
to symmetry breaking are regions centered around the point $L_1=
\ldots= L_p$. Moving along a ray from a point in these regions will
lead to symmetry restoration. In fact, we show in \cite{LT3} that
the renormalized mass will become positive whenever one of the
compactification lengths is large enough. The boundaries of the
shaded regions in Figures 9, 12, 13 are  the projections of the
hypersurfaces in $(\R^+)^p$ where $m_{T, \text{ren}}^{2\alpha}=0$ to
appropriate two-dimensional planes. Note that in all the graphs,
logarithm scales are used for the compactification lengths as we
think that this will better illustrate the symmetry between the
compactification lengths, i.e., the symmetry generated by $L_i
\leftrightarrow L_j$.

From Figures 9, 12, 13, we  notice that the $m_{T,
\text{ren}}^{2\alpha}<0$ region (shaded) is larger when $d-2\alpha$
tends towards the boundary of the range $(0, p)$. It becomes smaller
in the middle of the range $(0,p)$. We also observe a symmetry
between the region for $d-2\alpha= h$ and the region for $d-2\alpha=
p-h$. In fact, this is nothing but a direct consequence of the
reflection formula \eqref{eq104_5} of the Epstein zeta function.

As mentioned above, for $p=1, 2, 3, 4$, graphical results show that
the region that lead to symmetry breaking is a region that contains
the point $L_1=\ldots=L_p$. We also observed that these regions are
convex and connected. We give a mathematically rigorous proof in
\cite{LT3} that in fact for all $p\geq 2$ and all the values of
$d-2\alpha$ where symmetry breaking mechanism can exist, the region
of $(L_1, \ldots, L_p)$ where $m_{T, \text{ren}}^{2\alpha}<0$, is a
convex and therefore connected region containing the point
$L_1=\ldots=L_p$, when plotted using log scale.

\section{conclusion}
We have studied the problem of topological mass generation for a
quartic self-interacting fractional scalar Klein--Gordon field on
toroidal spacetime.  Our results show that the method used for
ordinary scalar field, namely the zeta regularization technique,
still applies with some appropriate modifications. We are able to
derive the one loop effective potential for such a system for both
the massless and massive case in terms of power series of
$\lambda\widehat{\varphi}^2$ with Epstein zeta functions as
coefficients. As usual in the zeta regularized method, there is a
dependence of the effective potential on an arbitrary scaling length
$\mu$. We proposed a scheme to renormalize the effective potential
so as to get rid of the dependence on $\mu$. We have  carried out a
detailed derivation of the renormalization counterterms. The results
of the renormalized topologically generated mass $m_{T,
\text{ren}}^{2\alpha}$ are given explicitly. We note that in the
massive case, $m_{T, \text{ren}}^{2\alpha}$ is always positive and
therefore there is no symmetry breaking in this case. For the
massless case, we  show that fixing the number of compactified
dimensions $p$, if $p\leq 9$, symmetry breaking appears if and only
if the combination $d-2\alpha$ of spacetime dimension $d$ and
fractional order $\alpha$ of Klein-Gordon field satisfies
$0<d-2\alpha\leq p$. However if $p\geq 10$, symmetry breaking only
exists when the value of $d-2\alpha$ lies in a proper subset  of
$(0, p]$. This subset becomes smaller when $p$ is increased. For all
$p\geq 2$,  whenever there exists symmetry breaking, symmetry
restoration also appears when suitably varying the compactification
lengths. Simulations are carried out to illustrate the dependence of
the renormalized mass $m_{T,\text{ren}}^{2\alpha}$ on $d-2\alpha$ as
well as the compactification lengths, when $p=1, 2, 3, 4$. Graphical
results show that regions that lead to symmetry breaking are always
convex regions containing the point $L_1=\ldots=L_p$ where all
compactification lengths are equal, agreeing with our theoretical
results in \cite{LT3}. It is interesting to note that we can obtain
essentially the same results if instead of considering fractional
scalar field of order $\alpha$ in a toroidal spacetime $T^p\times
\R^{d-p}$ with integer dimension $d$, we can equivalently employ an
ordinary scalar field $(\alpha=1 )$ in a toroidal spacetime
$T^p\times \R^{d+2-2\alpha-p}$ with fractional dimension
$d+2-2\alpha$.

    This paper is   our first attempt to explore the  fractional field theory with interactions.
    One possible extension of our discussion
is to include local structure like spacetime curvature in addition
to nontrivial global topology in our study. One expects the
generalization to finite temperature case will not pose difficulty
since in the Matsubara formalism, the thermal Green functions with
periodic boundary condition with period given by the inverse
temperature, have the same properties as the Green functions at zero
temperature with the imaginary time dimension compactified to a
circle of radius equals to the inverse of temperature. We can also
consider the extension of our results to the fractional gauge field
theory. As we mentioned in our introduction that Brownian motion
plays an important role in Feynman path integrals, we would like to
note that path integrals have been generalized to fractional
Brownian motion \cite{c49} and fractional oscillator process
\cite{c50}. Although fractional Brownian motion has found wide
applications in many areas in physics and engineering, so far it has
not really played a role in quantum theory yet, and no application
of these "fractional path integrals" have actually been carried out
so far. In view of the fact that fractional oscillator process can
be regarded as one-dimensional fractional Klein--Gordon field, with
fractional Brownian motion  its "massless" limit \cite{c51, c57}, it
will be interesting to extend such path integrals to fractional
Klein--Gordon fields and to exploit their possible uses. Finally, we
would like to mention that in many applications in condensed matter
physics, fractal or fractional processes have their limitations
since many phenomena considered are multifractal in nature. There
have already been works on multifractional Brownian motion
\cite{c52, c53}, multifractional Levy process \cite{c54} and
multifractional oscillator process \cite{c55}. In view of the
possible variable spacetime dimension, for example at the
sub-Planckian distance \cite{c7}, it would be interesting to
consider how our results can be generalized to Klein--Gordon field
with variable fractional order.

\vspace{1cm} \noindent \textbf{Acknowledgement}\; The authors would
like to thank Malaysian Academy of Sciences, Ministry of Science,
Technology  and Innovation for funding this project under the
Scientific Advancement Fund Allocation (SAGA) Ref. No P96c.

\appendix

\section{The generalized Epstein zeta function $Z_{E,N}(s; a_1,\ldots, a_N; c)$}
In this appendix, we summarize some facts about the generalized
Epstein zeta function \cite{Ep1, Ep2} which we have used in our
calculations. For details, we refer to \cite{EK1, E1, ET, K, a2, a3,
a4, a5, a6, d3, d7, d8,
 d14, d15, d5, d17, d18} and the references therein.

\subsection{Homogeneous Epstein zeta function}
First  consider the homogeneous Epstein zeta function $Z_{E,N}(s;
a_1,\ldots, a_N)$. For $N\geq 1$, it is  defined by the series
\begin{align}\label{eq104_1}
Z_{E,N}(s; a_1, \ldots,
a_N)=\sum_{\mathbf{k}\in\mathbb{Z}^N\setminus\{0\}}\left(\sum_{i=1}^N
[a_ik_i]^2\right)^{-s}
\end{align}when $\text{Re}\;\; s>\frac{N}{2}$. We extend the definition to $N=0$ by
defining $Z_{E,0}(s)=0$. For $N\geq 1$, $Z_{E,N}(s; a_1, \ldots,
a_N)$ has a meromorphic continuation to the complex plane with a
simple pole at $s=N/2$, and it satisfies a functional equation (also
known as reflection formula):
\begin{align}\label{eq104_5}
\pi^{-s}\Gamma(s)Z_{E,N}(s; a_1, \ldots , a_N)
=\frac{\pi^{s-\frac{N}{2}}}{\left[\prod_{j=1}^Na_j\right]}\Gamma\left(\frac{N}{2}-s\right)Z_{E,N}\left(\frac{N}{2}-s;
\frac{1}{a_1},\ldots, \frac{1}{a_N}\right).
\end{align}This formula relates the value of an Epstein zeta function at $s$ with
the value of its 'dual' at $N/2-s$. The Epstein zeta function
$Z_{E,N}(s; a_1,\ldots, a_N)$ behaves nicely under simultaneous
scaling of the parameters. Namely, for any $r\in\R^+$,
\begin{align}\label{eq917_10}
Z_{E,N}\left(s; r a_1,\ldots,r a_N\right) =r^{-2s}Z_{E,N}(s;
a_1,\ldots, a_N).
\end{align}Together with $Z_{E,N}(0;a_1,\ldots, a_N)=-1$, one gets
that
\begin{align}\label{eq914_1}
Z_{E,N}'\left(0; r a_1,\ldots, r a_N\right)=2\log r + Z_{E,N}'(0;
a_1,\ldots, a_N).
\end{align}

The function $\xi_{E,N}(s; a_1,\ldots,
a_N)=\Gamma(s)Z_{E,N}(s;a_1,\ldots, a_N)$ is also a meromorphic
function. For $N\geq 1$, it  has  simple poles at $s=0$ and $s=N/2$
 with residues \begin{align}\label{eq830_7}&\text{Res}_{s=0}\Bigl\{\Gamma(s)Z_{E,N}(s; a_1, \ldots,
a_N)\Bigr\}=-1\\
&\text{Res}_{s=\frac{N}{2}}\Bigl\{\Gamma(s)Z_{E,N}(s; a_1, \ldots,
a_N)\Bigr\}=\frac{\pi^{\frac{N}{2}}}{\left[\prod_{j=1}^N
a_j\right]}\nonumber
\end{align}and finite parts\begin{align}\label{eq830_8}
\text{PP}_{s=0} \Bigl\{\Gamma(s)Z_{E,N}\left(s; a_1, \ldots,
a_N\right)\Bigr\}=&Z_{E,N}'\left(0; a_1, \ldots, a_N\right)-\psi(1)
,\end{align}\begin{align*} &\text{PP}_{s=\frac{N}{2}}
\Bigl\{\Gamma(s)Z_{E,N}\left(s; a_1, \ldots,
a_N\right)\Bigr\}\nonumber\\=&\frac{\pi^{\frac{N}{2}}}{\left[\prod_{j=1}^N
a_j\right]}\Biggl[
Z_{E,N}'\left(0;\frac{1}{a_1},\ldots,\frac{1}{a_N}\right)
+2\log\pi-\psi(1)\Biggr]\nonumber
\end{align*}
respectively. Here $\psi(z)$ is the function $\Gamma'(z)/\Gamma(z)$.
Some special values of $\psi$ are $\psi(1)=-\gamma$, where $\gamma$
is the Euler constant; and for $k\geq 1$, $\psi(k+1)$ can be
computed recursively by the formula
\begin{align*}
\psi(k+1)=\psi(k)+\frac{1}{k}.
\end{align*}

One of the indispensable tools in studying the Epstein zeta function
is the Chowla--Selberg formula \cite{d19, d20}. One form of the
formula is
\begin{align}\label{eq830_9}
&\Gamma(s)Z_{E,N}(s; a_1,\ldots, a_N)\\
=\nonumber &2a_1^{-2s}\Gamma(s)\zeta_R(2s)
+2\sum_{j=1}^{N-1}\frac{\pi^\frac{j}{2}\Gamma\left(s-\frac{j}{2}\right)}
{a_{j+1}^{2s-j}\prod_{l=1}^{j}a_l} \zeta_R(2s-j) +4\pi^s
\sum_{j=1}^{N-1}
\frac{1}{\prod_{l=1}^{j}a_l}\times\\\nonumber&\sum_{\mathbf{k}\in
\Z^j\setminus\{\mathbf{0}\}}\sum_{p=1}^{\infty}\frac{1}{(pa_{j+1})^{s-\frac{j}{2}}}\left(\sum_{l=1}^j
\left[\frac{k_l}{a_l}\right]^2\right)^{\frac{s}{2}-\frac{j}{4}}K_{s-\frac{j}{2}}\left(
2\pi p a_{j+1}\sqrt{\sum_{l=1}^j
\left[\frac{k_l}{a_l}\right]^2}\right),
\end{align}which expresses the homogeneous Epstein Zeta function as a sum of
Riemann zeta function $\zeta_R$ plus a remainder which is a
multi-dimensional series that converges rapidly.  It can be used to
effectively compute the homogeneous Epstein zeta function to any
degree of accuracy.

\subsection{Inhomogeneous Epstein zeta function} For $N\geq 1$, $a_1, \ldots, a_N, c>0$,  the inhomogeneous Epstein zeta
function $Z_{E,N} (s;a_1,\ldots, a_N; c)$ is defined by
\begin{align*}
Z_{E,N} (s;a_1,\ldots, a_N; c)=\sum_{\mathbf{k}\in
\Z^N}\left(\sum_{j=1}^N[a_ik_i]^2+c^2\right)^{-s}
\end{align*}when $\text{Re}\,s>N/2$. When $N=0$, we define $$Z_{E,0}(s;c)=c^{-2s}.$$
$Z_{E,N}(s; a_1,\ldots, a_N;c)$ has a meromorphic continuation to
$\C$ given by
\begin{align}\label{eq23_1}
&Z_{E,N} (s;a_1,\ldots, a_N; c)
=\frac{\pi^\frac{N}{2}}{\left[\prod_{i=1}^N
a_i\right]}\frac{\Gamma\left(s-\frac{N}{2}\right)}{\Gamma(s)}c^{N-2s}\\&+\frac{2\pi^s}{\left[\prod_{i=1}^N
a_i\right]\Gamma(s)} \frac{1}{c^{s-\frac{N}{2}}}\sum_{\mathbf{k}\in
\Z^N\setminus\{ \mathbf{0}\}}\left(\sum_{i=1}^N
\left[\frac{k_i}{a_i}\right]^2\right)^{\frac{2s-N}{4}}
K_{s-\frac{N}{2}} \left(2\pi c \sqrt{\sum_{i=1}^N
\left[\frac{k_i}{a_i}\right]^2}\right).\nonumber
\end{align}
The second term is an analytic function of $s$. The first term shows
that $Z_{E,N}(s;a_1,\ldots, a_N; c)$ has simple poles at
$s=\frac{N}{2}-j, j\in \mathbb{N}\cup\{0\}$ if $N$ is odd, and at
$s=1, 2, \ldots, \frac{N}{2}$ if $N$ is even. On the other hand, one
can easily read from Eq. \eqref{eq23_1}  that the function
$\Gamma(s)Z_{E,N} (s;a_1,\ldots, a_N; c)$ has simple poles at
$s=\frac{N}{2}-j, j\in \mathbb{N}\cup\{0\}$ with residues
\begin{align}\label{eq26_14}
\text{Res}_{s=\frac{N}{2}-j}\left\{\Gamma(s)Z_{E,N} (s;a_1,\ldots,
a_N;
c)\right\}=\frac{(-1)^j}{j!}\frac{\pi^{\frac{N}{2}}}{\left[\prod_{i=1}^N
a_i\right]}c^{2j}
\end{align}and finite parts
\begin{align}\label{eq26_15}
&\text{PP}_{s=\frac{N}{2}-j}\left\{\Gamma\left(s\right)Z_{E,N}
\left(s;a_1,\ldots, a_N;
c\right)\right\}=\frac{(-1)^{j}}{j!}\frac{\pi^{\frac{N}{2}}}{\left[\prod_{i=1}^N
a_i\right]}c^{2j}\left(\psi(j+1)-2\log c\right)\\\nonumber
&+\frac{2\pi^{\frac{N}{2}-j}c^j}{\left[\prod_{i=1}^N a_i\right]}
\sum_{\mathbf{k}\in \Z^N\setminus\{ \mathbf{0}\}}\left(\sum_{i=1}^N
\left[\frac{k_i}{a_i}\right]^2\right)^{-\frac{j}{2}} K_{j}
\left(2\pi c \sqrt{\sum_{i=1}^N
\left[\frac{k_i}{a_i}\right]^2}\right)
\end{align}respectively.
From \eqref{eq23_1}, \eqref{eq26_14} and \eqref{eq26_15}, it can be
easily deduced that

\vspace{0.2cm}\noindent $\bullet$\;\; If
$s\notin\left\{\frac{N}{2}-j\,:\, j\in\mathbb{N}\cup\{0\}\right\}$,
then
\begin{align}\label{eq26_11}
\lim_{\substack{ a_i\rightarrow 0\\ 1\leq i\leq N}}
\left[\prod_{i=1}^N a_i\right] \Bigl\{ \Gamma(s)Z_{E,N}(s; a_1,
\ldots, a_N; c)\Bigr\}= \pi^\frac{N}{2}
 \Gamma\left(s-\frac{N}{2}\right) c^{N-2s}.
\end{align}

\vspace{0.2cm}\noindent $\bullet$\;\; If $s=\frac{N}{2}-j$ for some
$j\in\mathbb{N}\cup\{0\}$, then
\begin{align}\label{eq26_21}
\lim_{\substack{ a_i\rightarrow 0\\ 1\leq i\leq N}}
\left[\prod_{i=1}^N a_i\right] \text{Res}_{ s= \frac{N}{2}-j}\Bigl\{
\Gamma(s)Z_{E,N}(s; a_1, \ldots, a_N; c)\Bigr\}=
\frac{(-1)^j}{j!}\pi^{\frac{N}{2}}c^{2j}.
\end{align}

\begin{align}\label{eq26_22}
\lim_{\substack{ a_i\rightarrow 0\\ 1\leq i\leq N}}
\left[\prod_{i=1}^N a_i\right] \text{PP}_{ s= \frac{N}{2}-j}\Bigl\{
\Gamma(s)Z_{E,N}(s; a_1, \ldots, a_N; c)\Bigr\}=
\frac{(-1)^j}{j!}\pi^{\frac{N}{2}}c^{2j}\left(\psi(j+1)-2\log
c\right).
\end{align}
\section{Independence of $V_{\text{eff}}^{(\text{ren})}$ on $\mu$}
Here we give a sketch of the proof that the renormalized effective
potential $V_{\text{eff}}^{(\text{ren})}(\widehat{\varphi})$ is
independent of $\mu$. From the definition of the renormalized
effective potential \eqref{eq904_1} and the formula \eqref{eq905_5}
that determines the counterterms, we get
\begin{align*}
V_{\text{eff}}^{(\text{ren})}(\widehat{\varphi})=\frac{1}{2}
m^{2\alpha} \widehat{\varphi}^2+\frac{1}{4!} \lambda
\widehat{\varphi}^4+V_Q +\Lambda \Pi^{-1} T,
\end{align*}where
\begin{align*}
\Lambda =\begin{pmatrix} 1 & \frac{\widehat{\varphi}^2}{2!} &
\frac{\widehat{\varphi}^4}{4!}&\ldots
&\frac{\widehat{\varphi}^{2d_{\alpha}}}{(2d_{\alpha})!}\end{pmatrix},
\hspace{1cm} T=\begin{pmatrix} T_0 \\ T_1\\\vdots \\
T_{d_{\alpha}}\end{pmatrix}.
\end{align*}The terms containing $\log\mu^2$ can be extracted from
$V_Q$ (eq. \eqref{eq903_1} and \eqref{eq830_4}) and $T_j$ (eq.
\eqref{eq903_5} and \eqref{eq913_1}), with result given by
\begin{align}\label{eq913_2}
\Lambda S^{\mu}+\Lambda \Pi^{-1} T^{\mu},
\end{align}where

\vspace{0.3cm}\noindent $\bullet$\;\;in the massive case,
\begin{align*}S^{\mu}_k=\frac{m^{d}}{2^{d+1}\pi^{\frac{d}{2}}}\frac{(-1)^k(2k)!}{\Gamma(\alpha
k+1)}\left(\frac{\lambda}{2m^{2\alpha}}\right)^k\frac{(-1)^{\frac{d}{2}-\alpha
k}}{\left(\frac{d}{2}-\alpha k\right)!}\chi_{k;\alpha;
d},\end{align*}with\begin{align*} \chi_{k;\alpha;d} =\begin{cases}
1,
\hspace{1cm}&\text{if}\; \; k\in \mathcal{H}_{\alpha;d,0},\\
0, &\text{otherwise},\end{cases}\end{align*}and
\begin{align*}T^{\mu}_j=&\frac{(-1)^{j+1}  \lambda^j m^{d-2\alpha
j}}{2^{d+j+1}\pi^{\frac{d}{2}}} \sum_{ k\in
\mathcal{H}_{\alpha;d,j}}\frac{(-1)^{k+\frac{d}{2}-\alpha(k+j)}[2(k+j)]!}{[2k]!
\left[\frac{d}{2}-\alpha(k+j)\right]!\Gamma(\alpha(k+j)+1)}
\left(\frac{\lambda\widehat{\varphi}^2_j}{2m^{2\alpha}}\right)^k.
\end{align*}

\vspace{0.2cm} \noindent $\bullet$\;\; in the massless case,
\begin{align*}S^{\mu}_k
=\frac{(-1)^{\frac{d}{2\alpha}}\delta_{k,d_{\alpha}}\omega_{\alpha;d}}
{2^{d+1}\pi^{\frac{d}{2}}\Gamma\left(\frac{d+2}{2}\right)}(2k)!\left(\frac{\lambda}{2}\right)^k,\end{align*}and
\begin{align*}T_j^{\mu}=-\frac{\omega_{\alpha;d}}{2^{d+1}
\pi^{\frac{d}{2}}}\frac{(-1)^{\frac{d}{2\alpha}}}{\Gamma\left(\frac{d}{2}+1\right)}
\frac{\lambda^{j}}{2^{j}}
\left(\frac{\lambda\widehat{\varphi}_j^2}{2}\right)^{\frac{d}{2\alpha}-j}
\frac{\left(\frac{d}{\alpha}\right)!}{\left(\frac{d}{\alpha}-2j\right)!}.
\end{align*}

\vspace{0.5cm}\noindent In both cases, it is  easy to verify that
 $\Pi S^{\mu}=-T^{\mu}$, which shows that the term \eqref{eq913_2} is
identically zero and therefore
$V_{\text{eff}}^{(\text{ren})}(\widehat{\varphi})$ does not depend
on $\log \mu^2$.


\begin{thebibliography}{10}
\bibitem{c1}  B.~B. Mandelbrot, \emph{The fractal geometry of nature}, (W.H. Freeman,
New York, 1983).

\bibitem{c2} R.P.Feynman and A.R. Hibbs, \emph{Quantum mechanics and path
integrals}, (McGraw-Hill, New York, 1965).

\bibitem{c3} L.F. Abbot and M.B. Wise, \emph{Dimension of a quantum--mechanical
path}, Am. J. Phys. \textbf{49}, 37--39 (1981).

\bibitem{n1}
E. Nelson, \emph{Derivation of Schr\"odinger equation from Newtonian
mechanics}, Phys. Rev. \textbf{150}, 1079--1085 (1966).

\bibitem{n2} E. Nelson, \emph{Quantum fluctuations}, (Princeton
University Press, N. J., 1985).

\bibitem{c4} H. Kroger, \emph{Fractal geometry in quantum mechanics, field
theory and spin
       systems}, Phys. Rep. \textbf{323}, 81--181, 2000.


\bibitem{c5} H.~J. Rothe, \emph{Lattice gauge theories: An introduction}, 3rd edition
(World Scientific, Singapore, 2005).

\bibitem{c6} B.~J. Durhuus, J. Ambjorn, \emph{Quantum geometry: A statistical
field theory approach}, (Cambridge University, Cambridge, 1997).

\bibitem{c7}, O. Lauscher and M. Reuter, \emph{Fractal spacetime structure in
asymptotically safe gravity}, Preprint arXiv:hep-th/0508202, 2005.


\bibitem{c8} K.~S. Miller and B. Ross, \emph{An introduction to the fractional
calculus and fractional
      differential equations}, (John Wiley and Sons, New York, 1993).


\bibitem{c9} S. Samko, A.A. Kilbas and D.I. Maritchev, \emph{Integrals and
derivatives of the
       fractional order and some of their applications}, (Gordon and Breach,
       Armsterdam, 1993).

\bibitem{c10} I. Podlubny, \emph{Fractional differential equations}, (Academic Press,
New York, 1999).

\bibitem{c11} A.~A. Kilbas, H.~M. Srivastava and J.~J. Trujillo, \emph{Theory and
applications of
       fractional differential equations}, (Elsevier, Amsterdam, 2006).


\bibitem{c12} R. Hilfer ed., \emph{Applications of fractional calculus in
physics}, (World Scientific,
      Singapore, 2000).


\bibitem{c13} B.~J. West, M. Bologna and P. Grigolini, \emph{Physics of fractal
operators}, (Springer-
        Verlag, New York, 2003).


\bibitem{c14} R. Metzler  and J. Klafter, \emph{The restaurant at the end of the random
walk: Recent
      developments in the description of anomalous transport by fractional dynamics}, J
       Phys. A \textbf{37}, R161-R208 (2004).


\bibitem{c15} L.~M. Zelenyi   and A.~V. Milovanov, \emph{Fractal topology and strange
kinetics: from percolation theory to problems in cosmic
electrodynamics}, Phys. Uspekhi \textbf{47}, 749--788 (2004).

\bibitem{c16} G.~M. Zaslavsky, \emph{Hamiltonian chaos and fractional
dynamics}, (Oxford:
        Oxford University, 2005).


\bibitem{c17} Y. Hu and G. Kallianpur, \emph{Schrödinger equation with fractional
Laplacian}, Appl. Math. Optim. \textbf{42}, 281--290 (2000).

\bibitem{c18} N. Laskin, \emph{Fractals and quantum mechanics}, Chaos, \textbf{10}, 780--790
(2000).

\bibitem{c19} N. Laskin, \emph{Fractional Schrödinger equation}, Phys. Rev. E \textbf{66},
056108, (2002).

\bibitem{c20} M. Naber, \emph{Time fractional Schrödinger equation}, J. Math. Phys.
\textbf{45}, 3339--3352 (2004).

\bibitem{c21}
 X. Guo and M. Xu, \emph{Some applications of fractional Schrödinger
equation}, J. Math. Phys. \textbf{47}, 082104 (2006).

\bibitem{c22} S. Wong and M. Xu, \emph{Generalized fractional Schrödinger equation
with space-time fractional derivatives}, J. Math. Phys. \textbf{48},
043502 (2007).

\bibitem{c23} J. Dong and M. Xu,  \emph{Some solutions to the space fractional
Schrödinger equation using momentum representation method}, J. Math.
Phys. \textbf{48}, 072105 (2007).

\bibitem{c24} D. Baleanu and S.I. Muslih, \emph{About fractional supersymmetric
quantum mechanics}, Czech. J. Phys. \textbf{55}, 1063--1066 (2005).

\bibitem{c25} C.~G. Bollini and J.~J. Giambiagi, \emph{Arbitrary powers of
D'Alembertian and the Huygens' principle}, J. Math. Phys.
\textbf{34}, 610--621 (1993).

\bibitem{c26} C. Lammerzahl,  \emph{The pseudodifferential operator square root of
the Klein--Gordon equation},  J. Math. Phys. \textbf{34}, 3918--3932
(1993).

\bibitem{c27} M.~S. Plyushchay and M.~R. de Traubenberg, \emph{Cubic root of
Klein--Gordon equation}, Phys. Lett. B \textbf{477}, 276--284
(2000).

\bibitem{c28} Kh. Namsrai and H.~V. von Geramb, \emph{Square--root operator
quantization and nonlocality: a review}, Int. J. Theor. Phys.
\textbf{40}, 1929--2010 (2001).

\bibitem{c29} A. Raspini, \emph{Simple solution of fractional Dirac equation of
order 2/3}, Physica Scripta \textbf{64}, 20--22 (2001).


\bibitem{c30} P. Zavada, \emph{Relativistic wave equations with fractional
derivatives and pseudo-differential operators}, J. Appl. Math.
\textbf{2}, 163--197 (2002).

\bibitem{c31} R.~L.~P.~G. do Amaral and E.~C. Marino, \emph{Canonical quantization of
theories containing fractional powers of the d'Alembertian
operator}, J. Phys. A: Math. Gen. \textbf{25}, 5183--5200 (1992).

\bibitem{c32} D.~G. Barci, L.~E. Oxman, and M. Rocca, \emph{Canonical quantization of
non-local field equations}, Int. J. Mod. Phys. A \textbf{11},
2111--2126 (1996).

\bibitem{c33}  S.~C. Lim and S.~V. Muniandy, \emph{Stochastic quantization of nonlocal
fields}, Phys. Lett. A \textbf{324}, 396--405 (2004).

\bibitem{a}
 S. Albeverio H. Gottschalk and J.-L Wu, \emph{Convoluted generalized
white noise, Schwinger functions and their analytic continuation to
Wightman functions}, Rev. Math. Phys. \textbf{8}, 763--817 (1996).


\bibitem{b} M. Grothaus and L. Streit, \emph{Construction of relativistic quantum
fields in the framework of white noise analysis}, J. Math. Phys.
\textbf{40}, 5387--5405 (1999).

\bibitem{c34} S.C. Lim, \emph{Fractional derivative quantum fields at positive
temperature}, Physica A \textbf{363}, 269--281 (2006).

\bibitem{c35} C.~H. Eab, S.~C. Lim and L.~P. Teo, \emph{Finite temperature Casimir
effect for a massless fractional Klein-Gordon field with fractional
Neumann conditions}, J. Math. Phys. \textbf{48}, 082301 (2007).

\bibitem{c36} M.~E. Peskin and D.~V. Schroeder, \emph{An introduction to quantum field
theory}, (Addison-Wesley, Reading, 1995).

\bibitem{c37} E.~W. Kolb and M.~S. Turner, \emph{The early universe}, (Addison-Wesley,
Reading, 1990).

\bibitem{c38} J.Zinn-Justin, \emph{Quantum field theory and critical
phenomena}, (Clarendon
 Press, Oxford, 4th Edition, 2002).


\bibitem{c39} T. Dauxois and M. Peyrard, \emph{Physics of solitons},
(Cambridge University Press, Cambridge, 2006).

\bibitem{c40} L.~H. Ford and T. Yoshimura, \emph{Mass generation by self--interaction
in non-Minkowskian spacetimes}, Phys. Lett. A \textbf{70}, 89--91
(1979).

\bibitem{c41} D.~J. Toms, \emph{Symmetry breaking and mass generation by space-time
topology}, Phys. Rev. D \textbf{21}, 2805--2817 (1980).

\bibitem{c42} G. Denardo and E. Spallucci, \emph{Dynamical mass generation in $S^1\times R^3$},
Nucl. Phys. B \textbf{169}, 514--526 (1980).


\bibitem{c43} A. Actor, \emph{Topological generation of gauge field mass by toroidal
spacetime}, Class. Quantum Grav. \textbf{7}, 663--683 (1980).

\bibitem{c44} K. Kirsten, \emph{Topological gauge field mass generation by toroidal
spacetime}, J. Phys. A: Math. Gen. \textbf{26}, 2421--2435 (1993).

\bibitem{EK1} E. Elizalde and K. Kirsten, \emph{Topological symmetry breaking in
self-interacting theories on toroidal space-time}, J. Math. Phys.
\textbf{35}, 1260--1273 (1994).

\bibitem{a1}
S.~W. Hawking, \emph{Zeta function regularization of path integrals
in curved space time}, Comm. Math. Phys. \textbf{55}, 139--170
(1977).

\bibitem{E1}
E.~Elizalde, S.~D. Odintsov, A.~Romeo, A.~A. Bytsenko, and
S.~Zerbini,
  \emph{Zeta regularization techniques with applications}, (World Scientific
  Publishing Co. Inc., River Edge, NJ, 1994).

\bibitem{ET}
Emilio Elizalde, \emph{Ten physical applications of spectral zeta
functions},
  (Springer-Verlag,
  Berlin, 1995).


\bibitem{K}
K.~Kirsten, \emph{Spectral functions in mathematics and physics},
(Chapman \&
  Hall/ CRC, Boca Raton, FL, 2002).


\bibitem{EKZ}
E. Elizalde, K. Kirsten and S. Zerbini, \emph{Applications of the
Mellin--Barnes integral representation}, J. Phys. A: Math. Gen.
\textbf{28} (1995), 617--629.


\bibitem{LT1}
S.~C. Lim and L.~P. Teo, \emph{Finite temperature Casimir energy in
closed rectangular cavities: a rigorous derivation based on zeta
function technique},  J. Phys. A: Math. Theor. \textbf{40}  (2007),
11645-11674.

\bibitem{LT3}

S.~C. Lim and L.~P. Teo, \emph{On the minima and convexity of
Epstein Zeta function}, in preparation.


\bibitem{c49} K.~L. Sebastian, \emph{Path integral representation for fractional Brownian motion},
J. Phys. A: Math. Gen. 28  4305 (1995).

\bibitem{c50} C.~H. Eab and S.~C. Lim, \emph{Path integral representation of
fractional harmonic oscillator}, Physica A \textbf{371}, 303--316
(2006).

\bibitem{c51} S.~C. Lim, M. Li and L.~P.Teo, \emph{Locally self-similar fractional
oscillator processes}, Fluct. Noise Lett. \textbf{7}, L169--L179
(2007).

\bibitem{c57} S.~C. Lim and  C.~H. Eab, \emph{Riemann--Liouville and Weyl fractional
oscillator processes}, Phys. Lett. A \textbf{335},  87--93 (2006).

\bibitem{c52} R. Peltier and J. Levy Vehel, \emph{Multifractional Brownian motion:
Definition and preliminary results}, INRIA Report 2645 (1995).

\bibitem{c53} A. Benassi, S. Jaffard and D. Roux, \emph{Elliptic Gaussian random
processes}, Rev. Mat. Ibroamericana \textbf{13}, 19--90 (1997).

\bibitem{c54} C. Lacaux, \emph{Real harmonizable multifractional Levy motions},
Ann. Inst. Henri Poincare-Probab. Stat. \textbf{40}, 259--277
(2004).

\bibitem{c55} S.~C. Lim and L.~P. Teo, \emph{Weyl and Riemann--Liouville
multifractional Ornstein--Uhlenbeck processes}, J. Phys. A: Theo.
Gen. \textbf{40}, 6035--6060 (2007).



\bibitem{Ep1} P. Epstein, \emph{Zur Theorie allgemeiner Zetafunktionen},  Math. Ann. \textbf{56},  615--644 (1903).

\bibitem{Ep2}P. Epstein, \emph{Zur Theorie allgemeiner Zetafunktionen II}, Math. Ann. \textbf{65},
205--216 (1907).

\bibitem{a2} J. Jorgenson and S. Lang, \emph{Complex analytic properties of regularized products},
Lect. Notes Math. \textbf{1564}, (Springer-Verlag, 1993).

\bibitem{a3}
P. Sarnak, \emph{Determinants of Laplacians}, Comm. Math. Phys.
\textbf{110},  113--120 (1987).

\bibitem{a4}M. Spreafico, \emph{Zeta functions, special functions and the Lerch
formula}, Proc. Royal Soc. Ed. \textbf{136A}, 865--889 (2006).

\bibitem{a5}
I. Vardi, \emph{Determinants of Laplacians and multiple Gamma
functions}, SIAM J. Math. Anal. \textbf{19}, 493-507 (1988).

\bibitem{a6} A. Voros, \emph{Spectral functions, special functions and the Selberg
zeta function}, Comm. Math. Phys. \textbf{110}, 439-465 (1987).

\bibitem{d3}
E.~Elizalde and A.~Romeo, \emph{Regularization of general
multidimensional
  {E}pstein zeta-functions}, Rev. Math. Phys. \textbf{1} (1989), no.~1,
  113--128.

\bibitem{d7}
K. Kirsten, \emph{Inhomogeneous multidimensional {E}pstein zeta
functions},
  J. Math. Phys. \textbf{32}, no.~11, 3008--3014 (1991).

\bibitem{d8}
K. Kirsten, \emph{Generalized multidimensional {E}pstein zeta
functions}, J. Math.
  Phys. \textbf{35} (1994), no.~1, 459--470.

\bibitem{d14}
E. Elizalde, \emph{An extension of the {C}howla-{S}elberg formula
useful in
  quantizing with the {W}heeler-{D}e{W}itt equation}, J. Phys. A \textbf{27}, no.~11, 3775--3785
  (1994).

  \bibitem{d15}
E. Elizalde, \emph{Extension of the {C}howla-{S}elberg formula and
applications},
  Group theoretical methods in physics (Toyonaka, 1994), (World Sci. Publ.,
  River Edge, NJ, 1995), pp.~191--194.

\bibitem{d5}
E. Elizalde, \emph{Multidimensional extension of the generalized
{C}howla-{S}elberg
  formula}, Comm. Math. Phys. \textbf{198}, no.~1, 83--95  (1998).



\bibitem{d17}
E. Elizalde, \emph{Zeta functions: formulas and applications}, J.
Comput. Appl.
  Math. \textbf{118}, no.~1-2, 125--142, Higher transcendental functions
  and their applications (2000).

\bibitem{d18}
E. Elizalde, \emph{Explicit zeta functions for bosonic and fermionic
fields on a
  non-commutative toroidal spacetime}, J. Phys. A \textbf{34}, no.~14,
  3025--3035 (2001).

\bibitem{d19}
S.~Chowla and A.~Selberg, \emph{On {E}pstein's zeta function. {I}},
Proc. Nat.
  Acad. Sci. U. S. A. \textbf{35}, 371--374 (1949).

\bibitem{d20}
A. Selberg and S.~Chowla, \emph{On {E}pstein's zeta-function}, J.
Reine
  Angew. Math. \textbf{227}, 86--110 (1967).
\end{thebibliography}
\end{document}